\documentclass[a4paper,11pt]{article}
\pdfoutput=1 
\usepackage{jheppub}

\usepackage{graphicx}
\usepackage{dcolumn}
\usepackage{bm}

\newcommand{\pin}{\par\noindent}

\usepackage[T1]{fontenc} 

\title{Fermionic criticality is shaped by Fermi surface topology: a case study of the Tomonaga-Luttinger liquid
}

\author[a]{Anirban Mukherjee}
\author[a]{Siddhartha Patra}
\author[a,1]{Siddhartha Lal,\note{Corresponding author.}}

\affiliation[a]{Department of Physical Sciences, Indian Institute of Science Education and Research Kolkata,India}

\emailAdd{am14rs016@iiserkol.ac.in}
\emailAdd{sp14ip022@iiserkol.ac.in}
\emailAdd{slal@iiserkol.ac.in}

\abstract{We perform a unitary renormalization group (URG) study of the 1D fermionic Hubbard model. The formalism generates a family of effective Hamiltonians and many-body eigenstates arranged holographically across the tensor network from UV to IR. The URG is realized as a quantum circuit, leading to the entanglement holographic mapping (EHM) tensor network description. A topological $\Theta$-term of the projected Hilbert space of the degrees of freedom at the Fermi surface are shown to govern the nature of RG flow towards either the gapless Tomonaga-Luttinger liquid or gapped quantum liquid phases. This results in a nonperturbative version of the Berezenskii-Kosterlitz-Thouless (BKT) RG phase diagram, revealing a line of intermediate coupling stable fixed points, while the nature of RG flow around the critical point is identical to that obtained from the weak-coupling RG analysis. This coincides with a phase transition in the many-particle entanglement, as the entanglement entropy RG flow shows distinct features for the critical and gapped phases depending on the value of the topological $\Theta$-term. We demonstrate the Ryu-Takyanagi entropy bound for the many-body eigenstates comprising the EHM network, concretizing the relation to the holographic duality principle. The scaling of the entropy bound also distinguishes the gapped and gapless phases, implying the generation of very different holographic spacetimes across the critical point. Finally, we treat the Fermi surface as a quantum impurity coupled to the high energy electronic states. A thought-experiment is devised in order to study entanglement entropy generated by isolating the impurity, and propose ways by which to measure it by studying the quantum noise and higher order cumulants of the full counting statistics.}

\begin{document}

\maketitle
\flushbottom
\tableofcontents
\section{Introduction}
\pin
Electronic correlations in one dimension are generically observed to 
lead to exotic quantum liquids that either remain critical (and belong to the universality class of the Tomonaga Luttinger liquid (TLL)) or possess a gapped spectrum~\cite{gogolin2004bosonization,giamarchi2004quantum}. Careful studies involving a host of analytical and numerical methods reveal a variety of low-energy gapped quantum liquids including the Luther Emery (LE) liquid and the Mott insulating liquid (MI). One dimensional gapped quantum liquids can be classified via their emergent topological properties~\cite{gu2009,pollmann2010,chen2011,schuch2011,guo2011,pollmann2012,
gu2014,montorsi2017,verresen2019,zeng2019}, and are notably different from the tenfold classification of gapped non-interacting systems~\cite{altland1997,chiu2016}. Transitions from the TLL to various gapped phases are described by the Berezinski-Kosterlitz-Thouless universality class. Further, the universality has been shown to extend to a variety of models in 1D, e.g., the Schwinger model, massive Thirring model,  quantum sine-Gordon model etc. Extensive studies have shown that the BKT binding-unbinding transitions of vortices that describe instabilities of the TLL can be studied from the spin and charge vertex operators of the associated sine-Gordon field theory~ \cite{haldane1981luttinger,gogolin2004bosonization}. Topological aspects of the zero mode of these vertex operators~\cite{bogachek1990instanton} are understood from their connection to non-local gauge transformations~\cite{kohn1964theory,lieb1961two} arising from boundary condition changes on the Hilbert space~\cite{nakamura2002lattice}. Indeed, such boundary condition changes are associated with Berry phases that characterize the various phases in the BKT renormalization group (RG) phase diagram~\cite{nakamura2002lattice}. These insights inspire the following questions. Instabilities of a gapless state such as the TLL are outcomes of  divergent quantum fluctuations that begin with scatterings events at the Fermi surface, and lead to the gapping of its neighbourhood. Can we relate such scattering processes to topological properties of the fermionic Hilbert space at the Fermi surface? Can we build a skeletal phase diagram by studying this Fermi surface scattering problem, and if yes, how closely does this resemble the phase diagram obtained from a RG study of the divergent quantum fluctuations? An affirmative answer would indicate that the topological features of degrees of freedom at the Fermi surface can track the low-energy physics arising from UV-IR mixing.
\pin
At the same time, the nature of many-body entanglement analyzed via tensor network methods~\cite{orus2014practical,orus2019tensor,biamonte2017}, e.g., the matrix product state (MPS)~\cite{fannes1992,hastings2007,pollmann2010,schuch2011,fuji2016}
representation for gapped states and multiscale entanglement RG (MERA)~\cite{vidal2003entanglement,corboz2009} for gapless states, provides another important pathway for classifying low dimensional quantum liquids. The entanglement entropy (EE) of critical 1D quantum liquids has also been studied extensively using field theory methods~\cite{calabrese2004,casini2005,holzhey1994,its2005,jin2004}, and is consistent with other calculations for the  EE arising from a finite Fermi surface in higher spatial dimensions that show a violation of the area law~\cite{wolf2006,gioevklich2006,swingle2010entanglement}. Given the extensive developments in the understanding the physics of interacting fermions in 1D, a unified view is needed of the emergence of the effective low-energy Hamiltonians of quantum liquids and the associated entanglement content of their many-particle Hilbert space. This will surely provide deeper insights into the interplay of electronic correlations, low spatial dimensions and many-body entanglement. With this as our goal, we turn to the recently developed unitary  network RG (URG) for fermionic systems~ \cite{anirbanmotti,anirbanmott2,mukherjee2020,anirbanurg1,anirbanurg2}, as this method generates the RG flow of both Hamiltonians and their eigenbasis along the holographic RG direction~\cite{balasubramanian1999,de2000,heemskerk2009,lee2010,
casini2011,heemskerk2011,lee2014} via disentanglement of high-energy electronic states (UV) from their low-energy (IR) counterparts. More specifically, at every step of the URG, one UV degree of freedom is disentangled from the rest. The holographic organization of effective Hamiltonians and entanglement features (e.g., entanglement spectrum, entanglement entropy etc.) from UV to IR is carried out via a collection of unitary gate layers, and reveals the entanglement holographic mapping (EHM)~\cite{qi2013} tensor network representation of URG~\cite{mukherjee2020}. The boundary of the tensor network is in the UV, while the collection of layers comprising the bulk of the network are directed towards the IR.
\pin
In this manner, the URG can be realized as a quantum circuit~\cite{vidal2007,orus2014advances,witteveen2019,mukherjee2020}. Importantly, as the RG can be implemented on a 1D momentum-space lattice, we aim to clarify the role played by the topological properties of electronic states near the Fermi surface in guiding the RG flow towards a variety of gapped or gapless fixed points. The interplay between Fermi surface topology and electronic correlations has been recently demonstrated by us for the nested Fermi surface of the tight-binding model on the square lattice in a study of the 2D Hubbard model~\cite{anirbanmotti,anirbanmott2}. There, we took careful account of the singular nature of the Fermi surface, as it contained van Hove singularities at half-filling. Here, we aim is to study the simplest possible Fermi surface, i.e., the two-point Fermi surface of the tight-binding chain, in the hope of gleaning insights on higher dimensional systems with regular connected (non-singular)~\cite{anirbanurg2} or Dirac point-like~\cite{pal2019} Fermi surfaces. Even as we apply the URG to the 1D fermionic Hubbard Hamiltonian, our conclusions are relevant to the case of translation invariant four-fermionic interactions. Further, our analysis neither requires the linearisation of the electronic dispersion near the Fermi surface, nor relies on any emergent phenomenon such as the separation of charge and spin excitations.
\par\noindent
A variety of tensor network RG methods have been dicussed in the literature (see Ref.\cite{orus2019tensor} for a review), and we review briefly here those among them that are connected to the URG formalism. 
MERA is tensor network comprised of a collection of pairs of layers of quantum circuits: (i) the unitary gate layer has unit depth and is composed of a direct product of two-local unitary gates followed by, (ii) layer of isometries that removes the disentangled qubits from the bulk of the tensor network. The latter feature of MERA is distinct with regards to the EHM~\cite{qi2013} discussed earlier. By contrast, every unitary gate layer of the URG has a finite depth~\cite{mukherjee2020}, with every sublayer composed of a two-qubit gate that disentangles a given UV degree of freedom (with one electronic state) from all others, such that its quantum number becomes integral of motion. The finite depth feature is, however, similar to  the deep MERA (dMERA) construction of Ref.\cite{kim2017}. Unlike the variational determination of the parameters of the unitary gates in MERA~\cite{evenbly2009} and dMERA~\cite{kim2017}, there exists an exact construction of the unitary disentangler within the URG framework in terms of the electronic Hamiltonian~\cite{anirbanurg1}. Various real space renormalisation approaches similar to URG, in the sense that they lead to the iterative block diagonalization of Hamiltonian, have appeared in Refs.\cite{you2016entanglement,rademaker2016explicit,
monthus2016flow,rademaker2017many,ma1979sk,
aoki1982decimation,white2002numerical}. An important property of the URG tensor network~\cite{mukherjee2020} is that it satisfies the Ryu-Takayanagi bound~\cite{ryu2006aspects,ryu2006} for the entanglement entropy ($S$): this is a statement of the observation that $S(R)$ generated upon isolating a region $R$ located at the boundary of the tensor network (the UV) from the bulk (the IR) is bounded by the number of links connecting it to the rest of the system. This feature is also shared by other entanglement RG methods such as MERA~\cite{swingle2012b,swingle2012a} and EHM~\cite{qi2013}, providing a connection between entanglement scaling and holographic duality.  
\par\noindent 
The URG procedure discussed earlier generates a family of RG scale dependent effective Hamiltonians, from which we can extract the RG flow of various two-, four-, six-point and higher order vertices~\cite{anirbanmotti,anirbanmott2,anirbanurg1,anirbanurg2}. Further, the RG procedure reveals a multitude of energy scales ($\omega$) for quantum fluctuations that arise out of the non-commutivity between various terms of the Hamiltonian. Importantly, the RG equations of the renormalized vertices are nonperturbative in nature, as they involve 
the resummation of loops to all orders to obtain closed form analytic expressions~\cite{anirbanurg1}. The resummation process generically leads to denominators in the RG equations that are dependent on  the quantum fluctuation scale $\omega$, renormalized self- and correlation energies. In a study of the 2D Hubbard model in Ref.\cite{anirbanmotti}, we have also established an equivalence of the URG at weak coupling to the functional RG formalism~\cite{metzner2012}, and shown the origin of BCS, ZS, ZS$'$ diagrams~\cite{shankar1994} in our vertex RG equations. As a direct consequence of the nonperturbative RG equations, we obtain stable fixed points as a function of fluctuation scale $\omega$. Upon reaching gapped phases at stable fixed points, the effective Hamiltonian are written in terms of composite operators formed out of pairs of electronic states, and describe the condensation of these emergent degrees of freedom. Importantly, we have also demonstrated quantitative checks for the URG by benchmarking with high accuracy the ground state energy and double occupancy of the 2D Hubbard model at, and away from, half-filling~\cite{anirbanmotti} against various numerical methods~\cite{leblanc2015solutions,ehlers2017hybridDMRG,dagotto1992}. 
\pin
This leads us to ask the following questions in the present work. First, can the URG procedure show the emergence of nonlocal constraints associated with the Hilbert space topology of degrees of freedom residing at (and near the) Fermi surface, and how this leads to the condensation of composite degrees of freedom? As our starting point is a lattice version of the sine-Gordon model, we expect that an affirmative answer to this question will likely yield a non-perturbative version of the BKT phase diagram. The BKT phase diagram has separatrices that separate relevant from irrelevant flows, while the flows along the separatrices meet at a critical point~\cite{wilson1975renormalization}. Is it possible to reconcile the partitioning of the coupling space diagram obtained from analysis of topological objects living at the Fermi surface with the partitioning of the BKT RG phase diagram via the separatrices? Can we diagonalise the effective Hamiltonians obtained from the stable fixed points of the URG flow to obtain the low-energy spectrum and the associated eigenstates? If yes, we can perform a reverse URG on by the low-energy eigenstates of the fixed point Hamiltonian by re-entangling the UV degrees of feedom, generating thereby many-body eigenstates at higher energy scales. In Ref.\cite{anirbanmotti,mukherjee2020}, we have already demonstrated this scheme for the 2D Mott liquid (ML) ground state of the 2D Hubbard model, as well as its parent Marginal Fermi liquid (MFL) metallic state. In this way, we aim to obtain the EHM tensor network representation of the URG, and study the entanglement RG for the various phases in the BKT RG phase diagram. Will the critical and gapped fixed points reveal distinct entanglement entropy scaling features from such a study? We verified the Ryu-Takayanagi entanglement entropy bound~\cite{ryu2006,ryu2006aspects,mukherjee2020} for the ML and MFL ground states of the 2D Hubbard model, concretizing the connection between entanglement renormalization and holographic duality for the unitary quantum circuit/tensor network of the URG. In a similar fashion, can we construct the entanglement holographic mapping tensor network for the entanglement RG flow of correlated electrons in 1D, and verify the entropy bound in this case as well? Finally, can we provide a quantum circuit description of the RG flow in terms of two-local unitary disentangling maps? While a quantum circuit model for the entanglement scaling of massless Dirac fermions~\cite{witteveen2019} was recently achieved in the continuum field theory for gapless phases, we aim instead to construct quantum circuit models for both gapped as well as gapless phases arising from a Hamiltonian of lattice-based electrons.
\par\noindent 
We now present an outline of the rest of the work. In Sec.\ref{sec2}, we review the symmetries and topological properties of the Hilbert space for the degrees of freedom at the Fermi surface of the 1D tight binding chain. We follow this in Sec.\ref{sec3} by showing that, in the presence of a Fermi surface instability arising from the presence of inter-particle interactions, a topological property of the Fermi surface Hilbert space (the first Chern class) imposes a constraint on the condensation of four-fermion vertices, involving the formation of composite pseudospin $S=1/2$ degrees of freedom formed by the pairing of fermions. This is followed in Sec.\ref{sec4} by showing that the pseudospin backscattering vertices connecting the Fermi points $\pm k_{F}$ lead to a change in the c.o.m Hilbert space by satisfying a constructive interference condition for scattering processes. We then implement the URG formalism for the case of correlated electrons on the 1D tight-binding chain in Sec.\ref{sec5}, and demonstrate how the topological features of the Fermi surface Hilbert space guide the RG flows. In Sec.\ref{secholo}, we study the URG flow of the many-particle entanglement and unveil the role played by the Fermi surface in distinguishing between flows to gapless and gapped quantum liquid phases. We study dynamical spectral weight redistribution between the fundamental and emergent degrees of freedom in Sec.\ref{sec6}, showing the manifest unitarity of the RG formalism by tracking the flow of the Friedel's phase shift from the scattering matrix~\cite{2015lecturesweinberg}. In Sec.\ref{sec8}, we unveil signatures of topological order in some of the gapped quantum liquids attained from the URG flow, and design a thought experiment to track its features of from the properties of the degrees of freedom at the (erstwhile) Fermi surface. We end by summarising our results in Sec.\ref{sec9}, and presenting some future directions.
\section{Symmetries and Topology of the Fermi surface}\label{sec2}
We recapitulate the symmetries and topological properties of the Fermi volume and Fermi surface of non-interacting electrons in one spatial dimension in this section, as these will be critical in dealing with the case of interacting electrons in later sections. To begin with, the 1D tight binding Hamiltonian in momentum-space for spinful electrons is given by~\cite{Bloch1933}
\begin{equation}
H_{t}=-2t\sum_{k\sigma}\cos(k)~c^{\dagger}_{k\sigma}c_{k\sigma}~,
\end{equation}
and leads to the dispersion spectrum $E_{k\sigma}=-2t\cos k$ . The parity operation (P) $Pc^{\dagger}_{k\sigma}P^{\dagger}=c^{\dagger}_{-k\sigma},P\in Z_{2}$ and time reversal transformation (T) $Tc^{\dagger}_{k\uparrow}T^{\dagger}=c^{\dagger}_{k\downarrow},T\in Z_{2}$ leave the Hamiltonian invariant 
\begin{equation}
PH_{t}P^{\dagger}=H_{t}~,~TH_{t}T^{\dagger}=H_{t}~.
\end{equation}
Accordingly, the dispersion has the following symmetries: $E_{k\sigma}=E_{-k\sigma}$~,~ $E_{k\uparrow}=E_{k\downarrow}$. The Hamiltonian $H_{t}$ commutes with the number operator $\hat{N}=\sum_{k\sigma}\hat{N}_{k\sigma}$, $\left[H_{t}, \hat{N} \right] = 0$, admitting a global $U(1)$ phase-rotation symmetry of the Hamiltonian under the unitary operation $U(\theta)\in U(1)$:~$U(\theta)c^{\dagger}_{k\sigma}U^{\dagger}(\theta)=e^{i\theta}c^{\dagger}_{k\sigma}$. At zero temperature and with a chemical potential $\mu$, a description of the low-energy fermionic excitations about the sharp Fermi surface is obtained by linearisation of the dispersion around the two Fermi points, $k_{F}$ and $-k_{F}$:~$E_{ak}=-2t\cos k -E_{F}\approx v_{F}(k\pm k_{F})$~, where the Fermi energy and velocity are $E_{F}=-2t\cos k_{F}, v_{F}=-2t\sin k_{F}$ and where $a=R/L$. 
\pin
The retarded single-particle propagator (Green's function) for states in the window $W_{\Lambda_{0}}:[-\Lambda_{0}/\hbar v_{F},\Lambda_{0}/\hbar v_{F}]$ around the Fermi points ($-k_{F},k_{F}$) is given by 
\begin{equation}
G^{ret}_{a\sigma}(z,k)=(z-E_{ak})^{-1},Im(z)>0~,\label{retarded_green_function}
\end{equation}
and the advanced propagator by its complex conjugate: $G^{adv}_{a\sigma}(\bar{z},k)=(G^{ret}_{a\sigma}(z,k))^{*}$. The retarded Green's function has poles in the complex frequency plane at $z=\omega_{re}+i\omega_{im}$~, where $\omega_{im}=0+$, $\omega_{re,L}=v_{F}(k -k_{F})$ and $\omega_{re,R}=v_{F}(k +k_{F})$. For all non-zero momenta, the poles appear in pairs $(L,R)$ for $(\uparrow,\downarrow)$ spin-states due to the $Z_{2}^{P}\times Z_{2}^{T}$ symmetry mentioned above. For states at the Fermi energy ($k_{F},-k_{F}$), the poles appear at the points $z_{L}=0=z_{R}$. At this point, we recall a topological invariant ($N_{1}$) associated with number of states at the Fermi energy~\cite{luttinger1960fermi,volovik2009universe}. This arises from the $E=0$ poles of the non-interacting single-particle Green's function (eq.\eqref{retarded_green_function})
\begin{eqnarray}
N_{1}&=&\int d\omega \partial_{\omega}Tr(\ln\hat{G}_{0})~=~ Tr(\ln\hat{G}_{0}(0+i\eta))-Tr(\ln\hat{G}_{0}(0-i\eta))~.~\label{volovik's invariant}
\end{eqnarray}
For the 1D Fermi surface, $N_{1}=4$ (accounting for spin degeneracy).
\pin
We can represent the Hamiltonian of non-interacting electrons as a sum over a set of sub-Hamiltonians, each of which governs a group of four states $\tilde{k}=(k ,-k)\otimes (\uparrow,\downarrow)$ as follows
\begin{equation}
H_{t}=\sum_{\tilde{k}}H_{\tilde{k}}~,~H_{\tilde{k}}=E_{\tilde{k}}m^{\dagger}_{\tilde{k}}I_{4}m_{\tilde{k}}~,
\end{equation}
where $m_{\tilde{k}}=[c_{k\uparrow} ,c_{k\downarrow},c_{-k\uparrow},c_{-k\downarrow}]^{t}$, $E_{\tilde{k}}=E_{k\sigma}$, $I_{4}$ is the $4\times 4$ unit matrix, $I_{4}=I_{2}^{s}\times I_{2}^{L/R}$ and the two $I_{2}$ unit matrices are Casimir invariants of the $SU(2)$ groups for the spin (s) and chirality (L/R) sectors respectively. For the poles at $z=0$, the Fermi energy sub-Hamiltonian $H_{k=0}=0$ possesses a further particle-hole symmetry ($C \in Z_{2}$). This unites with the number conservation $U(1)$ symmetry, leading to an enhanced $SU_{C}(2)$ symmetry for states at the Fermi surface. Thus, the complete symmetry group for the two Fermi points is seen to be $SU_{s}(2)\times SU_{L/R}(2)\times SU_{C}(2)$. 
\pin
The $S=1/2$ representation of symmetry group SU(2) is associated with a topological $CP_{1}$ space, i.e., the Bloch sphere. The rotations on the Bloch sphere is generated by the set of unitary operations $U=aI+\boldsymbol{\sigma}\cdot\mathbf{n}$, where $\boldsymbol{\sigma}$ is the S=1/2 representative of the SU(2) group and the constraint $a^{2}+|\mathbf{n}|^{2}=1$ represents the compact space $S_{3}$. The homotopy group of the compact space SU(2) (with a constraint $\mathbf{n}^{2}=1$) is given by $\pi_{3}(SU(2))=Z$. The topological invariants of this space are Chern invariants. The existence of the Fermi surface as a sharp boundary in energy-momentum space at $T=0K$ can then be seen as the $N_{1}$ topological invariant (eq.\eqref{volovik's invariant}) arising out of a pole of the single-particle Green's function at zero frequency~\cite{volovik1988analog}. Further, $N_{1}$ is associated with a non-trivial homotopy group $\pi_{1}(S_{1})=Z$ characterizing the winding around the sharp Fermi surface.
\pin
A pole in the complex frequency plane of the single-particle Green's function $G^{ret}_{\sigma a}(z,k)$ at a given momentum $k$ near the left or right Fermi point ($a=L/R$) is also associated with a residue. This residue is picked up by the phase field $\Phi(k,z)$ of the single-particle Green's function $G(k,z)=|G(k,z)|e^{i\Phi(k,z)}$ by traversing a non-contractible closed path $P \equiv S_{1}$ around the singularity.  The number of electrons ($n_{\sigma, ak}$) $n_{\sigma, ak}\in Z$ at a given state $k\sigma$ belongs to the homotopy group of $P$, $\pi_{1}(S(1))=Z$, such that    
\begin{eqnarray}
n_{\sigma, a k}&=&\frac{1}{2\pi i}\int dz~ G^{ret}_{\sigma a}(z,k)~\partial_{z}\left[G^{ret}_{\sigma a}(z,k)\right]^{-1}~.\label{invariant}
\end{eqnarray}
The orientation of the two Fermi points~\cite{atiyah1963index,haldane2005luttinger} is defined as an integral over the momentum vector at $T=0K$ for a sharp Fermi surface within a window $2\Lambda_{0}/\hbar v_{F}$
\begin{eqnarray}
\nu_{\sigma,R}&=& \frac{1}{2\pi i}\int_{k_{F}-\frac{\Lambda_{0}}{\hbar v_{F}}}^{k_{F}+\frac{\Lambda_{0}}{\hbar v_{F}}} dk \frac{dn_{\sigma,R}(k)}{dk}~=-1~,~\nu_{\sigma, L}= \frac{1}{2\pi i}\int_{-k_{F}-\frac{\Lambda_{0}}{\hbar v_{F}}}^{-k_{F}+\frac{\Lambda_{0}}{\hbar v_{F}}} dk \frac{dn_{\sigma, L}(k)}{dk}~=~1~.\label{boundary_integral}
\end{eqnarray}
Using eq.\eqref{invariant} in eq.\eqref{boundary_integral}, along with compactifying the boundaries of the momentum-space $[\frac{-\Lambda}{\hbar v_{F}},\frac{\Lambda}{\hbar v_{F}}]$ and frequency-space $[-\infty +0i,\infty +0i]$ windows, the Fermi point singularities lead to first class Chern invariants on the frequency-momentum torus 
\begin{eqnarray}
\nu_{a\sigma}&=&\frac{1}{2\pi i}\oint dk\oint d\omega ~ Tr\bigg[\hat{G}_{\sigma a}(\omega ,k)\partial_{k}\hat{G}^{-1}_{\sigma a}(\omega ,k)\hat{G}_{\sigma a}(\omega ,k)\partial_{\omega}G^{-1}_{\sigma a}(\omega ,k)\bigg]~.\label{atiyah-singer-index}
\end{eqnarray}
It can be seen that $\nu_{\sigma L,R}=\pm 1\times C$, where $C=2S=1\in\pi_{3}(SU(2))$. The Atiyah-Singer index~\cite{atiyah1963index} is then obtained as $\nu_{\sigma L}-\nu_{\sigma R}=2$, and can be related to the net axial current ($\Delta J^{5}$) across the gapless Fermi points ($k_{F},-k_{F}$) in the presence of an electric field ($E$) 
\begin{eqnarray}
\Delta J^{5}_{\sigma}&=&J^{5}_{\sigma}(\frac{\Lambda_{0}}{\hbar v_{F}})-J^{5}_{\sigma}(-\frac{\Lambda_{0}}{\hbar v_{F}})=eE~,~ J^{5}_{\sigma}=J_{\sigma,L}-J_{\sigma,R}~,\nonumber\\
J_{\sigma,a}&=&eG_{a\sigma}^{-1}(k,\omega)\partial_{k}G_{a\sigma}(k,\omega)~,~(a=L/R)~,\label{ward-identity-breakdown}
\end{eqnarray}
where $e$ is the charge of the electron and $\Delta J^{5}$ is related to the anomaly of the axial charge $Q_{5}$~\cite{manton1985schwinger} 
\begin{equation}
\Delta J^{5}=\frac{dQ_{5}}{dt}=e\frac{d}{dt}(\int dx (\psi^{\dagger}_{L}\psi_{L}-\psi^{\dagger}_{R}\psi_{R}))~=~ \nu_{\sigma L}-\nu_{\sigma R}~.\label{axial anomaly} 
\end{equation}
\pin
For fermions on a 1D lattice, the axial current is generated by momentum imparted to the center of mass (c.o.m) spin ($s$) and charge ($c$) degrees of freedom (d.o.f) with positions $X_{s}=X_{\uparrow}-X_{\downarrow},X_{c}=X_{\uparrow}+X_{\downarrow}$ respectively by a twist operation~\cite{oshikawa1997magnetization,kohn1964theory, yamanaka1997nonperturbative}. Note that $X_{\sigma}=1/(2N)\sum_{j=1}^{N}j\hat{n}_{j\sigma}$(here $\hat{n}_{j\sigma}=\psi^{\dagger}_{j\sigma}\psi_{j\sigma}$, and we have kept a factor of $2$ for the spin degeneracy. The twist operator is defined in terms of the following unitary operations on the c.o.m 
\begin{equation}
\hat{O}_{s}=U_{\uparrow}U^{\dagger}_{\downarrow},~\hat{O}_{c}=U_{\uparrow}U_{\downarrow}~,~\label{twist_op_s_c}
\end{equation}
where $U_{\sigma}=\exp\left[i2\pi\phi X_{\sigma}\right]$ leads to twisted boundary conditions (TBC) in real space. Thus, the spin twist operator $\hat{O}_{s}$ changes boundary conditions for $\uparrow$ and $\downarrow$ fermions with opposite phases $\hat{O}_{s}\rightarrow \psi_{\uparrow}(j=N)=e^{i\pi\phi}\psi_{\uparrow}(j=0)~,~\psi_{\downarrow}(j=N)=e^{-i\pi\phi}\psi_{\downarrow}(j=0)$, causing a shift in center of mass momentum ($P_{cm}$) due to spin excitations of the Fermi sea 
\begin{equation}
\hat{O}_{s}\hat{T}\hat{O}_{s}^{\dagger}=Te^{i\phi (\pi/N)(NS-S^{z}_{tot})}~,
\end{equation}
where $S^{z}_{tot}=\sum_{j}S^{z}_{j}$ is the $z$-component of the total spin angular momentum and $\hat{T}=\exp\left[iP_{cm}\right]$ is the global translation operator. On the other hand, the charge twist operator $\hat{O}_{c}$ causes an equal amount of twist for $\uparrow$ and $\downarrow$ fermions, leading to a gain in momentum for the center of mass of the Fermi sea 
\begin{equation}
\hat{O}_{c}\hat{T}\hat{O}_{c}^{\dagger}=\hat{T}e^{i\phi(\pi/N)\hat{N}_{e}}~,
\end{equation}
where the total no. of electrons is given by $\hat{N}_{e}=\sum_{j\sigma}\hat{n}_{j\sigma}$~. The non-commutativity between $T$ and the twist operators $O_{c/s}$ defines Wilson loop operators on a two-tori~ \cite{altman1998haldane,lieb1961two, yamanaka1997nonperturbative} $ (\phi,X_{c,s})\mapsto \mathcal{T}^{2}\equiv S_{1}\times S_{1}$
\begin{equation}
\hat{Z}_{c/s}=\hat{O}_{c/s}\hat{T}\hat{O}_{c/s}^{\dagger}\hat{T}^{\dagger}\label{wilson_loop_operator}
\end{equation}
where $\log\hat{Z}_{c}=i\hat{N}_{e}/2N~,~\log\hat{Z}_{s}=i\hat{S}^{z}_{tot}/N-S$. The action of the Wilson loop operators $\hat{Z}_{c/s}$ on the c.o.m. state  $|P_{cm}=0,S^{z}_{tot}\rangle$ gives 
\begin{equation}
\hat{Z}_{c/s}|P_{cm}=0,S^{z}_{tot}\rangle = e^{i2\pi\nu_{c/s}}|P_{cm}=0,S^{z}_{tot}\rangle~,
\end{equation}
where $\nu_{c} =N_{e}/2N$ and $\nu_{s}=S^{z}_{tot}/N$ are the net flux in the $c$ and $s$ sectors respectively emanating from $\mathcal{T}^{2}$. By taking a state $|P_{cm}=0,S^{z}_{tot}\rangle$ around in a local patch (P) given by $P\equiv \lbrace \phi\in (0,\alpha_{1}),\hat{X}_{c/s}\in (0,\alpha_{2})\rbrace$, we obtain the twisted state 
\begin{equation}
|\Psi(\alpha_{1},\alpha_{2})\rangle =\hat{T}_{\alpha_{2}}\hat{O}_{c/s\alpha_{1}}|P_{cm}=0,S^{z}_{tot}\rangle~,
\end{equation}
where $\hat{T}_{\alpha_{2}}\hat{X}_{c/s}\hat{T}^{\dagger}_{\alpha_{2}}=\hat{X}_{c/s}+\alpha_{2}$ and $\hat{O}_{\alpha_{1}}\hat{P}_{cm}\hat{O}^{\dagger}_{\alpha_{1}}=\hat{P}_{cm}+\alpha_{1}$. 
\begin{figure}[h!]
\begin{center}
\includegraphics[scale=1.5]{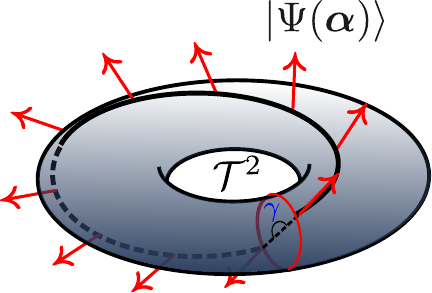}
\end{center}
\caption{The red vectors on the torus $\mathcal{T}^{2}$ represent the eigenstates $|\Psi(\boldsymbol{\alpha})\rangle$ that form a M\"obius strip for the centre of mass (c.o.m.) Hilbert space, where the angle between initial and final vector is $\gamma =\pi$.}\label{torus}
\end{figure}
In the space $\boldsymbol{\alpha}=(\alpha_{1},\alpha_{2})$, we define the persistent current vector $\mathbf{J}_{c/s\boldsymbol{\alpha}}=\langle\Psi(\boldsymbol{\alpha})|\boldsymbol{\nabla}|\Psi(\boldsymbol{\alpha})\rangle$, such that $\mathbf{J}_{c/s\boldsymbol{\alpha}}=(J_{1c/s\boldsymbol{\alpha}},J_{2c/s\boldsymbol{\alpha}})$ represents the zero mode of the charge/spin current generated on $\mathcal{T}^{2}$ due to TBC. The first Chern class~\cite{tao1986impurity,niu1984quantised} on the torus can now be computed in terms of the curl of the nonlocal current vector in the $\boldsymbol{\alpha}$ space
\begin{eqnarray}
\gamma &=&\frac{1}{2i}\int_{0}^{2\pi} \int_{0}^{2\pi}d^{2}\boldsymbol{\alpha} ~\hat{n}\cdot\boldsymbol{\nabla}\times\mathbf{J}_{c/s\boldsymbol{\alpha}} = 2\pi\frac{p}{q}~.\label{1stchernclass}
\end{eqnarray}
As shown in Fig.\ref{torus}, the ray field $|\Psi(\boldsymbol{\alpha})\rangle$ (represented by the red vector) gathers a phase as it traverses the torus $\mathcal{T}^{2}$ along a M\"obius strip. In the above relation, $p/q=(S-M)$ is the Luttinger's sum for spin excitations of the Fermi sea~\cite{lieb1961two,oshikawa1997magnetization} (where $S$ and $M$ are the total spin quantum number and magnetisation of the Fermi sea respectively), while $p/q=\nu$ is the Luttinger's sum for charge excitations~\cite{luttinger1960fermi,luttinger1960ground,
oshikawa2000commensurability,oshikawa2000topological}. In the presence of both time-reversal and particle-hole symmetries, $\nu=1/2$ (i.e., $M=0,N_{e}=N/2$) such that $\gamma =\pi$. 
\pin
We can now compute experimentally measurable spin/charge Drude conductivities $D_{c/s}$ from spectral flow arguments~\cite{kohn1964theory,scalapino1993insulator}. For this, we note that  the eigenvalue ($E_{c/s}(\phi)$) of the twisted tight-binding Hamiltonian is given by
\begin{equation}
\hat{O}_{c/s}\hat{H}_{t}\hat{O}^{\dagger}_{c/s}|\Psi\rangle=E_{c/s}(\phi)|\Psi\rangle~.
\end{equation}
The spin/charge Drude conductivity $D_{c/s}$~ \cite{kohn1964theory,scalapino1993insulator} can now be computed from the energy curvature $d^{2}E_{c/s}/d\alpha_{1c/s}^{2}=\hat{x}.\Delta \mathbf{J}(\alpha_{1},2\pi) /\Delta\alpha_{1c/s}$ (i.e., the change in persistent current in the ring $\hat{x}.\mathbf{J}(\alpha_{1},2\pi)=e\Delta P^{c/s}_{cm}$ under the twisting of boundary conditions)
\begin{eqnarray}
\hspace*{-0.2cm}
D_{c/s}=\lim_{\omega\rightarrow 0}\omega\sigma''_{s/c}(\omega)=\frac{d^{2}E_{c/s}}{d\alpha_{1c/s}^{2}}=e\frac{\Delta P^{c/s}_{cm}}{\Delta\alpha_{1c/s}}=e\frac{\gamma}{2\pi},
\end{eqnarray}
where $\sigma''_{s/c}(\omega)$ is the imaginary part of the longitudinal conductivity, $\omega$ the probing frequency, and the persistent current momentum generated in the ring is given by 
\begin{equation}
J_{c/s} \equiv e\Delta P^{c/s}_{cm}=e\hat{O}_{c/s}P_{cm}\hat{O}_{c/s}^{\dagger}-eP_{cm}~. 
\end{equation}
This relation shows that the Drude conductivity is related to the first Chern class $\gamma$ (eq.\eqref{1stchernclass}). Further, it can be shown that a net persistent current $\Delta J_{1c/s}=J_{1c/s}(\Phi_{0},0)-J_{1}(0,0)$ is accumulated by taking the c.o.m on a one-flux quantum circuit of $\phi=\Phi_{c/s}/\Phi_{0}\in [0,1]$ (where $\Phi_{c/s}$ is the AB flux), and is connected to the Atiyah-Singer index computed from the Chern numbers (eq.\eqref{atiyah-singer-index}) for the states $(k_{Fc/s},-k_{Fc/s})$ 
\begin{eqnarray}
\frac{1}{k_{Fc/s}}\Delta J_{1c/s}=(\nu_{L}-\nu_{R})=2 \equiv \Delta J^{5}~.
\label{bulk current boundary current}
\end{eqnarray}
In the above, $k_{Fc/s}$ represents the Fermi momentum of the charge ($k\uparrow,k\downarrow$) and spin ($k\uparrow, -k\downarrow$) excitations of the Fermi sea.
\begin{figure}[h!]
\centering
\includegraphics[scale=1]{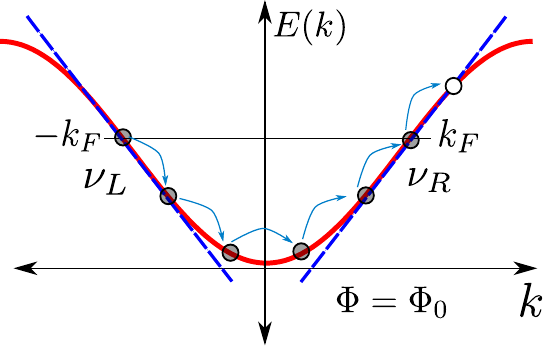}
\caption{Chern numbers $\nu_{L}$ and $\nu_{R}$ around the two Fermi energy points $\pm k_{F}$ of the 1D tight-binding electronic chain take part in the spectral flow of the c.o.m spectrum upon changing boundary conditions by a full flux quantum $\Phi=\Phi_{0}$. 
}\label{spectral_flow}
\end{figure}
\pin
The same thought experiment can also be carried out with the persistent current $J_{2}$, i.e., by now taking the c.o.m on a complete circuit of $X_{c/s}\in [0,2\pi]$. The above relation depicts that the net current accumulated by the c.o.m. is associated with an anomaly in the bulk of Fermi sea: the injection of one electron from the left Fermi point ($\nu_{L}=-1\equiv -C$), together with the ejection of one electron from the right Fermi point ($\nu_{R}=1\equiv C$), leads to a net transfer of $2C=2$ states upon tuning through a complete circuit. At the same time, as shown in Fig.(\ref{spectral_flow}), the relation can also be interpreted as the action of independent monopole sources at the L/R Fermi points (the Chern nos. $C$) in leading to an anomalous current $\Delta J^{5}$. This is the well-known phenomenon of anomalous axial-symmetry breaking~\cite{bell1969pcac,adler1969axial}, seen either from the perspective of the c.o.m or that of the Fermi surface. When taken together, these two anomalies cancel and  the apparent anomalous breaking of the symmetry is restored. Finally, the Atiyah-Singer index is directly related to the central charge ($c$) of the conformal field theory for a gapless 1D electronic system~\cite{stone1991edge,calabrese2004entanglement,swingle2010entanglement},~$c=\frac{1}{2}\left(\nu_{L}-\nu_{R}\right)$~. 
\pin
While the results presented in this section are for the case of a  two-point Fermi surface, the formalism adopted by us is equally applicable to Fermi surfaces of higher dimensional systems of noninteracting electrons. Further, the notion of adiabatic continuity guarantees the existence of a Fermi surface in Landau's formulation of the Fermi liquid even in the presence of electronic correlations. It is, thus, possible to explore various topological properties of the Fermi liquid in an analogous manner. In what follows, many of the ideas introduced in this section will be shown to be useful in understanding the interplay of interactions, symmetry and topology in shaping the many-body instabilities of the Fermi surface. Specifically, we will see in Sec.\ref{sec3} that, in the presence of a Fermi surface instability arising from the presence of inter-particle interactions, the first Chern class $\gamma=\pi$ imposes a topological constraint on the condensation of four-fermion vertices. This involves the formation of composite pseudospin $S=1/2=\gamma/2\pi$ d.o.fs formed by pairing $1/S=2$ fermions. We will follow this up in Sec.\ref{sec4} by showing that the pseudospin backscattering vertices connecting the Fermi points $\pm k_{F}$ lead to a change in the c.o.m Hilbert space by satisfying the constructive interference condition $2\gamma =2\pi$. 
\section{Topological constraints on condensation}\label{sec3}
\begin{figure}[h!]
\centering
\includegraphics[scale=1]{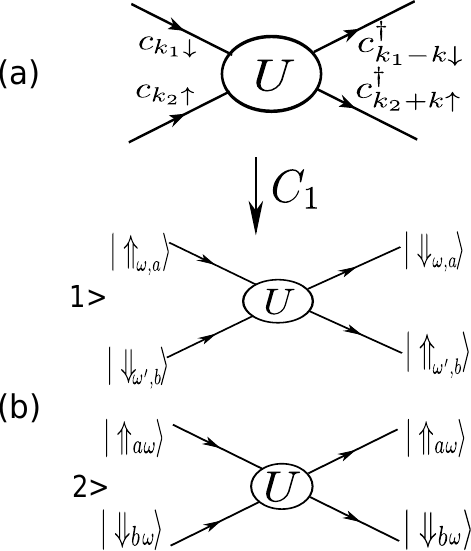}
\caption{(a) The four-fermionic vertex (designated by the $U$ within the oval) is composed of $c,c^{\dagger}\in\mathcal{F}_{4N}$. The constraint $C_{1}$ condenses pairs of fermions forming $S=1/2$ pseudo-spin $\mathbf{A}_{\omega_{0},\eta}\in SU(2)$. (b) The pseudo-spins scatter via spin-flip (1>) and non-spin-flip vertices (2>). }\label{Hilbert space Morphing}
\end{figure}

We will now look into the topological features that arise out of various instabilities of the Fermi surface (e.g., either the spin backscattering or the Umklapp scattering across the Fermi surface~\cite{giamarchi2004quantum}) upon adding the Hubbard term (with on-site repulsion strength $U$) to the tight-binding Hamiltonian 
\begin{equation}
H= -2t\sum_{k,\sigma}\cos(k)~c^{\dagger}_{k\sigma}c_{k\sigma} + U\sum_{i}n_{i\uparrow}n_{i\downarrow}~,
\end{equation}
where the local electronic density is $n_{i\sigma}=\psi^{\dagger}_{i\sigma}\psi_{i\sigma}$~. The Hubbard term has a spin backscattering vertex $c^{\dagger}_{k_{1}+q\uparrow}c^{\dagger}_{-k_{1}-q\downarrow}c_{-k_{1}\uparrow}c_{k_{1}\downarrow}~$, where $c^{\dagger}_{k\sigma}=1/\sqrt{2N}\sum_{j}e^{ikj}\psi^{\dagger}_{j\sigma}$. This scattering process operates in the low-energy subspace with zero momentum and opposite spin pairs respecting P and TR symmetries, as seen from the application of the following constraint on the many-body fermionic Fock space~\cite{anderson1958random}
\begin{equation}
C_{1}:~\hat{n}_{k\sigma}=\hat{n}_{-k-\sigma}~.\label{constraint_ang_mom}
\end{equation} 
\pin
This constraint can be classified in terms of matching of helicity ($\eta=sgn(k)sgn(\sigma)=\pm 1$) across the Fermi surface. Starting with the $SU(2)_{s}\otimes SU(2)_{L/R}\otimes U(1)_{C}$ symmetric many-body fermionic Fock space of the metal (see Sec.\ref{sec2}), the matching of helicities leads to a locking of spin $SU(2)_{s}$ and chirality $SU(2)_{L/R}$ projected Hilbert spaces (PHS) to form a new helicity $SU(2)_{s}\otimes SU(2)_{L/R}\rightarrow SU(2)_{\eta=\pm 1}$ PHS for the emergent degrees of freedom whose condensation characterizes the instability. Time reversal symmetry is respected via the formation of paired-fermion states ($N_{co}=2$) that are either occupied or unoccupied. The constraint $C_{1}$ ensures that each paired-fermion state is equivalent to a two-level system (i.e., isomorphic to $CP^{1}$). It is important to note the topological origin of $C_{1}$: the Chern class $\gamma =\pi$ (eq.\eqref{1stchernclass}) defining the c.o.m PHS topology of the metal is connected to constraint $C_{1}$ via the relation 
\begin{equation}
|\hat{n}_{k\sigma}+\hat{n}_{-k-\sigma}-1|=2\gamma/2\pi=N_{co}/2~.\label{topological_constraint}
\end{equation}
As a consequence of the constraint $C_{1}$, $\gamma=2\pi S$ leads to formation of SU(2) $S=1/2$ Hilbert spaces from the original many-body fermionic Fock space. Due to TRS, every $\uparrow$ state has a partner $\downarrow$. Hence, the $4N$ particle antisymmetrized Hilbert space of a 1D system of $2N$ lattice sites is defined as $\mathcal{F}_{4N}=A\mathcal{H}^{\otimes 4N}$, where $A$ is the antisymmetrizer and $\mathcal{H}$ is the single-particle projective Hilbert space (PHS) in which the many-body states are represented in the basis $B_{\mathcal{F}_{N}}:=\lbrace \prod_{l=1}^{N_{e}}c^{\dagger}_{k_{l}}|0\rangle, k_{l}\in [-\pi ,\pi]\rbrace$ and $N_{e}$ is the number of electrons. A subspace $\mathcal{F}_{8N_{\Lambda_{0}}}=A\mathcal{H}^{\otimes 4N_{\Lambda_{0}}}\subseteq\mathcal{F}_{4N}$ is constituted of $8N_{\Lambda_{0}}$ states in the window $W_{\Lambda_{0}}=[-\Lambda_{0},\Lambda_{0}]$ around the two Fermi points($k_{F},-k_{F}$). The momentum wave-vectors in the window are given by $k_{\Lambda}=k_{F\hat{s}}+\Lambda\hat{s}$, where $\Lambda$ is the normal distance from the two point Fermi surface and $\hat{s}$ denotes the orientation ($\hat{s}=1$ and $-1$ for right and left Fermi points respectively). 
The states within $\mathcal{F}_{8N_{\Lambda_{0}}}$ transform via imposition of the constraint $C_{1}$: as shown in Fig.(\ref{Hilbert space Morphing}), the constraint $C_{1}$ maps a subset of four-fermion scattering vertices involving zero-momentum pairs onto $S=1/2$ pseudospin vertices.
\pin
We now present a precise mathematical formulation of how the symmetries and topology of the Fermi surface guides the constraint $C_{1}$ in shaping the PHS of the emergent condensate. In order to carry out the transformation of the Hilbert space, we first obtain a compact notation for the one-electron states: $|\kappa\rangle \equiv |k_{\Lambda,\hat{s}},\sigma\rangle$. In this way, we define partial trace operators $Tr_{(j,l)}(...)$ that extract four-fermion vertices involving $|\kappa\rangle$ from the Hamiltonian $H$ 
\begin{eqnarray}
Tr_{\kappa}(c^{\dagger}_{\kappa}H)c_{\kappa}+h.c. &=& \frac{U}{V}\sum_{\kappa_{1},\kappa_{2},\kappa_{3}}c^{\dagger}_{\kappa}c^{\dagger}_{\kappa_{1}}c_{\kappa_{2}}c_{\kappa_{3}}~.
\end{eqnarray}
The single-electron states $|\kappa\rangle$ can be grouped into two helicity~($\eta$) classes, $\eta=1$ for $\kappa(\eta=1)=(\Lambda,1,1)$ or $(\Lambda,-1,-1)$ and $\eta=-1$ for $\kappa(\eta=-1)=(\Lambda,1,-1)$ or $(\Lambda,-1,1)$. The three normal distances $\Lambda_{i}'s$ and orientations $\hat{s}_{i}'s$ correspond to the momentum wave-vectors $k_{\Lambda_{1}\hat{s}_{1}}$, $k_{\Lambda_{2}\hat{s}_{2}}$ and $k_{\Lambda_{3}\hat{s}_{3}}$, such that $k_{\Lambda\hat{s}}+k_{\Lambda_{1}\hat{s}_{1}}=2n\pi + k_{\Lambda_{2}\hat{s}_{2}}+k_{\Lambda_{3}\hat{s}_{3}}$. Note that if $k_{F}=\pm\pi/2$, the Umklapp process is allowed, implying $n=0,1$. On the other hand, for  $k_{F}\neq\pm\pi/2$, we have only $n=0$. The three spin orientations are given as $(\sigma_{1},\sigma_{2},\sigma_{3})=(-\sigma,-\sigma,\sigma)$.
\pin
A subset of the four-fermion vertices containing the zero pair-momentum $p=0$ TRS-invariant pair ($\uparrow$,$\downarrow$) is extracted as follows 
\begin{eqnarray}
Tr_{\kappa}(c^{\dagger}_{\kappa}Tr_{\kappa'}(c^{\dagger}_{\kappa'}H))c_{\kappa'}c_{\kappa}+h.c.&=& \frac{U}{V}\sum_{\kappa_{1}}c^{\dagger}_{\kappa_{1'}}c^{\dagger}_{\kappa_{1}}c_{\kappa'}c_{\kappa}+h.c.
~,\label{pivot_projection}
\end{eqnarray}
where $\kappa =(\Lambda ,\hat{s},\sigma)$, $\kappa' =(\Lambda,-\hat{s},-\sigma)$, $\kappa_{1} =(\Lambda' ,\hat{s}',\sigma')$ and, $\kappa_{1}' =(\Lambda' ,-\hat{s}',-\sigma')$. We can easily see that the sum of momentum wave-vectors belonging to $\kappa$ and $\kappa'$ is $\mathbf{p}=0$. Summing over all the $\kappa$ electronic states, we obtain an effective model containing off-diagonal scattering terms involving $\mathbf{p}=0$ net-momentum electronic pairs   
\begin{eqnarray}
&&\sum_{\kappa}Tr_{\kappa}(c^{\dagger}_{\kappa}Tr_{\kappa'}(c^{\dagger}_{\kappa'}H))c_{\kappa'}c_{\kappa}+h.c.=\frac{U}{V}\sum_{\kappa,\kappa_{1}}c^{\dagger}_{\kappa}c^{\dagger}_{\kappa'}c_{\kappa_{1}'}c_{\kappa_{1}}~.\label{projected_HAM}
\end{eqnarray} 
The wavevectors $\kappa$, $\kappa'$, $\kappa_{1}$, $\kappa_{1}'$ are defined as earlier. We now define Nambu spinors~\cite{nambu1961dynamical} in the $CP^{1}$ representation $d^{\dagger}_{\Lambda,\eta}=(c^{\dagger}_{\kappa} ~ c_{\kappa'})$ for the $p=0$ TRS pairs, where $\eta=sgn(\hat{s})sgn(\sigma)$ is the helicity. 
From the spinors, we define Anderson pseudospins~\cite{anderson1958random} $\mathbf{A}_{\Lambda,\eta}$, with magnitude $|\mathbf{A}_{\Lambda,\eta}|^{2}=S(S+1)(\hat{n}_{\kappa}+\hat{n}_{\kappa'}-1)^{2}$. Here, $S=1/2$ as $A^{\pm 2}_{\Lambda,\eta}=0$. We also note that the Hartree processses $n_{\kappa}n_{\kappa_{1}}$ within the $p=0$ momentum subspace can be written purely in terms of $A^{z}_{\Lambda,\eta}A^{z}_{\Lambda',\eta'}$.
\par\noindent 
The instability in the electronic Hilbert space due to the backscattering processes results in the condensation of electronic pairs, as seen from eq.\eqref{constraint_ang_mom}. The resulting Anderson pseudo-spins~\cite{anderson1958random}~($\mathbf{A}_{\Lambda,\eta}\in SU(2)$) follow the $SU(2)$ algebra  
\begin{equation}
[A^{i}_{\Lambda,\eta},A^{j}_{\Lambda',\eta'}]=\epsilon_{ijk}\delta_{\Lambda,\Lambda'}\delta_{\eta,\eta'}A^{k}_{\Lambda,\eta}~,
\end{equation}
and the Casimir $|\mathbf{A}_{\Lambda,\eta}|^{2}=S(S+1)I$ (where $I=|\uparrow\rangle\langle\uparrow|+|\downarrow\rangle\langle\downarrow|$). $S=1/2$ leads to a two-level system ($|1_{k\uparrow},1_{-k\downarrow}\rangle,| 0_{k\uparrow},0_{-k\downarrow}\rangle$). This allows identification of the first Chern class $\gamma$ (eq\eqref{1stchernclass}) on the c.o.m torus $\mathcal{T}^{2}$ with a monopole of charge $2S=\gamma/\pi \Rightarrow \gamma =\pi$.  
\begin{figure}[h!]
\centering
\includegraphics[scale=1]{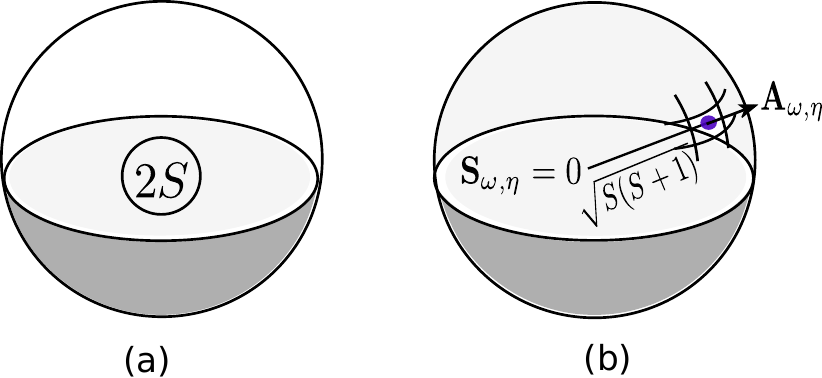}
\caption{(a) The unitary operations $\hat{L}$ ($\hat{L}\mathbf{A}_{\omega}\hat{L}^{\dagger}=\mathbf{A}_{\omega}$) span the space $S^{3}$ that wraps around a monopole of charge $2S$. (b) A representation of the geometry of the Hilbert space of the angular momentum vectors $\mathbf{A}_{\omega}$, $\mathbf{S}_{\omega}$. }\label{wrappings}
\end{figure}
\pin
Within the PHS enforced via eq.\eqref{constraint_ang_mom} arises an emergent symmetry $SU(2)^{\otimes 2N_{\Lambda_{0}}}$ of the resulting projected Hamiltonian eq.\eqref{projected_HAM}. Here, $\Lambda_{0}$ is the width around both the Fermi points within which the putative condensation takes place, and $N(\Lambda_{0})$ is the number of electronic states within the window. We will demonstrate this via the unitary RG prescription in a later section. Further, this space of SU(2) transformations is spanned by 
\begin{equation}
\hat{L}=\prod_{j=1}^{N_{\Lambda}}\exp[i\theta_{j} \mathbf{S}_{\Lambda_{j}}\cdot\mathbf{n}_{j}]~,~\mathbf{S}_{\Lambda_{j},\eta}=S\tilde{c}^{\dagger}_{\Lambda_{j},\eta}\boldsymbol{\sigma}\tilde{c}_{\Lambda_{j},\eta}~,
\end{equation}
where $\tilde{c}^{\dagger}_{\Lambda_{j},\eta}=(c^{\dagger}_{\Lambda_{j},\eta}~c^{\dagger}_{\Lambda_{j'},\eta})$~,~$S=1/2$ and with a parameter space $S^{3}\rightarrow(\theta_{j},\mathbf{n}_{j})$. As shown by Fig.(\ref{wrappings}a), the $S^{3}$ spheres explore the topology of the enclosed null vectors at the origin of the angular momentum sphere: $|\mathbf{S}_{\Lambda,\eta}|=S(S+1)(\hat{n}_{k_{\Lambda,\hat{s}}
\uparrow}-\hat{n}_{k_{\Lambda,-\hat{s}}\downarrow})^{2}=0$. The non-trivial topology of the symmetry group $SU(2)^{\otimes 2N_{\Lambda}}$ is seen, therefore, from the homotopy group $\pi_{3}(SU(2))=Z$, and is a reflection of the existence of angular momentum spheres of radius ($|\mathbf{A}_{\Lambda,\eta}|=\sqrt{S(S+1)}$) in the Fock space $\mathcal{F}_{4N}$. As shown in Fig.(\ref{wrappings}b), the geometry of the angular momentum vector $\mathbf{A}_{\Lambda,\eta}$ traces a sphere $S^{2}$ centred around a monopole of charge $2S=\gamma /\pi$ coinciding with the null vector $\mathbf{S}_{\omega}$. This symmetry in the PHS reflects in the invariance of the basis $B\equiv \lbrace \otimes |A^{z}_{\Lambda}\rangle\rbrace$ under unitary operation: $\hat{L} B =B$.  
\pin
For $\mathbf{n}_{j}=(0,0,n_{z}),\theta_{j}=\delta k_{Fj}$, the $U(1)^{\otimes 2N_{\Lambda_{0}}}$ group generated by the c.o.m. translation operators is 
\begin{eqnarray}
U_{z}=\exp(i\sum_{j}k_{j}S^{z}_{j}) &:& \hat{X}_{cm}^{\Lambda_{0}}\rightarrow \hat{X}_{cm}^{\Lambda_{0}}+2\pi S,~\hat{X}_{cm}^{\Lambda}=(1/4N_{\Lambda})\sum_{js\sigma}j\hat{n}_{\Lambda_{j},\hat{s},\sigma}~,
\end{eqnarray}
where $U_{z}\in S^{1}$ and $k_{j}=k_{F}+\Lambda_{j}$, and the number of states within window $W_{\Lambda_{0}}$ is given by $N_{\Lambda}$. Further, $U_{z}|\Psi\rangle =|\Psi\rangle$ implies the presence of parity, and leads to a vanishing total current for composite objects: $2eP_{cm}^{\Lambda}=0$. Therefore, a vanishing c.o.m. kinetic energy,  $P_{cm}^{\Lambda2}/2I=0$, ensures the stability of the composite objects. A shift of the c.o.m momentum, $P_{cm}^{\Lambda}\neq 0$, can be generated by applying twisted boundary conditions for the states in the window $W_{\Lambda_{0}}$ via the twist operator $\hat{O}_{c}$~(eq.\eqref{twist_op_s_c}). The twist operator imparts a net momentum $p$ to every electronic pair, resulting in the pseudospins 
$\mathbf{A}_{p,\Lambda,\eta} = \hat{O}_{c}\mathbf{A}_{\Lambda,\eta}\hat{O}^{\dagger}_{c}$ being constituted of electronic states $(\Lambda+p,+1,\sigma)$ and $(\Lambda,-1,-\sigma)$. This momentum gain creates a collective Cooper-pair persistent current $P_{cm}=2N_{\Lambda p}$ with a modified constraint $C_{1}:n_{k,\sigma}=n_{-k+p,-\sigma}$.
\pin
The dynamics of the modified PHS, $\mathcal{H}_{tot}= (CP^{1})^{\otimes 2N_{\Lambda_{0}}}$, is governed by the projected Hamiltonian
\begin{equation}
H^{p=0}=v_{F}\sum_{\omega}\Lambda A^{z}_{\Lambda,\eta}+\frac{U}{2N_{\Lambda}}\mathbf{A}_{+1}\cdot\mathbf{A}_{-1} \label{bcs_ham}~, 
\end{equation}
where $\mathbf{A}_{\eta}$ is the total pseudospin vector given by $\mathbf{A}_{\eta}=\sum_{\Lambda}\mathbf{A}_{\Lambda,\eta}$. The zero mode of the Hamiltonian is given by
\begin{equation}
H^{p=0}_{\Lambda=0}=\frac{U}{2N_{\Lambda}}\mathbf{A_{+1}}\cdot\mathbf{A_{-1}}~,
\end{equation}
and possesses eigenstates and eigenvalues given by~\cite{van2008spontaneous}
\begin{eqnarray}
&&|A=p;A_{+1}=A_{-1}=n+1/2\rangle~,~E_{n,m,p}=\frac{U}{2N_{\Lambda_{0}}}[p(p+1)-(2n+1)(n+\frac{3}{2})]~.~~~~~~~
\end{eqnarray}
The gap $\Delta E$ around the Fermi energy $E_{F}=0$ between the highest negative energy state ($E_{0}=|U|/(8N_{\Lambda})$, a pseudo-spin triplet ($|1;1/2,1/2\rangle$)) and the lowest positive energy state ($E_{1}=-3|U|/(8N_{\Lambda})$, a pseudo-spin singlet ($|0;1/2,1/2\rangle$)) is given by $\Delta E =E_{1}-E_{0}=U/2N_{\Lambda}$. This shows that $\Delta E$ gaps the spin excitations around $E_{F}$. We will show subsequently that $\Delta E$ survives after taking account of divergent fluctuations in a renormalization group (RG) formalism. Unlike backscattering processes, forward scattering physics is given by $\mathbf{A}_{+1}^{2}+\mathbf{A}_{-1}^{2}$ and does not cause the coupling of helicities. Hence, the latter does not lead to instabilities of the Fermi surface. The RG irrelevance of forward scattering events in gap opening of the 1D Fermi surface has been confirmed by RG~\cite{shankar1994renormalization,polchinski1992effective} and bosonization~\cite{giamarchi2004quantum,gogolin2004bosonization} methods.  
\pin
A similar analysis of the Umklapp (charge) scattering instability leads to the constraint $C_{2}: n_{\omega}=n_{\bar{\omega}}$, where $\omega=(\Lambda,\hat{s},\sigma)$ and $\bar{\omega}=(-\Lambda,\hat{s},-\sigma)$, and which can be classified in terms of the chirality ($\hat{s}=+1/-1$). The Fermi momentum satisfies the condition $2k_{F}=\pi$, such that the particle-hole symmetry enforces the first Chern class for the torus $\mathcal{T}^{2}$ in the c.o.m Hilbert space, $\gamma=\pi$. In turn, this leads to pairs of fermions from the same side of the Fermi sea forming a $SU(2)$ PHS of spins with $S=1/2=\gamma /2\pi$.  This PHS is again associated with a $S=1/2$ representation of the $SU(2)$ group for pseudospin operators $\mathbf{A}_{\Lambda,\hat{s}}=S(c^{\dagger}_{\omega}~c_{\bar{\omega}})\boldsymbol{\sigma} (c^{\dagger}_{\omega}~c_{\bar{\omega}})^{\dagger}$, and where $\hat{s}=+1/-1$ represents the two chiralities $L/R$. 
\pin
The projection mechanism leads to the condensation of spinors representing  pairs of fermions with total momentum $2ak_{F}$. The Umklapp instability Hamiltonian is governing the dynamics in the PHS $SU(2)^{\otimes 2N_{\Lambda_{0}}}$ is given by
\begin{equation}
H^{p=\pm 2k_{F}}_{\Lambda_{0}}=\frac{U}{2N_{\Lambda}}\mathbf{A}_{L}\cdot\mathbf{A}_{R}~,\label{mott_ham}
\end{equation} 
where $\mathbf{A}_{\hat{s}}=\sum_{\Lambda}\mathbf{A}_{\Lambda,\hat{s}}$ is the total pseudospin vector that acts as the generator of global rotations in the PHS. The Hamiltonian $H^{\pm 2k_{F}}$ has eigenvalues and eigen vectors given by 
\begin{eqnarray}
&&|A_{\Lambda}=p;A_{L}=A_{R}=n+1/2\rangle~,~ E_{n,m,p}=\frac{U}{2N_{\Lambda}}[p(p+1)-(2n+1)(n+\frac{3}{2})]~.
\end{eqnarray}
The gap around the Fermi energy for charge excitations exists between pseudospin singlet and triplet states: $\Delta E= U/2N_{\Lambda}$. The robust nature of this charge gap will be confirmed via RG in a later section. For the sake of generality, we will study the anisotropic version of this Hamiltonian
\begin{equation}
H^{p=\pm 2k_{F}}_{\Lambda_{0}}=\frac{J_{||}}{2N_{\Lambda}}A^{z}_{L\Lambda}A^{z}_{R\Lambda}+\frac{J_{\perp}}{4N_{\Lambda}}\left[A^{+}_{L\Lambda}A^{-}_{R\Lambda}+h.c.\right] \label{anisotropic_form}~.
\end{equation}
In the next section, we will see the formation of topological objects (with $\gamma=\pi$) at the Fermi surface arising from the constraints $C_{1}$ or $C_{2}$. Subsequently, we will study their effect on the c.o.m Hilbert space due to the ensuing instability.
\section{Structure of the Fermi surface pseudospin Hilbert space}\label{sec4}
\begin{figure}[h!]
\centering
\includegraphics[scale=1]{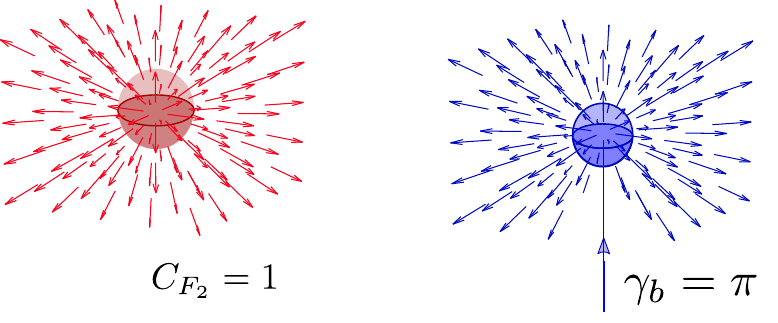}
\caption{(Left) Topological object in subspace $F_{2}$, the WZNW topological term. (Right) Topological object in subspace $F_{1}$, the $\Theta$-term for a monopole with a Dirac string.}\label{top_objects}
\end{figure}

In this section, we will see how the instability associated with constraints $C_{1}$ and $C_{2}$ form $SU(2)$ PHS at the Fermi surface (FS), supporting topological objects like Dirac strings and magnetic monopoles. Furthermore, the first Chern class $\gamma=\pi$ on the c.o.m Hilbert space will be seen to characterize the topological objects at FS. The states at the FS are given by 
\begin{eqnarray}
F&\equiv& \lbrace |\uparrow^{a}_{F;s,c}\rangle = |1_{k_{F}\uparrow}1_{\mp k_{F}\downarrow}\rangle, |\downarrow^{a}_{Fs,c}\rangle =|0_{k_{F}\uparrow}0_{\mp k_{F}\downarrow}\rangle \rbrace \otimes
\nonumber\\
&& \lbrace |\uparrow^{b}_{F;s,c}\rangle = |1_{k_{F}\downarrow}1_{\mp k_{F}\uparrow}\rangle, |\downarrow^{b}_{Fs/c}\rangle = |0_{k_{F}\downarrow}0_{\mp k_{F}\uparrow}\rangle \rbrace
\end{eqnarray}
belonging to PHS $SU(2)_{a}\otimes SU(2)_{b}$. Here, the indices $(a/b)\rightarrow s:=(+/-)$ and the index $c:=(R/L)$ correspond to the pseudospin PHS of spin and charge instability sectors respectively. The sub-Hamiltonian ($H_{F}$) operating on $F$ is given by
\begin{eqnarray}
 H_{F}~(\delta k_{F}=0)=\frac{U}{2N_{\Lambda}}\mathbf{A}_{Fa}\cdot\mathbf{A}_{Fb} ~, 
 \label{effHam1}
 \end{eqnarray}
and possesses a resonant backscattering at the FS in terms of the action of the pseudospin-flip piece $1/2(A^{+}_{Fa}A^{-}_{Fb}+h.c.)$ on the subspace $F_{1}\equiv \lbrace |\uparrow_{Fa}\downarrow_{Fb}\rangle ,|\downarrow_{Fa}\uparrow_{Fb}\rangle \rbrace$ containing 2 electrons ($\hat{N}_{F}=\sum_{a=\pm, \sigma}\hat{n}_{ak_{F},\sigma} =2$). The backscattering leads to helicity ($\eta =+/-$) symmetry breaking: $SU(2)_{a}\times SU(2)_{b}\rightarrow SU(2)_{a+b}$.  The subspace $\hat{N}_{F}=2\equiv A^{z}_{Fa}+A^{z}_{Fb}=A^{z}_{F}=0$ allows the identification $\hat{N}_{F}=2C_{F_{1}}$, where $C_{F_{1}}=2S$ is the Chern number of the effective monopole charge associated with the homotopy group $\pi_{3}(SU(2)_{F_{1}})=Z$ and $SU(2)_{F_{1}}\subset SU_{a}(2)\times SU_{b}(2)$. The PHS $SU(2)_{F_{1}}$ is composed of the states in $F_{1}$, and is associated with a topological $CP^{1}$ space. 
\pin
As seen from the c.o.m. PHS, a doubled twist operator $\hat{O}^{2}_{c/s}$ causes the total momentum ($P_{cm}$, defined on the compact space $S^{1}:P_{cm}\in S^{1}:(0,2\pi]$) to shift by a reciprocal lattice vector $P_{cm}\rightarrow P_{cm}+2\gamma = P_{cm}+2\pi$ due to the transfer of 2 electrons~\cite{nakamura2002lattice,yamanaka1997nonperturbative}. The relation $2\gamma =2\pi$ is equivalent to the Lieb-Schultz-Mattis criterion~\cite{lieb1961two}, and allows either a gapless unique ground state or a doubly-degenerate gapped state of matter. The gapped state of matter is associated with constructive interference between paths on a non-simply connected $P_{cm}$ Hilbert space manifold, and will be discussed further in Sec.\ref{sec6}. This leads us to conclude that, in basis $F_{1}$, the pseudospin flip term (corresponding to a vertex operator in the equivalent sine-Gordon theory) is equivalent to the double-twist operator $\hat{O}^{2}_{c/s}$~\cite{nakamura2002lattice}
\begin{equation}
(A^{+}_{Fa}A^{-}_{Fb}+h.c.) \equiv \hat{O}^{2}_{c/s}+h.c.~.\label{spin-flip operation}
\end{equation} 
\pin
Tunnelling between the two degenerate levels of the subspace $F_{1}$ can be studied via the effective Hamiltonian  
\begin{eqnarray}
H_{F_{1}}=-U/2N_{\Lambda}D_{z}^{2}+ U/2N_{\Lambda}D_{x}~,\label{backscattering_FS}
\end{eqnarray}
 where $\mathbf{D}=(\gamma /2\pi) A^{\dagger}_{F}\boldsymbol{\sigma}A_{F}$, and the spinor $A^{\dagger}_{F}=(A^{+}_{Fa} A^{+}_{Fb})$ etc. $H_{F_{1}}$ is invariant under a Unitary transformation $U_{1}=\exp\left[i2\pi D_{z}\right]\in Z_{2}$ \& $U_{2}=\exp\left[i\theta D_{x}\right]\in U(1)$. In the parameter space $S^{2}$ of radius $R=|U/2N_{\Lambda}|$, a closed path is traced such that the eigenstate $|\psi(\mathbf{R})\rangle \in CP^{1}$ traces a great circle on the Bloch sphere described by Fubini-Study metric~\cite{anandan1990geometry,braunstein1994statistical}
\begin{eqnarray}
ds^{2}&\equiv &\langle D\psi| D\psi\rangle = S^{2}r^{2}(d\theta^{2}+\sin^{2}\theta d\phi^{2})=S^{2}d\phi^{2}~,~(r=1,\theta =\pi/2, \phi=(0,2\pi])~,~~~~
\end{eqnarray} 
 where $|D\psi\rangle =|\psi\rangle -|\psi\rangle\langle\psi |d\psi\rangle$ is a gauge-invariant derivative on the PHS. The great circle $S_{1}\rightarrow ds^{2}=d\phi^{2}$ traced by $U_{2}$ winds around the Dirac-string associated with the monopole of charge $2S=2\gamma /2\pi$ ($R=0\rightarrow E_{+}=E_{-}$) arising from the energy degeneracy point. The integrated two-form, $d\mathbf{F}=\langle D\psi|\wedge |D\psi\rangle =\mathcal{F}_{\mu\nu}dx_{\mu}\wedge dx_{\nu}$, can be written in terms of the Berry curvature~\cite{berry1984quantal} $\mathcal{F}_{\mu\nu}=Im(\langle D_{\mu}\psi|D_{\nu}\psi\rangle)$ for a hemisphere of the Bloch sphere (with solid angle $\Omega =2\pi$). Parametrizing the Bloch sphere by the angles $(\theta,\phi)$ gives the non-commutativity between the unitary operators $\hat{T}_{1}=\exp\left[i(\pi/2)\hat{D}_{\theta}\right]$ and $\hat{T}_{2}=\exp\left[i2\pi\hat{D}_{\phi}\right]$ 
\begin{equation}
\hat{T}_{1}\hat{T}_{2}\hat{T}_{1}^{\dagger}\hat{T}_{2}^{\dagger}=e^{i\gamma_{b}}~,
\end{equation} 
where, following Schwinger~\cite{schwinger1952angular}, we define the generators $D_{\theta}=(1/2i)(D_{+}e^{-i\phi}-e^{i\phi}D_{-})$ and $D_{\phi}=(1/i)\partial_{\phi}$. The topological phase accrued by the closed circuit ($\gamma_{b}$) is then given by 
\begin{eqnarray}
\gamma_{b}&=&\frac{\gamma}{2\pi}\int_{\theta <\pi/2}  d\mathbf{F} =\frac{\gamma}{2\pi}\Omega =\gamma~.
\end{eqnarray}
 This shows that the nesting instability associated with  $\gamma_{b}$ arises from a monopole of charge $2S=\gamma/\pi$~\cite{dirac1931quantised}, where $\gamma$ is the first Chern class of the gapless FS. This leads to the equivalence: $\hat{T}_{1}\hat{T}_{2}\hat{T}_{1}^{\dagger}\hat{T}_{2}^{\dagger}=e^{i\gamma_{b}} \equiv    \hat{O}_{c/s}\hat{T}\hat{O}_{c/s}^{\dagger}\hat{T}^{\dagger}=e^{i\gamma}$. The Dirac string~\cite{dirac1931quantised} (see Fig.(\ref{top_objects}) (right)) carries an effective flux $\Phi =\gamma /2\pi\Phi_{0}$, punctures the XZ plane from the North/South pole ($\gamma_{bN}=-\gamma_{bS}$), and is revealed by using Gauss' law 
 \begin{eqnarray}
 \gamma_{b}&=&\pm\oint dA ,\quad dA=d\mathbf{l}\cdot\hat{\phi} \frac{\gamma}{2\pi}~.\label{Theta-term}
\end{eqnarray}   
The associated $\Theta$-term, $\gamma_{b}/\gamma\in \pi_{1}(U(1))=Z$, acts as a half-flux quantum $\Phi_{0}/2$ Dirac string for the M\"obius strip PHS encircling the great circle of the Bloch sphere made by the subspace $F_{1}$.
\pin
By making the two-spin interaction of the Hamiltonian eq.\eqref{effHam1} anisotropic
\begin{equation}
H_{F}=\frac{J_{||}}{2N_{\Lambda}}A^{z}_{Fa}A^{z}_{Fb}+\frac{J_{\perp}}{4N_{\Lambda}}(A^{+}_{Fa}A^{-}_{Fb}+h.c.)~,\label{FP_anisotropic}
\end{equation}
we extend our arguments to a two-dimensional coupling space: $(J^{||},J^{\perp})$. Now, $H_{F}$ lowers the symmetry associated with its PHS from $SU(2)_{a}\otimes SU(2)_{b}\rightarrow U(1)_{a+b}
\otimes Z_{2 a+b}$. Along two special lines on the ($J_{||},J_{\perp}$) plane, $J_{||}=\pm J_{\perp}$, the Hamiltonian $H_{F}$ has the enhanced symmetry $SU(2)_{a+b}$. Further, precisely at the crossing point of the two lines, $J^{||}=\pm J^{\perp}$, the Hamiltonian vanishes ($H_{F}=0$) and the symmetry group is emergently restored to $SU(2)_{a}\otimes SU(2)_{b}$. The $H_{F}$ Hamiltonian can be block decomposed as $H_{F}= H_{F_{1}}\oplus H_{F_{2}}$, where $H_{F_{1}}$ is given in eq.\eqref{backscattering_FS} and the Hamiltonian $H_{F_{2}}=J^{||}S^{2} I_{F_{2}}$ is represented by subspace $F_{2}\equiv \lbrace |\uparrow_{a}\uparrow_{b}\rangle ,|\downarrow_{a}\downarrow_{b}\rangle\rbrace \subset F$ and $I_{F_{2}}=|\uparrow_{a}\uparrow_{b}\rangle\langle\uparrow_{a}\uparrow_{b}|+|\downarrow_{a}\downarrow_{b}\rangle\langle\downarrow_{a}\downarrow_{b}|$. The family of unitary operations that keep $H_{F_{2}}$ invariant is $U_{\delta k_{F}=0,F_{2}}(\theta,\hat{n})=\exp\left[i\theta\mathbf{S}_{1}\cdot\hat{n}\right]\in S^{3}$, where
$\mathbf{S}_{1}=S\bar{A}^{+}_{F}\boldsymbol{\sigma}\bar{A}^{-}_{F}$,$\bar{A}^{+}_{F}=(A^{+}_{Fa} A^{-}_{Fb})$ and $S=1/2$. This $S^{3}$ wraps the FS metric singularity of the PHS $F_{2}$: $ds^{2}=S^{2}r^{2}(d\theta^{2} +\sin^{2}\theta d\phi^{2})=0$ as $r=0$. The density matrix $\rho_{F_{2}}=\frac{1}{2}[I+\mathbf{r}\cdot\boldsymbol{\sigma}]|_{r=0}$ lies at the origin of the Bloch sphere, reaffirming the $SU(2)$ symmetry of $F_{2}$. The $\hat{n}$ vectors form a unit sphere $S^{2}$ 
homeomorphic to $CP^{1}$ (i.e., once again a Bloch sphere  with a monopole of charge $2S$). The topological term associated with the homotopy group of this $SU(2)$ PHS is the $0+1D$ Wess-Zumino-Novikov-Witten term~\cite{fradkin2013field}
\begin{eqnarray}
C_{F_{2}}=2S\int d^{2}\mathbf{x} \mathbf{n}\cdot(\partial_{\theta}\mathbf{n}\times\partial_{\phi}\mathbf{n})~,\label{chern}
\end{eqnarray}
where the Chern numbers $C_{F_{2}}\in\pi_{3}(SU(2)_{A^{z}_{F}=\pm 1})\in Z$. The monopole charge $2S=1$ is shown in Fig.(\ref{top_objects}(left)), and arises from an equivalence of the PHS $F_{2}$ to a two-level system formed by pairing of two pseudospins of $S=1/2$ representation.
\pin
We have seen earlier that $H_{F_{1}}=-\frac{J^{||}}{4}D_{z}^{2}+\frac{J^{\perp}}{2}D_{x}$ acts in sub-basis $F_{1}$ for which we have computed the topological $\Theta-term$  associated with a Dirac string is given by $\gamma_{b}=\gamma\frac{\Omega}{2\pi},\Omega =2\pi(1-\cos\theta),\theta =\pi/2$. By varying the coupling values ($J^{||},J^{\perp}$), the four energy eigenvalues $(E_{\uparrow_{a}\uparrow_{b}}=E_{\downarrow_{a}\downarrow_{b}}=J^{||}/4),(E_{\uparrow\downarrow\pm \downarrow\uparrow}=-J^{||}/4\pm J^{\perp}/2)$ rearrange themselves. Thus, the existence of topological objects like magnetic monopoles or Dirac strings in the lowest energy subspace ($LES:\lbrace \psi_{i}\rbrace\quad$ such that $\quad H_{F}\psi_{i}=E\psi_{i}~,~E=min[E_{\uparrow_{a}\uparrow_{b}},E_{\downarrow_{a}\downarrow_{b}},E_{\uparrow\downarrow + \downarrow\uparrow},E_{\uparrow\downarrow - \downarrow\uparrow}]$) is determined by whether $LES\in F_{1}$ or $F_{2}$. A compact form for $\gamma_{b}$ can be written down in terms of ($J^{||},J^{\perp}$) 
\begin{eqnarray}
\gamma_{b} =f+sgn(f)(\gamma -|f|)\label{berry_phase_dirac}
\end{eqnarray}
where $f=sgn(J^{\perp})(\gamma/2(1+sgn(r))sgn(J^{||})+\gamma/4(1-sgn(r))(1+sgn(J^{||}))$ and $r=J^{\perp 2}-J^{|| 2}$.
\begin{figure}[h!]
\centering
\includegraphics[width=0.7\textwidth]{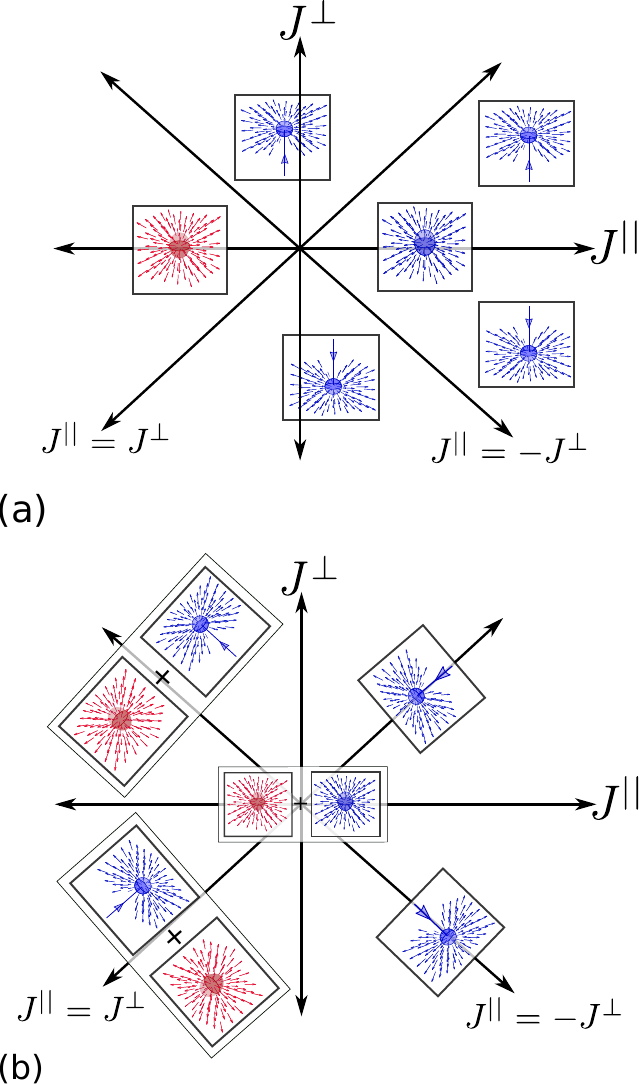}
\caption{Fig.(a) corresponds to topological objects lying between the separatrices of the coupling $(J^{||},J^{\perp})$ phase diagram obtained by studying the effective problem at the Fermi surface. Fig.(b) corresponds to the topological objects living on the separatrices, and at the critical point (the origin). Together, these two diagrams reveal the shape of the BKT RG phase diagram. See main text for detailed discussions.}\label{topological_objects}
\end{figure}
\pin
Fig.(\ref{topological_objects}a) is a skeletal phase diagram that depicts the topological objects residing in the LES of the Fermi surface PHS for anisotropic couplings characterized by $r\neq 0$. For $J_{||}<0$ and $r<0$, the $LES =F_{2}$ is associated with a monopole with strength given by Chern number $C_{F_{2}}=2S$ (eq.\eqref{chern}). On the other hand, for either $r>0$ or $J_{||}>0$, the $LES=F_{1}$ posesses a Dirac string whose $\Theta$-term coefficient is given by $\gamma_{b}$ (eq.\eqref{Theta-term})). Two special half-lines $J^{\perp}=\pm J^{||}=J<0$ where $r=0$ lead to the $LES$ 
\begin{eqnarray}
LES_{\gamma_{b}=\pm \pi,J<0}&=&\lbrace |\uparrow_{Fa}\uparrow_{Fb}\rangle ,|\downarrow_{Fa}\downarrow_{Fb}\rangle ,|\gamma_{b}\rangle\rbrace ~~ \text{where} \nonumber\\
|\gamma_{b}\rangle &=& |\mathbf{A}_{F}=1/2(1+e^{i\frac{(\gamma -\gamma_{b})}{2}}),A^{z}_{F}=0\rangle ~.~\label{LES_FS}
\end{eqnarray}
This special $LES$ contains states belonging to both $F_{1}$ and $F_{2}$. As depicted in the skeletal phase diagram of Fig.(\ref{topological_objects}b), the topological objects describing this LES is a monopole associated with PHS $F_{2}$ (with Chern number $C_{F_{2}}=2S=1\in \pi_{3}(SU(2)_{F_{2}})$) and a Dirac string associated with $F_{1}$  (with $\Theta$ term $\gamma_{b}/\gamma\in\pi_{1}(U(1)) $). The co-existence of two topological objects in the LES for $r=0,J_{||}<0$ is an outcome of an emergent $SU(2)_{a+b}$ seen from a $2\otimes 2 = 3\oplus 1$ block decomposition of the Hamiltonian 
\begin{equation}
H_{F}=J/4 I^{\gamma_{b}=\pm\pi}_{LES}-3J/4|\gamma_{b}\rangle\langle \gamma_{b}|~,
\end{equation}
where $I^{\gamma_{b}=\pm\pi}_{LES}$ is a $n\times n,n=3 $ Casimir invariant under $SU(2)$ rotations of spin representation $S=(n-1)/2=1$. For $J>0,r=0$, the $LES_{\gamma_{b}=\pm \pi,J<0}$ switches to $LES_{\gamma_{b}=\mp \pi,J>0}=\lbrace |\gamma_{b}\rangle\rbrace$. Precisely at the critical point $J^{||}=J^{\perp}=0$, the $LES =F_{1}\cup F_{2}=F$ and possesses both the Chern numbers $C_{F_{1}}\in\pi_{3}(SU(2)_{F_{2}})$ and $C_{F_{2}}\in\pi_{3}(SU(2)_{F_{1}})$ given earlier, displaying the enhanced symmetry of $SU(2)_{a}\times SU(2)_{b}$. In the following section, we will treat the instabilities of the Fermi surface via a renormalization group (RG) procedure. We will see there that the RG flows are characterized by the quantity $r$ shown above, such that the relevance or irrelevance of a flow is decided by the quantities $(sgn(J_{||}),sgn(r),sgn(J_{\perp}))$.  In this way, RG flows will be dictated by the discontinuous changes in boundary conditions accounted for by the topological object at the FS ($\gamma_{b}$).  
\section{URG for Fermi surface instabilities: a topological viewpoint}\label{sec5}
In Sec.\ref{sec3}, we showed that the constraints $C_{1}$ and $C_{2}$ lead to the condensation of four-fermion vertices into $SU(2)^{\otimes 2N_{\Lambda_{0}}}$ PHS associated with the putative BCS and Mott instabilities respectively. This way, the first Chern class $\gamma =\pi$ associated with the Fock space $\mathcal{F}_{8N_{\Lambda_{0}}}$ of the gapless FS became the Chern number belonging to the homotopy group of the SU(2) condensate PHS, i.e., $C=2S=2\gamma /2\pi \in \pi_{3}(SU(2))\equiv Z$. Further, in Sec.\ref{sec4}, the first Chern class $C$ was observed to be a monopole of charge ($2\gamma/\pi$), possessing a Dirac string given by the $\Theta$-term $\gamma_{b}=\gamma$ acting on the Fermi surface PHS $F$. In this section, we will show how FS topology shapes the instabilities via a renormalization group (RG) formalism. For this, we start with an electronic model for strong correlations, and perform the unitary renormalization group (URG) method by decoupling electronic states starting from the UV. In this way, we show the emergence of IR fixed points governed by the topological constraints $C_{1},C_{2}$ of Sec.\ref{sec3}. Furthermore, we show that the essential features of the  RG phase diagram obtained via URG is dictated by a topological term $\gamma_{b}$, and reflects the qualitative accuracy of the skeletal phase diagrams shown in Figs.(\ref{topological_objects}a) and (b).
\\
\par\noindent{\bf \textit{URG algorithm}} 
\pin 
In the URG scheme~\cite{anirbanmotti,anirbanmott2,mukherjee2020,anirbanurg1,anirbanurg2}, the Hamiltonian $H$ is iteratively block-diagonalized by a succession  of unitary maps $U_{(N)}$, $U_{(N-1)}$,$\ldots$, $U_{(j)}$, leading to the flow equation
\begin{eqnarray}
H_{(j-1)}=U_{(j)}H_{(j)}U^{\dagger}_{(j)},~[H_{(j)},\hat{n}_{j}]=0~,
\end{eqnarray}
where $j$ is the $j$-th electronic state that is disentangled and $H_{(N)}=H$ is the bare Hamiltonian.  The occupancies of the disentangled electronic states become good quantum numbers, as they commute with the Hamiltonian, $[H_{(j)},\hat{n}_{i}]=0$ for $i\geq j$. The $N$ electronic states are labelled as $j\in [1,N]$, in increasing order of bare one-particle energy $\epsilon_{1}\leq \epsilon_{2}\leq\ldots\leq \epsilon_{N}$. Electronic states in the UV are disentangled first, eventually scaling towards the IR. Concomitantly, this involves the entanglement renormalization within the eigenstates $|\Psi_{(j)}\rangle$ of $H_{(j)}$
\begin{eqnarray}
|\Psi_{(j-1)}\rangle = U_{(j)}|\Psi_{(j)}\rangle~,~\hat{n}_{i}|\Psi_{(j-1)}\rangle=|\Psi_{(j-1)}\rangle~.
\end{eqnarray}
This guarantees the preservation of the many-body eigenspectrum
\begin{eqnarray}
H_{(j)}|\Psi_{(j)}\rangle =E(|\Psi\rangle)~|\Psi_{(j)}\rangle,
\end{eqnarray}
where the initial (bare) eigenstate is $|\Psi\rangle = |\Psi_{(N)}\rangle$. The unitary operation $U_{(j)}$~\cite{anirbanmotti,anirbanmott2,mukherjee2020,anirbanurg1,anirbanurg2} is given by
\begin{eqnarray}
U_{(j)}&=&\frac{1}{\sqrt{2}}(1+\eta_{(j)}-\eta^{\dagger}_{(j)})~,~\eta^{\dagger}_{(j)}=\frac{1}{\hat{\omega}_{(j)}-Tr_{j}(H_{(j)}^{D}\hat{n}_{j})\hat{n}_{j}}c^{\dagger}_{j}Tr_{j}(H_{(j)}c_{j})~.\label{UnitaryOpDef}
\end{eqnarray}
The operators $\eta_{(j)}$, $\eta^{\dagger}_{(j)}$ in the above expression satisfy the algebra $\lbrace\eta_{(j)},\eta^{\dagger}_{(j)}\rbrace=1$, $[\eta_{(j)},\eta^{\dagger}_{(j)}]=1-2\hat{n}_{(j)}$. The operator $\hat{\omega}_{(j)}$ accounts for the residual quantum fluctuations due to renormalized off-diagonal blocks, and is defined as 
\begin{eqnarray}
\hat{\omega}_{(j)}=H^{D}_{(j-1)}+H^{X}_{(j-1)}-H^{X}_{(j)}~. \label{omegaOpdef}
\end{eqnarray}
$H^{D}_{(j)}$ and $H^{X}_{(j)}$ represents the diagonal and the off-diagonal components of $H_{(j)}$. Note that eq.\eqref{omegaOpdef} is essentially a rewriting of the Hamiltonian RG flow equation: $\Delta  H_{(j)}= H_{(j-1)}-H_{(j)}=\hat{\omega}_{(j)}-H^{D}_{(j)}$. In terms of $\eta{(j)}$, the Hamiltonian renormalization $\Delta H_{(j)}$ is given by
\begin{eqnarray}
\Delta H_{(j)}=\tau_{j}\lbrace Tr(c^{\dagger}_{j}H_{(j)})c_{j},\eta^{\dagger}_{j}\rbrace~,\label{Hflow}
\end{eqnarray}
where $\lbrace A,B\rbrace=AB+BA$ represents the anticommutator and $\tau_{j}=n_{j}-\frac{1}{2}$ represents the disentangled degree of freedom. Associated with the quantum fluctuation operator $\hat{\omega}_{(j)}$ are eigenstates $|\Phi(\omega_{(j)})\rangle$ and eigenvalues $\omega_{(j)}$. These constitute the natural quantum energyscales arising from the off-diagonal blocks. For each of the $\omega_{(j)}$s, an effective Hamiltonian RG flow is obtained, describing the renormalization of a sub-part of the many-body energy spectrum. The condition for reaching a RG fixed point is obtained from vanishing of the matrix element
\begin{eqnarray}
\langle\Phi_{(j)}|(\hat{\omega}_{(j)}-H^{D}_{(j)})|\Psi_{(j)}\rangle = \langle\Phi_{(j)}|(\omega_{(j)}-H^{D}_{(j)})|\Psi_{(j)}\rangle = \langle\Phi_{(j)}|\Delta H_{(j)}|\Psi_{(j)}\rangle=0~. 
\end{eqnarray} 
This implies that the projected subspace generated by $|\Phi_{(j)}\rangle\langle\Phi_{(j)}|$, $\hat{\omega}_{(j)}$ is number diagonal: $[\omega_{(j)}|\Phi_{(j)}\rangle\langle\Phi_{(j)}|,H^{D}_{(j)}]=0$.
\\
\par\noindent{\bf\textit{URG study of the BCS and Mott instabilities 
of 1D correlated electrons}}
\pin
We adapt the URG procedure to a 1D model of strongly correlated electrons. We first address the BCS instability by starting from the model
\begin{eqnarray}
H_{1}&=&\sum_{\kappa}\epsilon_{\kappa}\hat{n}_{\kappa}+\sum_{\kappa,\kappa',p,\eta}\bigg[Kc^{\dagger}_{\kappa(p,\eta)}c^{\dagger}_{\kappa'(p,\eta)}c_{\kappa'_{1}(p,-\eta,)}c_{\kappa_{1}(p,-\eta)}\nonumber\\
&+&Vc^{\dagger}_{\kappa(p,\eta)}c^{\dagger}_{\kappa'(p,\eta)}c_{\kappa'_{1}(p,\eta,)}c_{\kappa_{1}(p,\eta)}\bigg]~,
\label{H-BCS}
\end{eqnarray}
where $p$ is the pair momentum and $\eta$ is the helicity. We have included the spin-backscattering process (with coupling $K$) and forward scattering processes for the various opposite spin, $p$ momentum pairs (with coupling $V$).
The redefinition of momentum wavevectors $k_{\Lambda\hat{s}}=k_{F\hat{s}}+\Lambda\hat{s}$ in Sec.\ref{sec3} unveils a natural labelling scheme for states in terms of normal distances from Fermi points $\Lambda_{N}>..>\Lambda_{j}>\Lambda_{j-1}>..>0$. The RG transformations then disentangle electronic states farthest from the Fermi points, while scaling gradually towards it. At RG step $j$, two electronic states with spins $(\uparrow,\downarrow)$ at a distance $\Lambda_{j}$ from both the $L$ and $R$ Fermi points are simultaneously disentangled. The net unitary transformation at the step $j$ is $U_{(j)}=\prod_{\hat{s}=\pm 1,\sigma=\pm 1}U_{j,\hat{s},\sigma}$, where $U_{j,\hat{s},\sigma}$~(eq.\eqref{UnitaryOpDef}) disentangles the electronic state $|k_{\Lambda_{j}\hat{s}},\sigma\rangle$. The rotated Hamiltonian $H_{(j-1)}=U_{(j)}H_{(j)}U_{(j)}^{\dagger}$ thus obtained is off-diagonal with respect to electronic states at distances $\Lambda<\Lambda_{j}$.
\pin
Next, from the Hamiltonian RG flow equation, we extract the vertex RG flow equations for all the $p=0$ and $p\neq 0$ pair momentum vertices (see Appendix~\ref{BCS_RG} for details) 
\begin{eqnarray}
\Delta K^{(j)}(p)&=&\frac{K^{(j)}(p)V^{(j)}(p)}{\omega-\frac{1}{2}(\epsilon_{\kappa_{j}}+\epsilon_{\kappa'_{j}})-\frac{V^{(j)}(p)}{4}}~,~\Delta V^{(j)}(p)=\frac{(K^{(j)}(p))^{2}}{\omega-\frac{1}{2}(\epsilon_{\kappa_{j}}+\epsilon_{\kappa'_{j}})-\frac{V^{(j)}(p)}{4}}~,\label{RGset1}
\end{eqnarray}
where $\kappa_{j}=(\Lambda_{j},\hat{s},\sigma)$ and $\kappa'_{j}=(\Lambda_{j}+p,-\hat{s},-\sigma)$. In obtaining the above equations, we have chosen the intermediate configuration $\tau_{\kappa_{j}}=\tau_{\kappa'_{j}}=\frac{1}{2}$. It is important to note that for arriving at the above RG equations, we have accounted only for the quantum fluctuation energy scales $\omega$ at the one-particle level. Also, we have not accounted for the feedback of the renormalized vertices in the $\omega$'s~(eq.\eqref{omegaOpdef}). We investigate the RG equation in the regime
\begin{equation}
\omega<\frac{1}{2}(\epsilon_{\kappa_{j}}+\epsilon_{\kappa'_{j}})~,\label{regime-1}
\end{equation}
as it corresponds to the vicinity of Fermi energy. Note that the only set of non-trivial fixed points of the above RG equations (eq.\eqref{RGset1}) exists for $V,V^{(j)}(p)<0$. Further, in those cases, the two-particle vertices scattering $p=0$ momentum pairs constitute the dominant RG flows
\begin{eqnarray}
&& |\omega-\hbar v_{F}(\Lambda_{j}+\frac{p}{2})|>|\omega - \hbar v_{F}\Lambda_{j}|\nonumber\\
&&\implies |\Delta K^{(j)}(p)|<|\Delta K^{(j)}(0)|,~|\Delta V^{(j)}(p)|<|\Delta V^{(j)}(0)|~.~~~~~~~~
\end{eqnarray}
Here, the net kinetic energy for the pair of electronic states near the Fermi surface (about $E_{F}$) is given by 
$(\epsilon_{\kappa_{j}}+\epsilon_{\kappa'_{j}})\approx\hbar v_{F}(2\Lambda_{j}+p)$~. We have additionally reasoned that $|\omega-\hbar v_{F}(2\Lambda_{j}+p)|>|V|$; as $V\propto O(1/L)$ (where $L$ is the system dimension), the relation is naturally satisfied for large system sizes. Thus, we study the dominant vertex ($p=0$ momentum vertices) RG flow equations in the continuum limit
\begin{eqnarray}
\frac{d K_{1}}{d\log\frac{\Lambda}{\Lambda_{0}}}&=&K_{1}(1-\omega G(\omega,\Lambda))+\frac{K_{1}V_{1}}{1-V_{1}/4},\nonumber\\
~\frac{d V_{1}}{d\log\frac{\Lambda}{\Lambda_{0}}}&=&V_{1}(1-\omega G(\omega,\Lambda))+\frac{K_{1}^{2}}{1-V_{1}/4}~,\label{BCS_RG1}
\end{eqnarray} 
where $G(\omega,\Lambda)=\frac{1}{\omega-\hbar v_{F}\Lambda}$. In obtaining the continuum RG equations, we have replaced the discrete difference $\Delta\log\frac{\Lambda_{j}}{\Lambda_{0}}$ by the differential $d\log\frac{\Lambda}{\Lambda_{0}}$, and defined the couplings $K_{1}=K(\Lambda)/(\omega-\hbar v_{F}\Lambda)$ and $V_{1}=V(\Lambda)/(\omega-\hbar v_{F}\Lambda)$. In the regime of eq.\eqref{regime-1}, the signatures of $K_{1}$ and $V_{1}$ are related to those of $K$ and $V$, as $sgn(K_{1})=-sgn(K)$ and $sgn(V_{1})=-sgn(V)$.   For $\omega<0$ and upon scaling towards the Fermi surface $\Lambda\to 0$, the ratio $\hbar v_{F}\Lambda/\omega\to 0-$, such that $\omega G(\omega,\Lambda)=(1-\hbar v_{F}\Lambda/\omega)^{-1}\to 1$. This ensures that both RG equations are eventually dominated by the second term, i.e.,  
\begin{eqnarray}
\frac{d K_{1}}{d\log\frac{\Lambda}{\Lambda_{0}}}&=&\frac{K_{1}V_{1}}{1-V_{1}/4}~,~\frac{d V_{1}}{d\log\frac{\Lambda}{\Lambda_{0}}}=\frac{K_{1}^{2}}{1-V_{1}/4}~.
\end{eqnarray} 
\par\noindent
Similarly, in order to study the Mott instability, we study the Hamiltonian that includes Umklapp scattering processes (with coupling $K'$) along with forward scattering processes (with coupling $V'$)
\begin{eqnarray}
H_{2}&=&\sum_{\kappa}\epsilon_{\kappa}\hat{n}_{\kappa}+\sum_{\kappa,\kappa',p,\eta}\bigg[K'c^{\dagger}_{\kappa(p,\hat{s})}c^{\dagger}_{\kappa'(p,\hat{s})}c_{\kappa'_{1}(p,-\hat{s},)}c_{\kappa_{1}(p,-\hat{s})}\nonumber\\
&+&V'c^{\dagger}_{\kappa(p,\hat{s})}c^{\dagger}_{\kappa'(p,\hat{s})}c_{\kappa'_{1}(p,\hat{s},)}c_{\kappa_{1}(p,\hat{s})}\bigg]~,\label{H-Mott}
\end{eqnarray}
where $\kappa(p,\hat{s})=(\Lambda,\hat{s},\sigma)$ and $\kappa'(p,\hat{s})=(-\Lambda+p,\hat{s},-\sigma)$. The pair of electronic states $|\kappa\rangle$ and $|\kappa'\rangle$ scatter onto the opposite side of the Fermi surface, such that net momentum transfer is $2\pi$. The resulting electronic states $|\kappa_{1}\rangle$ and $|\kappa_{1}'\rangle$ are given by
$\kappa_{1}=(\Lambda,-\hat{s},\sigma)$, $\kappa_{1}'=(\Lambda,-\hat{s},\sigma)$ and $\kappa_{1}=(-\Lambda+p,-\hat{s},-\sigma)$. By constructing the unitary maps $U_{(j)}$ (eq.\eqref{UnitaryOpDef}), we obtain the coupling RG equations from eq.\eqref{Hflow} as
\begin{eqnarray}
\Delta K^{'(j)}(p)&=&\frac{K^{'(j)}(p)V^{'(j)}(p)}{\omega-\frac{1}{2}(\epsilon_{\kappa_{j}}+\epsilon_{\kappa'_{j}})-\frac{V^{'(j)}(p)}{4}},~\Delta V^{'(j)}(p)=\frac{(K^{'(j)}(p))^{2}}{\omega-\frac{1}{2}(\epsilon_{\kappa_{j}}+\epsilon_{\kappa'_{j}})-\frac{V^{'(j)}(p)}{4}}.~~~~
\end{eqnarray} 
We again make the choice of electronic states in the vicinity of the Fermi surface, $\frac{1}{2}(\epsilon_{\kappa_{j}}+\epsilon_{\kappa'_{j}})\approx - \hbar v_{F}p$, and investigate the above RG equations in the regime $\omega+\hbar v_{F}p>0$, where $V,V^{(j)}(p)>0$ leads to non-trivial fixed points. In this case, the pairs with net momentum $k_{\Lambda,\hat{s}}+k_{-\Lambda,\hat{s}}=\pi$ dominate the RG flows as seen below
\begin{eqnarray}
\omega<\omega+\hbar v_{F}p~\Rightarrow ~|\Delta K^{'(j)}(p)|<|\Delta K^{'(j)}(0)|~,~|\Delta V^{'(j)}(p)|<|\Delta V^{'(j)}(0)|~. 
\end{eqnarray} 
In the continuum limit, the RG equations attain the form
\begin{eqnarray}
\frac{d K_{1}'}{d\log\frac{\Lambda}{\Lambda_{0}}}&=&\frac{K_{1}'V_{1}'}{1-V'_{1}/4},~\frac{d V'_{1}}{d\log\frac{\Lambda}{\Lambda_{0}}}=\frac{(K'_{1})^{2}}{1-\frac{V'_{1}}{4}}~,\label{Mott_RG}
\end{eqnarray}
where $K'_{1}=K'(\Lambda)/\omega$, $V'_{1}=V'(\Lambda)/\omega$.
\\
\par\noindent{\bf \textit{RG and Fermi surface topology}} 
\pin
The continuum RG equations for the Mott (eq.\eqref{Mott_RG}) and BCS instabilities (eq.\eqref{BCS_RG1}) can be written in a compact form as 
\begin{eqnarray}
\frac{dJ_{||}}{d\ln\frac{\Lambda}{\Lambda_{0}}} &=& \frac{J^{2}_{\perp}}{1-\frac{J_{||}}{4}},~\frac{dJ_{\perp}}{d\ln\frac{\Lambda}{\Lambda_{0}}} = \frac{J_{\perp}J_{||}}{1-\frac{J_{||}}{4}}~,\label{anisotropic_RG_eqns}
\end{eqnarray}
with $J_{||}=V_{1}'$ or $V_{1}$ and $J_{\perp}=K_{1}'$ or $K_{1}$ for the Mott and BCS cases respectively. These equations have the same form as the Berezinskii-Kosterlitz-Thouless (BKT) RG equations~ \cite{berezinskii1971destruction,kosterlitz1978two}, and are precisely identical to them at weak coupling (i.e., for $J_{||}\to 0$). They also possess the same RG invariant labelling each RG trajectory, $r=J_{\perp}^{2}-J_{||}^{2}$. However, the presence of the $(1-J_{||}S^{2})^{-1}$ term in the denominator of both RG equations represents a new non-perturbative feature obtained from the URG formalism, and will be seen to be responsible for the RG flows reaching stable fixed points at intermediate coupling~\cite{anirbanmotti,anirbanmott2,mukherjee2020,anirbanurg1,anirbanurg2}. 
\begin{figure}[h!]
\centering
\includegraphics[width=0.7\textwidth]{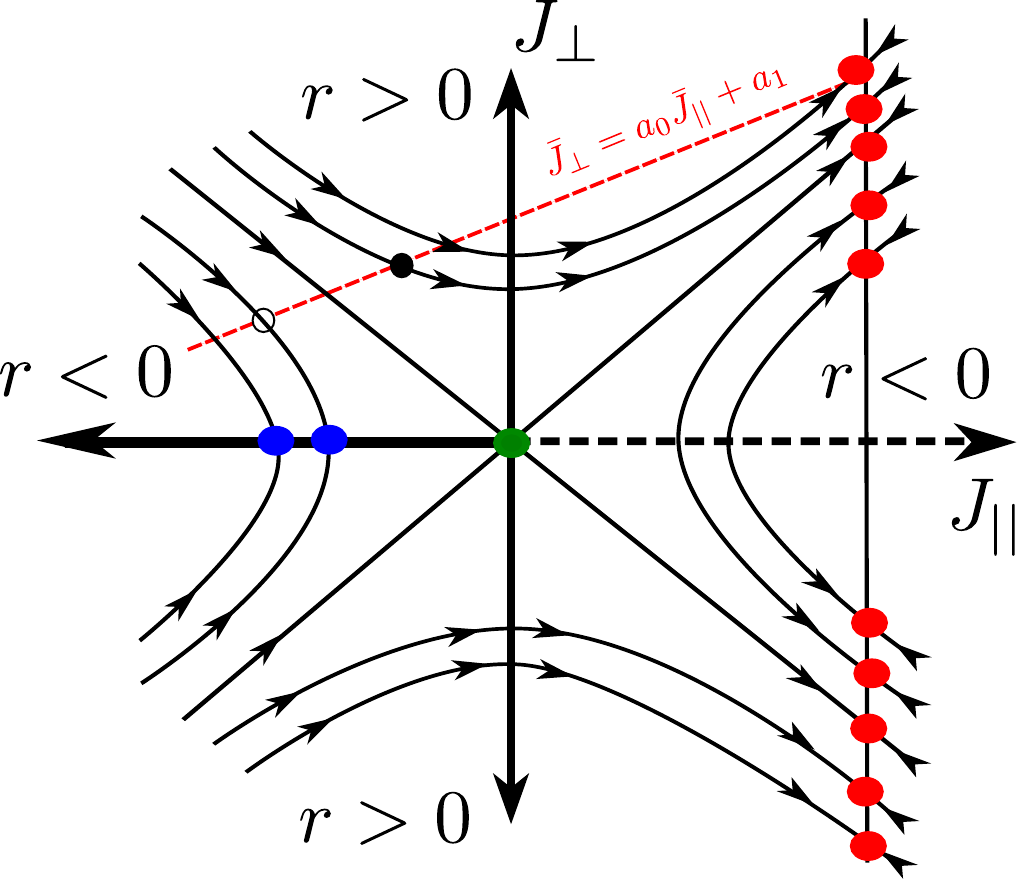}
\caption{Nonperturbative BKT RG phase diagram for the LE/Mott instabilities of the TLL. The WZNW transitions are depicted by the separatrices (i.e., the $r=0$ trajectories). The red line drawn across trajectories shows the transition from the TLL fixed points (blue circles) to the LE/Mott phases (red circles). The dashed line (including the origin) corresponds to a line of interaction driven Fermi surface topology changing (Lifshitz) transitions of the Fermi surface.}\label{RG-topology}
\end{figure}
The critical and intermediate coupling stable fixed point features of the RG phase diagram are depicted in Fig.(\ref{RG-topology}). Remarkably, the essential structure of this RG phase diagram is captured by the skeletal phase diagrams presented earlier in Fig.(\ref{topological_objects}a) and (b). We recall that the diagrams in Fig.(\ref{topological_objects}a) and (b) were obtained purely from the consideration of the topological features of the Fermi surface Hamiltonian.
\pin
We can now use the RG invariant $r$ to write the two RG equations in a combined fashion 
\begin{eqnarray}
\frac{dJ_{\perp}}{d\ln\Lambda} &=&\frac{\text{sgn}(J_{\perp})\text{sgn}(J_{||})\sqrt{J_{\perp}^{2}-|r|sgn(r)}}{1-\frac{\text{sgn}(J_{||})}{4}\sqrt{J_{\perp}^{2}-|r|sgn(r)}}~.
\end{eqnarray}
Using eq.\eqref{berry_phase_dirac}, it is possible to simplify this RG equation for the case of the WZNW lines ($r=0,J_{||}=\pm J_{\perp}\equiv J$) in terms of the topological properties of the Fermi surface, i.e., the first Chern class $\gamma$ and the $\Theta$ term $\gamma_{b}$    
\begin{eqnarray}
\frac{d|J|}{ d\ln\Lambda }&=& e^{i(\gamma -\gamma_{b})/2}\frac{|J|^{2}}{1-e^{i(\gamma -\gamma_{b})/2}\frac{|J|}{4}}~,
\label{RGWZNW}
\end{eqnarray}
where particle-hole/time reversal symmetry leads to $\gamma =\pi$ (eq.\eqref{1stchernclass}). We recall that a change in $\Theta$-term (eq.\eqref{berry_phase_dirac}) with the sign of $J$ coupling, i.e., $\gamma_{b}=\gamma\text{sgn}(J)$, is associated with a change in the Fermi surface LES from $LES_{\gamma_{b},J<0}$ to $LES_{-\gamma_{b},J>0}$ (eq.\eqref{LES_FS}). Indeed, as seen from RG flow eq.\eqref{RGWZNW}, this change in $\gamma_{b}$ triggers the back-scattering instability, leading to an irrelevant coupling turning dangerously relevant. Further, the RG equation eq.\eqref{RGWZNW} has a $Z_{2}$ helicity-inversion symmetry, $(J_{||},J_{\perp})\mapsto(J_{||},-J_{\perp})$, given by the unitary operation: $\hat{K}\equiv\hat{U}^{1/2}_{1}=\exp\left[i\pi /2(A^{z}_{\Lambda a}-A^{z}_{\Lambda b})\right])$. 
\par \noindent
The enhanced symmetry of the critical point $r=0,J_{\perp}=0$ can now be seen as follows. For $\gamma_{b}=-\gamma$, $d|J| /d\log\Lambda<0$ (irrelevant RG flows) and $\Lambda \rightarrow 0$ leads to the coupling sequence $\lbrace |J_{\Lambda_{n_{0}}}|>|J_{\Lambda_{n_{0}-1}}|>\ldots >|J_{\bar{n}}|>|J_{\bar{n}-1}|=0\rbrace$.  This tracks the passage through a space of $SU(2)_{a+b}$ symmetric theories ending at a $SU(2)_{a}\otimes SU(2)_{b}$ fixed point theory. The enhanced symmetry of the critical point is associated with two Chern invariants $C_{F_{1}}$ and $C_{F_{2}}$ in the pseudo-spin composite operator basis $F$ (see Sec.(\ref{sec4})). Remarkably, it is also a consequence of the independent conservation laws for the chiral currents $J_{L}$ and $J_{R}$ for the gapless Fermi surface. We recall that in eq.\eqref{atiyah-singer-index}, these conserved currents gave rise to Chern invariants $\nu_{a}$ and $\nu_{b}$. In the presence of an electric field, this gapless spectrum will again display the axial anomaly seen earlier the non-interacting metal (eq.\eqref{axial anomaly}). Similarly, the relation observed earlier between the central charge ($c$) of the associated conformal field theory and the Atiyah-Singer indices ($\nu_{L}-\nu_{R}$) holds here as well.
\\
\par
\begin{figure}[h!]
\includegraphics[scale=2]{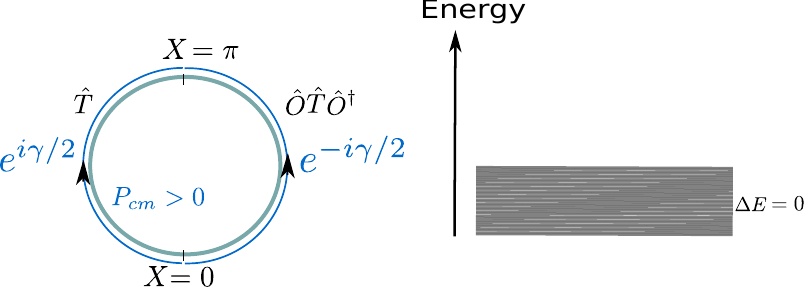}
\caption{Destructive interference in the scattering processes involving the center of mass Hilbert space (left) leads to a gapless TLL metal (right). The interference shows up between two opposing trajectories involving the twist ($\hat{O}$) and translation ($\hat{T}$) operations of the c.o.m. position $X$, with phases $\pm\gamma/2$ respectively.}
\label{LSM-1}
\end{figure}

\noindent{\bf \textit{Topological structure of the RG phase diagram}}
\pin
For RG invariant $r\neq 0$, the $\Theta$-term ($\gamma_{b}=f$ in eq.\eqref{berry_phase_dirac}) changes from $\gamma_{b}=\gamma$ (for either $r>0 \text{ or }(r<0,J_{||}>0)$) to $\gamma_{b}=0$ $(r<0,J_{||}<0)$. This change shows up as a distinction between irrelevant ($\gamma_{b}=0$) and relevant ($\gamma_{b}=\gamma$) RG flows (above and below the $r=0$ WZNW separatrices respectively) in the equation 
\begin{eqnarray}
\frac{d|J_{\perp}|}{d\log\Lambda} =e^{i(\gamma-\gamma_{b})}\frac{\sqrt{J_{\perp}^{2}-r}}{1-\frac{1}{4}\sqrt{J_{\perp}^{2}-r}}~.\label{RGanisotropic}
\end{eqnarray}
By combining the isotropic (eq.\eqref{RGWZNW}) and anisotropic (eq.\eqref{RGanisotropic}) RG flows, we obtain the RG equation
\begin{eqnarray}
\frac{d|J_{\perp}|}{d\log\Lambda} =e^{i\varpi(r)}\frac{\sqrt{J_{\perp}^{2}-r}}{1-\frac{1}{4}\sqrt{J_{\perp}^{2}-r}}
\end{eqnarray}
where $\varpi(r)$ is given by
\begin{eqnarray}
\varpi(r) &=& (\gamma -\gamma_{b})~, r\neq 0~,\nonumber\\
           && (\gamma-\gamma_{b})/2~, r=0~. \label{topological angle}
\end{eqnarray}
Changing $\varpi$ from $\theta :0\rightarrow\gamma$ leads to constructive interference $2\varpi =2\pi$ between clockwise \& anticlockwise paths in the c.o.m PHS $ P_{cm}:[0,2\pi)$. This causes vortices and anti-vortices in momentum-space, $\hat{O}_{c/s}^{2}=e^{i(2\pi/\gamma)X_{c/s}}$ and $\hat{O}_{c/s}^{2\dagger}=e^{-i(2\pi/\gamma)X_{c/s}}$ respectively, to bind via the spin/charge backscattering term $J_{\perp}(\hat{O}_{c/s}^{2}+h.c.) \equiv J_{\perp}(A^{+}_{a}A^{-}_{b}+h.c.)$~\cite{nakamura2002lattice}. The red dots in Fig.(\ref{RG-topology}) for the $J_{\perp}>0$ and $J_{\perp}<0$ regime indicate stable fixed points at intermediate coupling, and lead to (charge/spin) vortex-antivortex pseudospin singlet and triplet condensates which are odd and even under (helicity/chirality) exchange respectively. This is equivalent to the unbinding of real space vortex-anti vortex pairs, as is well known for the BKT transition~\cite{berezinskii1971destruction,kosterlitz1978two}. Instead, for the cases $(r<0,J_{||}<0)$, there is no Dirac string in the LES (see Fig.(\ref{topological_objects}a): $\gamma_{b}=0$, and the $e^{i\gamma}=-1$ phase factor in the RG equation, and will lead to destructive interference between paths traversed in clockwise and anticlockwise senses in the c.o.m PHS $P_{cm}\in [0,\pi)$ (as shown in Fig.(\ref{LSM-1})). The irrelevant RG flows then lead to a line of blue dots in Fig.(\ref{RG-topology}), corresponding to gapless Tomonaga-Luttinger liquid (TLL) metallic system~\cite{tomonaga1950remarks,luttinger1960fermi}. The dashed line in Fig.(\ref{RG-topology}) corresponds to a line of Lifshitz transitions of the Fermi surface labelled by the topological angle $\varpi(r)$ (eq.\eqref{topological angle}). Finally, note that the unitary RG procedure generates effective Hamiltonian at the gapless fixed points (blue dots) and gapped fixed points (red dots) in Fig.\ref{RG-topology}. In the next section we present the mathematical forms of the effective Hamiltonians, as well as their eigenstates and eigenvalues. 
\\
\pin{\bf \textit{Effective Hamiltonians obtained at the stable fixed points of the RG flow}}
\pin
We just saw that the appearance of a $\Theta$-term with $\gamma_{b}$ (eq.\eqref{berry_phase_dirac}) in the RG eq.\eqref{RGWZNW} governs the nature of stable fixed points, i.e., $\gamma_{b}=\pi$ corresponds to non-critical phases and $\gamma_{b}=0$ corresponds to the critical phases. From the coupling RG flows for these two cases, we can construct the effective Hamiltonians $H^{*}_{0}$ and $H^{*}_{\pi}$ corresponding to the critical and gapped fixed points (blue dots and red dots respectively in Fig.(\ref{RG-topology}) as
\begin{eqnarray}
H^{*}_{0}&=&J^{*}_{1,||}A^{z}_{a}A^{z}_{b}~,\label{critical}\\
H^{*}_{\pi}&=&J^{*}_{||}A^{z}_{a}A^{z}_{b}+\frac{J^{*}_{\perp}}{2}(A^{+}_{a}A^{-}_{b}+h.c.)~.\label{non-critical}
\end{eqnarray}
 $J^{*}_{1,||},J^{*}_{||}$ and $J^{*}_{\perp}$ are the magnitudes of the couplings for the Ising and pseudospin scattering terms at the fixed points, and can be reconstructed from their definitions below eqs.\eqref{BCS_RG1} and \eqref{Mott_RG}. Note that in obtaining the above fixed point Hamiltonians, we have only accounted for only the dominant RG flows.
\par\noindent 
The values of the fixed point couplings at the Luther-Emery~(LE) and Mott liquid~(ML) phases arising from the RG flow of Hamiltonians $H_{1}$~eq.\eqref{H-BCS} and $H_{2}$~eq.\eqref{H-Mott} are tabulated in Table-\ref{table-coup}.
\begin{table}
\centering
\begin{tabular}{ |c|c|c| } 
 \hline
 $\Theta$-term~($\gamma_{b}$) & LE & ML \\
 \hline 
 $0$ & $J^{*}_{1,||}=-|r|$ &  $J^{*}_{1,||}=-|r|$\\ 
 \hline
 $\pi$ & $J^{*}_{||}=-4|\omega-\hbar v_{F}\Lambda^{*}
 |$ &  $J^{*}_{||}=4\omega$\\ 
 \hline
\end{tabular}~.
\caption{Magnitude of couplings at critical and non-critical fixed points.}\label{table-coup}
\end{table}
Along the WZNW lines ($r=0$ and with $\gamma_{b}=\pi$), the eigenstates at the stable fixed point are given by
\begin{eqnarray}
|\Psi\rangle_{\pi} =|A_{a}+A_{b}=m, A^{z}_{a}+A^{z}_{b}=0, A_{a}=\frac{N_{*}}{2}, A_{b}=\frac{N_{*}}{2}\rangle~,\label{off-cric-states}
\end{eqnarray}  
where $N_{*}$ is the number of pseudospins within each emergent window $[-\Lambda^{*},\Lambda^{*}]$ centered about the Fermi points. From $H_{\pi}$ (eq.\eqref{non-critical}), we obtain the corresponding eigenspectrum as $E(m)=J^{*}\left[m(m+1)-2N_{*}(N_{*}+1)\right]$, where $0<m<2N_{*}$ and $J^{*}=-4|\omega-\hbar v_{F}\Lambda^{*}|$ for the LE liquid, and $J^{*}=4\omega$ for the Mott liquid phase. As the LE fixed point is reached in the attractive regime $J^{*}<0$, the eigenfunctions are determined completely from $m=2N_{*}$; on the other hand, $m=0$ for the ML fixed point.  
Similarly for the critical phases governed by $H^{*}_{0}$ (eq.\eqref{critical}), the many-body eigenfunctions are given by
\begin{eqnarray}
|\Psi\rangle_{0}= |A_{a}=N_{*}, A_{b}=N_{*},A^{z}_{a}=m,A^{z}_{a}=m,A^{z}_{b}=m\rangle~,\label{cric-states}
\end{eqnarray}
and the corresponding eigenspectrum is $E(m)=J_{1,||}^{*}m^{2}$.
In subsequent sections, we will show that the electronic states at the Fermi surface are witness to the low energy features of the LE and ML phases.
\section{Holographic entanglement scaling towards the Fermi surface}\label{secholo}
\begin{figure}[h!]
\centering
\includegraphics[width=0.7\textwidth]{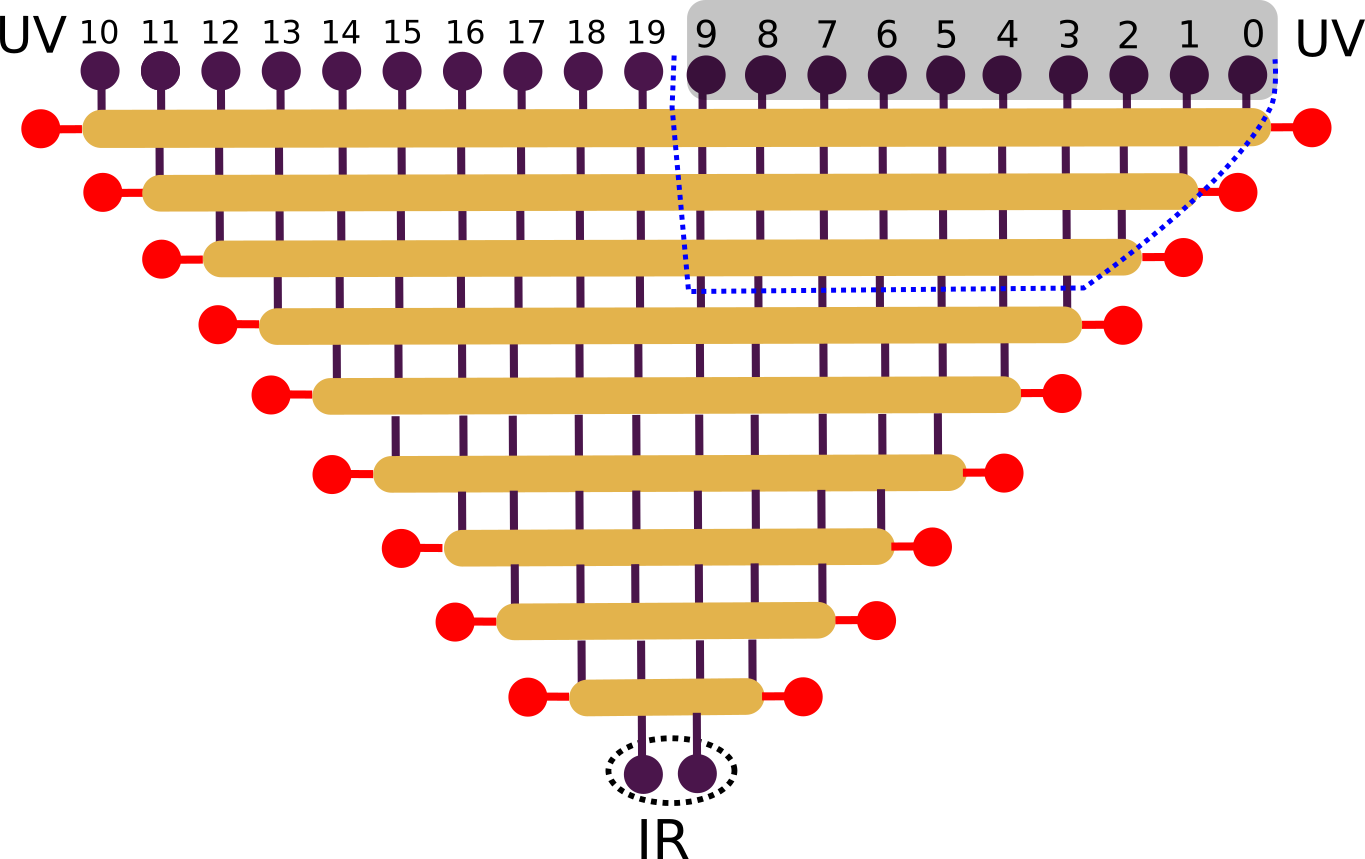}
\caption{EHM representation of URG flow towards the LE ground state. The black nodes $0,\ldots,9$ represent pseudospins with helicity $\eta=+1$, while $10,\ldots,19$ represent psuedospins with $\eta=-1$. Together, they comprise the holographic boundary in the UV. Pseudospins are labelled in descending order of energy, such that the pair $(0,9)$ is associated with the highest energy. The energy of the pairs $(1,10)$, $(2,11)$,$\ldots$, forms a monotonically decreasing sequence such that the $(9,19)$ pair is located at the Fermi energy. The yellow block represents the unitary map $U_{(j)}$ that, at every step, disentangles the highest energy pair of pseudospins (represented by red nodes in the bulk of the EHM). The dotted circle around the pair of black nodes $(9,19)$ comprise the entangled ground state emergent in the IR. The grey box represents the region (nodes 0 to 9) in the UV that is being isolated by the blue dotted \textit{minimal curve} from within the bulk of the network.}\label{entangle}
\end{figure}  
The Entanglement entropy~(EE) $S(R)$ of an interacting quantum system is a measure of many-particle quantum entanglement that is generated upon isolating a region $R$ from the rest of the system. It quantifies the information lost with regards to quantum correlations between degrees of freedom in region $R$ and its complement. Earlier works based on real space entanglement RG~\cite{vidal2007entanglement} revealed distinct scaling features of EE for gapped as against gapless phases. Using the URG formalism, some of us have recently studied entanglement RG flow towards various IR fixed points of the 2D Hubbard model, distinguishing thereby the entanglement scaling features of the normal and Mott insulating states~\cite{mukherjee2020}. The URG represents the nonlocal unitary disentanglement transformations as a product of two-local unitary operations, providing thereby a entanglement holographic mapping~(EHM)~\cite{qi2013,lee2016} or tensor network representation of URG. The URG method generates Hamiltonian and entanglement RG along the holographic scaling direction of the EHM~\cite{evenbly2011,swingle2012b,qi2013}. Further, Ref.\cite{mukherjee2020} shows that the many-body states generated by the entanglement renormalisation respects the Ryu-Takyanagi EE bound~\cite{ryu2006aspects,ryu2006}, i.e., the entanglement entropy generated upon isolating region $R$ from its complement is bounded from above by the number of links between them. In this section, we will demonstrate that the RG flow of EE and the holographic EE bound possess distinct features in the TLL phase and Luther Emery phases.
\par\noindent
As noted above, upon scaling from UV towards the IR fixed point, the URG generates effective Hamiltonians at each step. Starting with the ground state at the IR fixed point, $\Psi_{(j^{*})}$ (eq.\eqref{off-cric-states} and eq.\eqref{cric-states}) of the effective Hamiltonian $H_{(j^{*})}$ (eq.\eqref{non-critical} and eq.\eqref{critical} respectively), and performing reverse URG steps using the unitary maps $U^{\dagger}$ leads to the reconstruction of the states in the UV energy scale:
\begin{eqnarray}
\text{UV}\xrightarrow{\text{URG}} \text{IR}:&& H_{(N)}\xrightarrow{U_{(N)}}H_{(N-1)}\xrightarrow{U_{(N-2)}}\ldots\ldots\xrightarrow{U_{(j^{*}+1)}}H_{(j^{*})}~,\nonumber\\
\text{IR}\xrightarrow{\text{reverse URG}} \text{UV}:&& |\Psi_{(N)}\rangle\xleftarrow{U^{\dagger}_{(N)}}\ldots\xleftarrow{U^{\dagger}_{(j^{*}+2)}}|\Psi_{(j*+1)}\rangle\xleftarrow{U^{\dagger}_{(j^{*}+1)}}|\Psi_{(j*)}\rangle~.
\label{reverse}
\end{eqnarray}
We now discuss some important features of this scheme. Recall that at each step of the URG, two pseudospin degrees of freedom of opposite helicities $\mathbf{A}_{\Lambda_{j},+1}$ and $\mathbf{A}_{\Lambda_{j},-1}$ are disentangled, generating a Hamiltonian flow towards, say, the Luther-Emery fixed point (see Table-\ref{table-coup}). On the other hand, along the reverse RG flow, two pseudospins re-entangle at each step, enabling the reconstruction of the eigenstates at high energy scales. Fig.\ref{entangle} displays the EHM construction for the entanglement RG flow from the UV towards the LE ground state in the IR. Pseudospin states are labelled in descending order of energy, such that pair of pseudospins $|0\rangle$, $|10\rangle$ with opposite helicities~($\eta$) are located farthest from the Fermi points and naturally associated with the highest electronic pair-energy. Thus, a succession of pseudospin pairs ($|1\rangle$, $|11\rangle$), ($|2\rangle$, $|12\rangle$), $\ldots$ carry a decreasing sequence of net electronic energy, such that the pair $|9\rangle$, $|19\rangle$ is located at the Fermi points. The dotted oval in Fig.\ref{entangle} represents the entangled groundstate $|\Psi_{\pi}\rangle$ of the LE phase (eq.\eqref{off-cric-states}). Comprised of the pseudospins $|9\rangle$ with helicity $\eta=+1$ and $|19\rangle$ with helicity $-1$, it has the form $|\Psi_{\pi}\rangle=|A=1,A_{z}=0,A_{+1}=\frac{1}{2},A_{-1}=\frac{1}{2}\rangle$. The complete ground state at the IR fixed point can be obtained by performing a tensor product of $|\Psi_{\pi}\rangle$ with the disentangled pseudospin states labelled $0$ to $8$ and $11$ to $18$, $|\Psi_{(j*)}\rangle=|\Psi_{\pi}\rangle\otimes_{i=0}^{8}|s_{i}\rangle\otimes_{i=11}^{18}|s_{i}\rangle$. Here, the $s_{i}$ are the $|\Uparrow\rangle$ and $|\Downarrow\rangle$ configurations of the individual pseudospins. The reverse unitary operations $U^{\dagger}$ (yellows blocks in Fig.\ref{entangle}, eq.\eqref{reverse}) map the ground state from IR to UV (with RG step $j^{*}=1$ being the fixed point): $U^{\dagger}_{1}|\Psi\rangle$, $U^{\dagger}_{2}U^{\dagger}_{1}|\Psi\rangle$, $U^{\dagger}_{3}U^{\dagger}_{2}U^{\dagger}_{1}|\Psi\rangle$, $\ldots$, generating thereby the EHM tensor network.
\begin{figure}
\includegraphics[width=\textwidth]{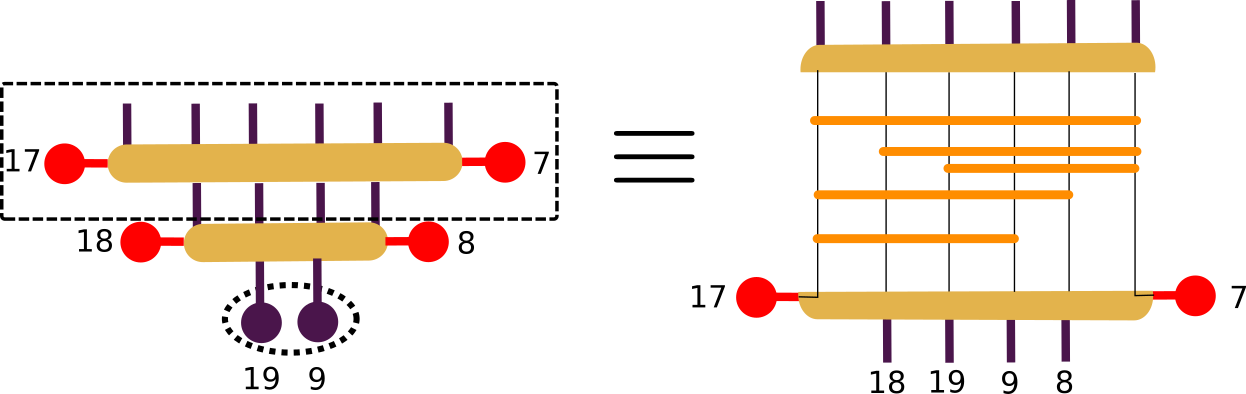}
\caption{Decomposition of nonlocal unitary disentanglement map $U_{2}$ into two-qubit disentanglers represented as orange blocks (within the right figure). $U_{2}$ decouples the pseudospins/qubits $|17\rangle$ and $|7\rangle$ (red nodes) from the pseudospins $|8\rangle$, $|9\rangle$, $|18\rangle$, $|19\rangle$ and has depth (d=number of orange blocks acting sequentially) $d=5$. An orange block operates on the two pseudospins demarcated by the lines crossing its edges.}\label{EHMcircuit}
\end{figure}
\par\noindent
An essential feature of the EHM network is that it can be represented entirely as a product of two-qubit disentangling gates~\cite{qi2013}. Such a decomposition will allow us to interpret the URG as a quantum circuit renormalization group~\cite{mukherjee2020}. For the present EHM network shown in Fig.\ref{entangle}, we show some aspects of the equivalent quantum circuit in Fig.\ref{EHMcircuit}. The second last yellow block $U_{3}$ (connecting states $|\Psi_{2}\rangle$ and $|\Psi_{3}\rangle=U^{\dagger}_{3}|\Psi_{2}\rangle$) can be decomposed as follows
\begin{eqnarray}
U_{3}=U_{17,9}U_{17,8}U_{7,19}U_{7,18}U_{7,17}~,\label{structGate}
\end{eqnarray}
where $U_{i,j}$ disentangles the pair of pseudospins $|i\rangle$ and $|j\rangle$ (orange block in Fig.\ref{EHMcircuit}). Note that in the parent Hamiltonian for the RG analysis (eq.\eqref{H-BCS}), we have only accounted for the backscattering diagrams that couple pairs of electronic states with opposite helicity ($\eta$). In the second last RG step, the collection of unitary disentanglers $U_{7,19}U_{7,18}U_{7,17}$ decouple the pseudospin $|7\rangle$ with helicity $\eta=+1$ from the pseudospins $|17\rangle$, $|18\rangle$, $|19\rangle$ of opposite helicity $\eta=-1$. This is followed by the next set of unitary maps $U_{17,9}U_{17,8}$ that disentangle pseudospin $|17\rangle$ of helicity $\eta=-1$ from $|8\rangle$ and $|9\rangle$ with helicity $\eta=+1$. The depth $d$ of the orange block equals the number of $U_{a,b}$ operations carried out sequentially (in eq.\eqref{structGate}) to complete the disentanglement operation $U_{j}$. For $U_{3}$, as shown in Fig.\ref{EHMcircuit}, the circuit depth $d=5$. The form of the individual qubit disentanglers $U_{a,b}$ in eq.\eqref{structGate} are constrained by the analytical form of the complete unitary map $U_{j}$ (eq.\eqref{UnitaryOpDef}) determined by the renormalized Hamiltonian $H_{(j)}$. Many-body states $U_{7,17}|\Psi_{3}\rangle$, $U_{7,18}U_{7,17}|\Psi_{3}\rangle$, $\ldots$ generated at each sublayer of the circuit composing $U_{3}$ (see Fig.\ref{EHMcircuit}) involve the vanishing of mutual information $I(i:j)=S(i)+S(j)-S(i,j)$~\cite{qi2013,hyatt2017} for the pair of qubits $(7,17)$, $(7,18)$ and so on~\cite{mukherjee2020}. $S(i),S(j)$ is the entanglement entropy associated with isolating qubit $i$ and $j$ from the system and $S(i,j)$ is the joint entropy associated with the isolating pair of qubits $(i,j)$.
\par
\begin{figure}
\centering
\includegraphics[width=\textwidth]{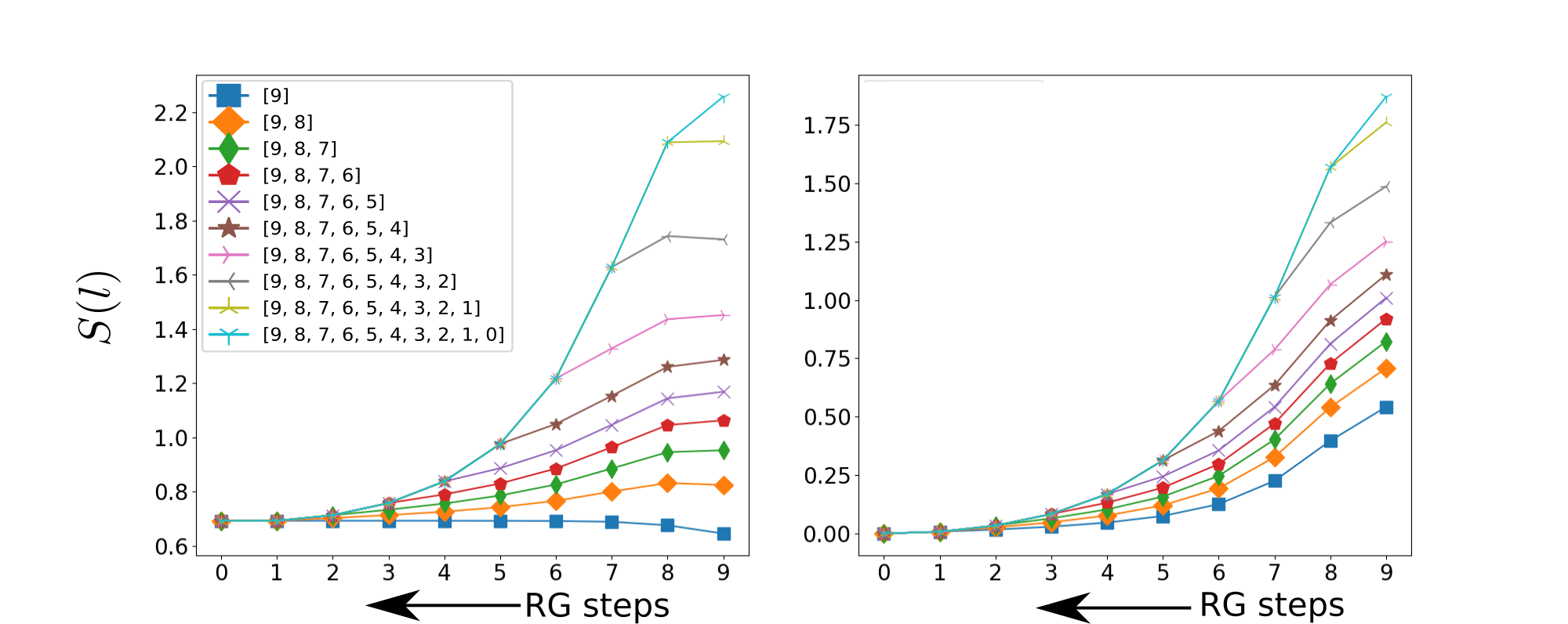}
\caption{Entanglement Entropy $S(l)$ renormalization for varying sizes of momentum-space blocks $(9),(9,8),(9,8,7),\ldots, (9,8,\ldots,0)$~(i.e.g, block size $l$ ranging from $1$ to $10$) for (a) the LE and (b) the TLL phases. See discussion in main text.}\label{Entropy}
\end{figure}

\noindent 
Upon implementing the reverse URG, we isolate at each RG step $j$ momentum-space blocks of increasing lengths $(9)$,$(9,8)$, $(9,8,7)$, \ldots $(9,8,\ldots,0)$, and obtain therefrom the EE $S_{j}(l)$
\begin{eqnarray}
S_{j}(l)=-Tr(\rho_{j}(l)\log\rho_{j}(l)),~\rho_{j}(l)=Tr_{9,\ldots, 10-l}(|\Psi_{(j)}\rangle\langle\Psi_{(j)}|)~,
\end{eqnarray}
where $l\in[1,10]$ is the size of the momentum-space block and $\rho_{j}(l)$ is the reduced density matrix. For this, the state $|\Psi_{(j)}\rangle$ is generated from $|\Psi_{(j^{*})}\rangle$ by doing reverse URG: $|\Psi_{(j)}\rangle=U^{\dagger}_{j}U^{\dagger}_{j-1}\ldots U_{j^{*}}|\Psi_{(j^{*})}\rangle$. We also provide a reverse URG formulation for the TLL phase starting from the state $|\Psi_{0}\rangle=|A^{z}_{+1}=\frac{1}{2}, A^{z}_{-1}=\frac{1}{2}\rangle$ (eq.\eqref{cric-states}). This allows us to construct a EHM quantum circuit realization of the URG for the TLL similarly to Figs. (\ref{entangle}) and (\ref{EHMcircuit}) given above, generating the URG flow for the block entanglement entropy of the TLL. The left and right panels of Fig.\ref{Entropy} show the entanglement entropy RG flow for different momentum-space block sizes in the LE and TLL phases respectively. Along the forward RG flow, the successive disentanglement of pseudospins leads to a gradual reduction of block EE: the disentanglement of UV degrees of freedom reduces the entanglement sharing between IR and UV degrees of freedom. It is important to note, however, that in the left panel of Fig.\ref{Entropy} the EE for the lowest block size $9$ at the Fermi point increases along the RG flow, terminating at a final value of $S=\log 2$. This shows that the entanglement between Fermi points at the IR fixed pointis enchanced, leading to formation of the maximally entangled state  
$|\Psi_{\pi}\rangle=\sqrt{2^{-1}}[|\Uparrow_{+1}\Downarrow_{-1}\rangle-|\Downarrow_{+1}\Uparrow_{-1}\rangle]$. On the other hand, for the TLL phase, the momentum-space block EE is observed to decrease monotonically to zero along the RG flow for \textit{all} block sizes, suggesting perfect disentanglement of the pseudospins. In this way, we observe distinct scaling features for the TLL and LE phases.
\par\noindent
Finally, we turn to display the holographic feature of the EHM: when computed from the bulk of the EHM tensor network, the EE associated with region $R$ possesses an upper bound related to the number of degrees of freedom in $R$ ($n(R)$) that are entangled~\cite{swingle2012b,evenbly2011,qi2013}. To compute the upper bound, we multiply $n_{j}(R)$ with the maximum one-pseudospin entanglement entropy $S_{j}(i)$ at RG step $j$, such that
\begin{eqnarray}
S_{j}(R)\leq n_{j}(R)\times \max_{i\in R}(S_{j}(i))~.\label{bound}
\end{eqnarray} 
This leads to the Ryu-Takyanagi formula for entanglement entropy~\cite{ryu2006aspects,ryu2006}. The grey block ranging from pseudospin $0$ to $9$ in Fig.\ref{entangle} represents the region $R$ at UV, while the dotted blue line is the minimal surface, or, equivalently, the number of links cut to isolate $R$ from deep within the EHM. Evidently, the minimal surface shrinks as we proceed deeper into the bulk of the EHM such that $n_{j}(R)$ reduces. In the present case of Fig.\ref{entangle}, $n_{j-1}(R)=n_{j}(R)-1$.
 \begin{figure}
 \centering
 \includegraphics[width=0.7\textwidth]{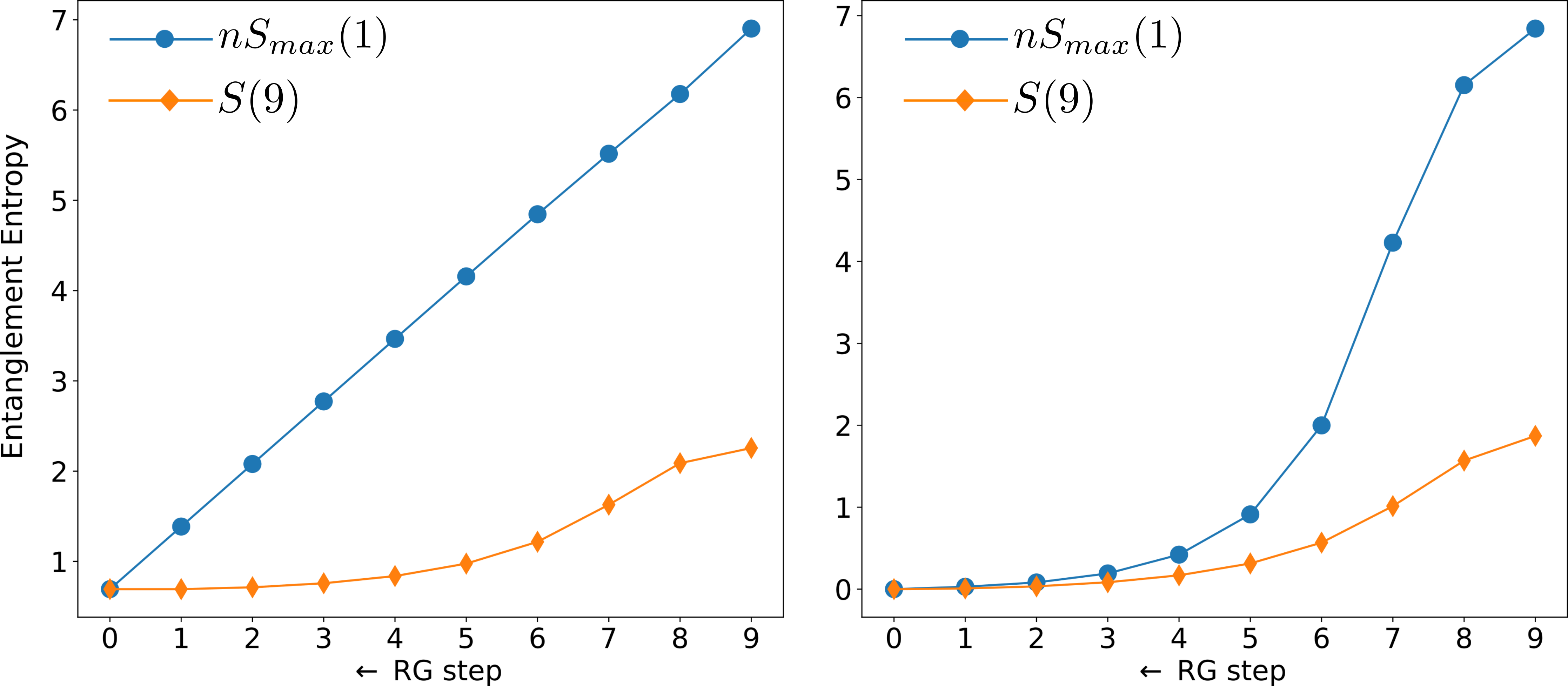}
 \caption{Blue and orange curves show the RG scaling of the holographic entanglement entropy bound and the entropy scaling of the the momentum-space block $(9,\ldots ,0)$. Left panel represents the curves for the LE phase, and right panel for the TLL phase. The legend $S(9)$ is the entanglement entropy generated by isolating block of size $9$ units, $S_{max}(1)$ is the maximum single-pseudospin EE and $n$ the size of the minimal surface in the bulk of the EHM (varies with the RG step). See discussion in main text.}\label{entropyScaling}
 \end{figure}
In Fig.\ref{entropyScaling}, we confirm the holographic entanglement entropy relation (eq.\eqref{bound}) with $R=\lbrace 0,1,\ldots,9\rbrace$ for both the LE and TLL phases. The blue curve shows the holographic entropy upper bound, while the orange curve shows the entropy computed for the region $R$ from the bulk of the EHM. Importantly, the distinct holographic entanglement entropy scaling features imply that it is a witness to the entanglement phase transition between the critical TLL phase (blue dots in the phase diagram Fig.\ref{RG-topology}) and the gapped LE phase (red dots of Fig.\ref{RG-topology}). First, we note that the $S_{max}(1)$ leading to the EE upper bound for the LE and TLL phases are distinct: while it arises entirely from the degree of freedom at the Fermi surface (pseudospin $9$) along the RG trajectory (step $9$ to step $0$) for the LE phase, it shifts gradually from UV (pseudospin $0$) to the IR (Fermi surface pseudospin $9$) for the TLL as the RG proceeds. Second, the shape of the upper bound for the two phases is quite different: while it scales linearly with the logarithmic RG step size for the LE phase, it is clearly non-linear for the TLL phase. While the latter is reminiscent of the rapid expansion of holographically generated entanglement spacetime expected for a gapless phase (i.e., the AdS-CFT conjecture for continuum field theories~\cite{maldacena1999large,witten1998anti}, and its discrete counterpart in lattice field theories~\cite{swingle2012a}), this will need further investigation and will be presented elsewhere. Though we have not shown the analysis here, precisely similar results to those shown for the LE phase are also obtained for the gapped ML phase described earlier. 
\section{Observing the instability of the Fermi surface}
\label{sec6}
\pin{\bf \textit{Dynamical spectral weight transfer}}
\pin
We recall that a topological constraint (eq.\eqref{topological_constraint}) ensured that the FS backscattering vertex acts on the $2$-electron subspace 
\begin{eqnarray}
N_{F}=\int_{-\infty}^{\infty} \frac{d\omega}{i\pi}\frac{\partial Tr_{F_{1}}(ln(\hat{G}_{0}))}{\partial\omega}=2~,
\end{eqnarray}
where $\hat{G}_{0}$ is the non-interacting Greens' function given in eq\eqref{retarded_green_function}. The vertex is described by $H_{F_{1}}$ (eq.\eqref{backscattering_FS}), and satisfies the criterion~\cite{lieb1961two} 
\begin{eqnarray}
N_{F}\gamma =2\pi~,~\gamma =2\pi\nu~,~2\pi S=\pi~,
\end{eqnarray}
where $\nu=\gamma /2\pi=1/2$ and $S=1/2$ are the Luttinger sum for the charge and spin excitations (eq.\eqref{1stchernclass}, in the presence of p-h and TRS symmetries) respectively. As seen earlier, backscattering leads to the formation of composite objects with spectral weight given by Friedel's sum rule~\cite{weinberg2015lectures}, relating the scattering phase-shift ($\delta$) to the scattering matrix ($\hat{S}_{F_{1}}$)
\begin{eqnarray}
2\delta &=&2Tr_{F_{1}}(ln(\hat{S}_{F_{1}}))~,~
\hat{S}_{F_{1}}= 1-2\pi \hat{G}_{0F_{1}}\frac{H_{F_{1}}}{1-\hat{G}_{0F_{1}}H_{F_{1}}}~,\label{scattering_matrix}
\end{eqnarray}
where $H_{F_{1}}$ is given in (eq.\eqref{backscattering_FS}). The scattering matrix $\hat{S}_{F_{1}}$ has the following matrix representation in the basis $(1/\sqrt{2})\left[|\uparrow_{Fa}\downarrow_{Fb}\rangle \pm|\downarrow_{Fa}\uparrow_{Fb}\rangle\right]$
\begin{eqnarray}
\hat{S}_{F_{1}}=\left[\begin{array}{cc} e^{i\delta_{s}} & 0 \\
										    0 & e^{i\delta_{t}}
					  \end{array}\right]~,
\end{eqnarray}
where the pseudospin singlet/triplet scattering phase shifts are given respectively by $\delta_{s}=\tan^{-1}(3\pi J/4),\delta_{t}=\tan^{-1}(-\pi J/4)$. We recall that the change in c.o.m momentum under twist via a full flux-quantum is related to the Luttinger sum $\gamma/2\pi$ (eq.\eqref{1stchernclass}), $\Delta P_{cm}=\gamma/2\pi$~. Due to Kohn's theorem~\cite{kohn1961cyclotron}, this relation holds even in the presence  of electronic interactions. Thus, in the presence of a putative instability of the FS, the total spectral weight associated with the FS subspace $F_{1}$ within the window $L_{n^{*}}$ (i.e., of the composite objects, together with that of any remnant fermionic degrees of freedom) is given by a generalized Luttinger's sum rule~\cite{langer1961friedel, martin1982fermi},
\begin{eqnarray}
\frac{2\pi}{\gamma}&=&Tr(ln(\hat{G}_{F_{1}}(0+i\eta)))-Tr(ln(\hat{G}_{F_{1}}(0-i\eta))) + Tr(\ln(\hat(S)_{F_{1}}))~,
\end{eqnarray}
where the fermionic single-particle Greens' function is given by $\hat{G}_{F_{1}}(E)=(\hat{G}_{0F_{1}}(E)^{-1}-H_{F_{1}})^{-1}$~. The RG equation for the pseudo-spin singlet scattering phase-shift $\delta_{s}$ along the WZNW line $J_{||}=J_{\perp}=J$ is then given by 
\begin{equation}
\frac{d\delta_{s}}{d\ln\Lambda} = \frac{e^{i(\gamma -\gamma_{b})/2}}{1+\frac{9\pi^{2}}{4}J^{2}}\frac{d|J|}{d\ln\Lambda} ~.
\end{equation}
This relation shows that the Friedel's phase shift changes non-analytically across the critical point $J=0$ for the WZNW flows as $\gamma_{b}$ changes from $-\gamma$ to $\gamma$. Further, for $\gamma_{b}=\gamma$, unitarity dictates that the increasing dynamical spectral weight transfer between fermions and composite degrees of freedom within the LEB be obtained from
\begin{equation}
\frac{d}{d\ln\Lambda}Tr(ln((\hat{S})_{F_{1}}))+\frac{d}{d\ln\Lambda}Tr(ln(\hat{G}_{F_{1}}))=0 ~.
\end{equation}
The dynamical spectral weight transfer stops eventually at the IR fixed point $J^{*}=4$, where the scattering phase shift $\delta_{s}^{*} =\tan^{-1}(3\pi J^{*}/4)=\tan^{-1}(3\pi)$. For the $N_{F}=2$ particle subspace at the Fermi surface, we define a quantity $\bar{N}_{1}$~\cite{gurarie2011single} that tracks spectral weight redistribution in terms of $N_{F}$ and the Friedel's scattering phase shift ($\delta_{s}$)~\cite{martin1982fermi}
\begin{eqnarray}
\bar{N}_{1}&=&\int_{-\infty}^{\infty} \frac{d\omega}{i\pi}\frac{\partial Tr_{F_{1}}(ln(\hat{G}))}{\partial\omega}=N_{F}-\frac{2\delta_{s}}{\pi} = N_{F}-\frac{2Tr(ln\hat{S}_{F_{1})}}{\pi} ~.~\label{spectral weight redistribution}
\end{eqnarray}
\\
\par
\begin{figure}[h!]
\centering
\includegraphics[width=0.5\textwidth]{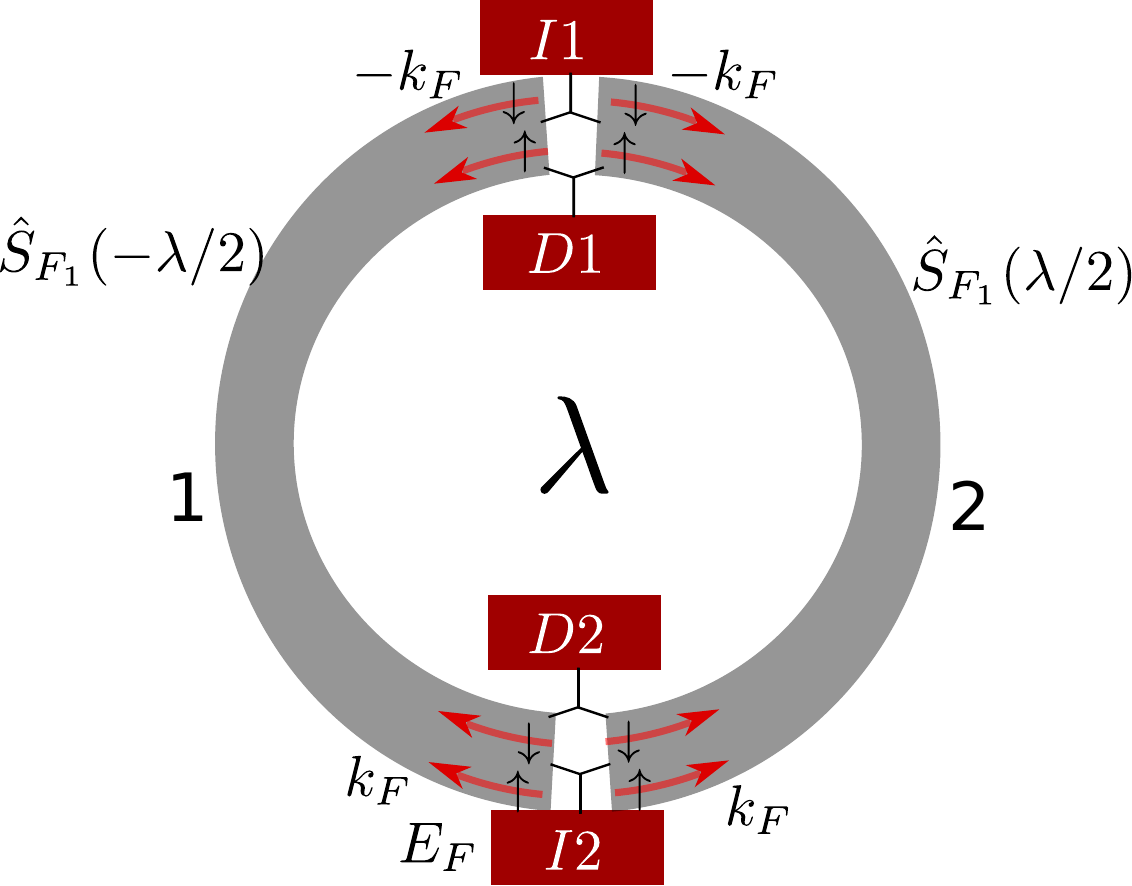}
\caption{Thought experiment for probing the entanglement entropy related to two-particle scattering through transport measurements. Two open 1D systems $1$ and $2$ (at the brink of LE/Mott instability) are coupled to electron injectors $I1$,$I2$ and detectors $D1$, $D2$ via Y junctions. The ring geometry encloses a flux $\lambda$. The Y junctions of injectors $I1$ and $I2$ feed two electrons ($|\uparrow_{Fa}$ and $\downarrow_{Fb}\rangle$) to either system $1$ or $2$. This results in a superposition state $\Psi^{F}_{s}(\lambda)$ created by the scattering process described by matrices $S_{F_{1}}(\lambda/2)$ and $S_{F_{1}}(-\lambda/2)$. See main text for discussions.}\label{ring_geom_exp}
\end{figure}

\noindent{\bf \textit{Measuring full counting statistics for Fermi surface electrons}}
\pin
We have seen in earlier sections that the electronic degrees of freedom at the FS possess topological attributes that track the two- particle scattering induced instabilities. In this spirit, we now propose a ``two-path" thought experiment for a ring-like geometry of the interacting 1D electronic system. This gedanken aims to measure the various moments of the spin/charge backscattering dynamics of FS electrons (full counting statistics (FCS)) in the presence of a putative instability. As shown in Fig.(\ref{ring_geom_exp}), the setup has two open identical 1D systems that are tunnel coupled to injectors $I1$ and $I2$ and detectors $D1$ and $D2$ in a ring geometry enclosing a flux $\lambda=\Phi/\Phi_{0}$. The injectors and detectors are momentum resolved, such that they inject and extract electrons in a resonant manner at the Fermi energy $E_{F}$, and with well defined helicity/chirality. Further, the injection and extraction events involve 1-particle superpositions across the two arms of the ring.
\pin
 In order to track the BCS instability in the two 1D systems, injectors $I1$ and $I2$ simultaneously inject an electron each with a given helicity $\eta =\pm 1$, but with oppositely directed momenta $-k_{F}$ and $+k_{F}$ respectively. The injectors are switched off immediately after the injection, and the detectors are switched on simultaneously. The injected electrons suffer backscattering in each of the arms, and reach the two detectors through trajectories that involve two-particle interfering pathways between the two arms that together enclose the AB flux $\lambda$. This two-particle interference thus provides information related to the correlations accrued from the backscattering processes in the two arms. For the Mott instability, two electrons are instead injected from $I1$ in superposition between the two arms and are extracted at $D2$ (while $I2$, $D1$ are deactivated). This thought experiment is, therefore, a two-particle fermionic Hanbury-Brown-Twiss setup whose purpose is to expose the interplay of interparticle correlations and Fermi surface topology. An Andreev scattering variant of this gedanken would involve injection of electrons and extraction of holes.
\pin
The two-electron scattering matrix at the Fermi surface eq.\eqref{scattering_matrix} in the basis $F_{1}=\lbrace|\uparrow_{Fa}\downarrow_{Fb}\rangle, |\downarrow_{Fa}\uparrow_{Fb}\rangle\rbrace$ of the composite pseudospin instability operator (see Sec.(\ref{sec4})) is given by
\begin{eqnarray}
\hat{S}_{F_{1}}=\frac{1}{2}\left[\begin{array}{cc}
r & t\\
t & r
\end{array}\right]~,
\end{eqnarray} 
where $\delta_{s},\delta_{t}$ are the pseudospin singlet/triplet phase shifts defined earlier for the Fermi surface electrons, and $r=e^{i\delta_{s}}+e^{i\delta_{t}}$,$t=e^{i\delta_{s}}-e^{i\delta_{t}}$ are the reflection and transmission coefficients respectively. The $r$ and $t$ coefficients can be represented as follows
\begin{eqnarray}
r&=&\sqrt{\frac{\gamma_{s}}{2\pi}}e^{i\theta_{r}}~,~t=\sqrt{\frac{2\pi -\gamma_{s}}{2\pi}}e^{i\theta_{t}}~,\nonumber\\
\theta_{r}&=&\arctan\left(\frac{\sin\delta_{s}+\sin\delta_{t}}{\cos\delta_{s}+\cos\delta_{t}}\right)~,~
\theta_{t} = \arctan\left(\frac{\sin\delta_{s}-\sin\delta_{t}}{\cos\delta_{s}-\cos\delta_{t}}\right)~,
\end{eqnarray}
where the two-electron wavefunction upon scattering in PHS $F_{1}$ acquires a geometric phase
\begin{equation}
\gamma_{s}=\gamma(1-\cos(\delta_{s}-\delta_{t}))\label{scattering_geometric_phase}
\end{equation}
associated with the monopole of charge $2S=\gamma/\pi$ (see discussion in Sec.(\ref{sec4})). The geometric phase $\gamma_{s}$ is acquired as follows. The initial state $|\uparrow_{Fa}\downarrow_{Fb}\rangle$ of the injected electrons is an equal amplitude superposition of the two arms $1$ and $2$. For instabilities in the the two arms and in the absence of the AB flux $\lambda$, this initial state scatters to
\begin{eqnarray}
|\psi(\omega)\rangle &=&\hat{S}_{F_{1}}|\uparrow_{Fa}\downarrow_{Fb}\rangle = e^{i\theta_{r}}(|r||\uparrow_{Fa}\downarrow_{Fb}\rangle +e^{i\omega}|t||\downarrow_{Fa}\uparrow_{Fb}\rangle )~.
\end{eqnarray}
This scattered state then traverses a closed loop over $\omega=\theta_{t}-\theta_{r}\in [0,2\pi)$, acquiring a scattering berry-phase $\gamma_{s}$ computed from the berry potential $\mathbf{A}=i\hat{\phi}\langle\psi(\omega)|\partial_{\omega}|\psi(\omega)\rangle$ as follows,
\begin{eqnarray}
\gamma_{s}=\oint d\boldsymbol{\omega}\cdot\mathbf{A}~,~ \boldsymbol{\omega}=\omega\hat{\phi}~.
\end{eqnarray}
\pin
We define flux-resolved momentum space Wilson loop operators that encode the amount of effective flux observed by charge/spin degrees of freedom
\begin{eqnarray}
\hat{\bar{Z}}^{\lambda}_{c/s}&=&\hat{\bar{T}}_{c/s}\hat{\bar{O}}_{c/s}^{\lambda}\hat{\bar{T}}^{\dagger}_{c/s}\hat{\bar{O}}_{c/s}^{\lambda\dagger}~,~\nonumber\\
\hat{\bar{Z}}^{\lambda}_{c}&=&\exp\left[i\gamma\lambda(\hat{n}_{k_{F}\uparrow}+\hat{n}_{k_{F}\downarrow}-1)\right]~,~ 
\hat{\bar{Z}}^{\lambda}_{s}= \exp\left[i\gamma\lambda(\hat{n}_{k_{F}\uparrow}-\hat{n}_{k_{F}\downarrow})\right]~.
\end{eqnarray}
In the above expressions for $\hat{\bar{Z}}^{\lambda}_{c/s}$, we have used the spin ($\hat{\bar{T}}_{s}$) and charge ($\hat{\bar{T}}_{c}$) translation operators given by
\begin{eqnarray}
\hat{\bar{T}}_{s}&=&\exp[i\sum_{|k|<|k_{F}|}k(\hat{n}_{k\uparrow}-\hat{n}_{k\downarrow})]~,~
\hat{\bar{T}}_{c} = \exp[i\sum_{|k|<|k_{F}|}k(\hat{n}_{k\uparrow}+\hat{n}_{k\downarrow}-1)]~,
\end{eqnarray}
and the spin ($\hat{\bar{O}}_{s}$) and charge ($\hat{\bar{O}}_{c}$) twist operators defined as
\begin{eqnarray}
\hat{\bar{O}}_{c}=\exp[i\hat{X}_{cm}]~,~\hat{\bar{O}}_{s}=\exp[i(N/N_{e})\hat{X}_{cm}]~.
\end{eqnarray}
These Wilson loop operators are the momentum-space duals of the real-space variants shown in eq.\eqref{wilson_loop_operator} (and which led to the first Chern class $\gamma$ eq\eqref{1stchernclass}). Further, they track boundary condition changes at the Fermi surface for the spin/charge degrees of freedom. In the scattering problem between the two Fermi points ($\pm k_{F}$), backscattering processes can be visualized on the ring in terms of interfering clockwise (1) and a anticlockwise (2) paths (see fig\ref{ring_geom_exp}). Along path-1, a flux is accrued by FS electrons via the Wilson loop operator $\hat{\bar{Z}}^{\lambda/2}$, leading to the modified S-matrix 
\begin{eqnarray}
S_{F_{1}}(\lambda /2)=\hat{\bar{Z}}^{\lambda/2}_{c/s}S_{F_{1}}\hat{\bar{Z}}^{\dagger\lambda/2}_{c/s}.
\end{eqnarray}
This leads to a scattered two-particle wavefunction: $|\psi^{F}_{c/s}(\lambda/2)\rangle=S_{F_{1}}(\lambda/2)|\uparrow_{Fa}\downarrow_{Fb}\rangle$. Along path-2, the modified S-matrix is 
\begin{equation}
S_{F_{1}}(-\lambda /2)=\hat{\bar{Z}}^{\dagger\lambda/2}_{c/s}S_{F_{1}}\hat{\bar{Z}}^{\lambda/2}_{c/s}~,
\end{equation}
and the scattered state is given by $|\psi^{F}_{c/s}(-\lambda/2)\rangle =S_{F_{1}}(-\lambda/2)|\uparrow_{Fa}\downarrow_{Fb}\rangle$. Superposing these two scattered wavefunctions gives the total wavefunction as $|\Psi^{F}_{c/s}(\lambda)\rangle =|\psi^{F}_{c/s}(\lambda/2)\rangle +|\psi^{F}_{c/s}(-\lambda/2)\rangle$, and leads to the following two-particle interference pattern for spin/charge currents carried by the pseudospin composite degrees of freedom (see Sec.\ref{sec3})
\begin{eqnarray}
\langle\Psi^{F}_{c/s}(\lambda)|\Psi^{F}_{c/s}(\lambda)\rangle &=& 2+2\text{Re}\langle \psi^{F}_{c/s}(-\lambda /2)|\psi^{F}_{c/s}(\lambda/2)\rangle)~,~\nonumber\\
						 &=&2+2|r|^{2}+2|t|^{2}\cos\gamma\lambda ~.~	\label{two-electron-interference}
\end{eqnarray}
We can now define a moment generating function, $M^{F}_{c/s}(\lambda)$, containing the FCS for the two-particle back-scattering at the FS
\begin{eqnarray}
\hspace*{-0.3cm}M^{F}_{c/s}(\lambda)&=&\langle \psi^{F}_{c/s}(-\lambda/2)|\psi^{F}_{c/s}(\lambda/2)\rangle =\langle\psi^{F}_{c/s}(0)|\hat{\bar{Z}}^{\lambda}_{s/c}|\psi^{F}_{c/s}(0)\rangle ~.
\end{eqnarray}  
The real part of the moment generating function, $Re(M^{F}_{c/s}(\lambda))$, can be determined from the two-electron interference pattern (eq.\eqref{two-electron-interference}). Alternatively, the quantity $\langle\Psi^{F}_{c/s}(\lambda)|\Psi^{F}_{c/s}(\lambda)\rangle$ is a measure of the expectation value of the charge operator 
\begin{eqnarray}
\hat{Q}^{F}_{c/s}(\lambda)&=&\frac{e}{S}\Psi^{F\dagger}_{c/s}(\lambda)\Psi^{F}_{c/s}(\lambda)~,~(S=1/2)\nonumber\\
\langle\hat{Q}^{F}_{c/s}\rangle &=&Re(M^{F}_{c/s}(\lambda)) =2|r|^{2}+2|t|^{2}\cos\gamma\lambda~,~
\end{eqnarray}
where $\Psi^{F\dagger}_{c/s}(\lambda)$ is given by
\begin{equation}
\Psi_{Fc/s}^{\dagger}(\lambda)=\hat{S}_{F_{1}}(\lambda/2)A^{+}_{Fa}+\hat{S}_{F_{1}}(-\lambda/2)A^{+}_{Fa}~,
\end{equation}
and, as shown earlier,  $A^{+}_{Fa}=c^{\dagger}_{k_{F}\uparrow}c^{\dagger}_{\pm k_{F}\downarrow}~,~A^{+}_{Fb}=c^{\dagger}_{-k_{F}\uparrow}c^{\dagger}_{\mp k_{F}\downarrow}$~.
Similarly, $Im(M^{F}_{c/s}(\lambda))$ is evaluated from the expectation value of the composite current operator $\langle\hat{I}_{c/s}(\lambda)\rangle$ defined as 
\begin{eqnarray}
\hat{I}^{F}_{c/s}(\lambda)&=&\frac{e}{S}\left[\psi^{F\dagger}_{c/s}(\frac{\lambda}{2})\psi^{F}_{c/s}(-\frac{\lambda}{2})-\psi^{F\dagger}_{c/s}(-\frac{\lambda}{2})\psi^{F}_{c/s}(\frac{\lambda}{2})\right]~,\nonumber\\
\langle\hat{I}^{F}_{c/s}(\lambda)\rangle &=&Im(M^{F}_{c/s}(\lambda)) =2|t|^{2}\sin\gamma\lambda~.
\end{eqnarray}
Thus, in terms of the charge ($\langle\hat{Q}_{Fc/s}(\lambda)\rangle$) and current ($\langle\hat{I}_{Fc/s}(\lambda)\rangle$) observables, the two-particle FCS generating function can be written as
\begin{eqnarray}
M^{F}_{c/s}(\lambda)=\langle\hat{Q}^{F}_{c/s}(\lambda)\rangle +i\langle\hat{I}^{F}_{c/s}(\lambda)\rangle~.~
\end{eqnarray}
A cumulant generating function can now be constructed from the moment generating function: $G^{F}(\lambda)=\ln M^{F}_{c/s}(\lambda)$~. By varying the counting field $\lambda$, we can generate all cumulants of the spin or charge distributions at the FS. Further, using the Klich-Levitov formula~\cite{klich2009quantum} together with eq.\eqref{scattering_geometric_phase}, the even cumulants yield the entanglement entropy $S_{FS}$ between the helicity/chirality sectors (see Sec.(\ref{sec3}))
\begin{eqnarray}
S_{FS}&=&\sum_{m>0}\frac{\alpha_{2m}}{(2m)!}\frac{d^{2m}}{d\lambda^{2m}}G^{F}(\lambda)|_{\lambda =0}\nonumber\\
	  &=&-\frac{\gamma_{s}}{2\pi}\ln\frac{\gamma_{s}}{2\pi} -(1-\frac{\gamma_{s}}{2\pi})\ln(1-\frac{\gamma_{s}}{2\pi})\nonumber\\
	  &=&-\ln\frac{\gamma}{2\pi}+\frac{\gamma}{2\pi}\cos(\delta_{s}-\delta_{t})\ln\tan\frac{\delta_{s}-\delta_{t}}{2}~,\label{FCS_entropy}	
\end{eqnarray}
where we have used eq\eqref{scattering_geometric_phase} in the final line. The first piece of this entanglement entropy is purely topological, i.e., $S_{top}=-\ln\frac{\gamma}{2\pi}=\ln 2$ depends purely on the first Chern class $\gamma$ and accounts for dimension  of PHS $F_{1}$ ($d=2$). We will see in the next section that this topological piece of the entanglement entropy ($S_{top}$) reappears in the c.o.m. Hilbert space. The second piece ($S_{dyn}$) is a function of the pseudospin singlet and triplet scattering phase shifts, and is dependent on the geometry of the PHS $F_{1}$. This reflects on the additional entanglement content of the many-body wavefunction arising from the dynamical spectral weight transfer. This second piece renormalizes incrementally as the instability hits the Fermi surface ($\gamma_{b}=\gamma$, and $d|J|/d\ln\lambda >0$), leading to a value at the IR stable fixed point ($J^{*}=4$) given by  
\begin{equation}
S_{dyn}^{*}=\frac{\gamma}{2\pi}\cos(\phi)\ln\lbrace\tan\frac{\phi}{2}\rbrace~,~\phi =\tan^{-1}(3\pi)-\tan^{-1}(-\pi)~.
\end{equation}
Note that $S_{FS}\to 0$ as $\delta_{s}=\pi/2=-\delta_{t}$ (unitary limit). Interestingly, for $J=4/(\sqrt{3}\pi)$, $\delta_{s}-\delta_{t}=\pi/2$, such that $S_{FS}=-\ln\frac{\gamma}{2\pi}$ is purely topological. In this way, a measurement of the FCS should enable that of the associated many-particle entanglement entropy~\cite{klich2009quantum}.
\section{Topological order and its observables}\label{sec8}
\pin
We begin by clarifying how the c.o.m. Hilbert space topology is shaped by the instability. Recall that for $J_{||}<0$, $\text{sgn}(r)$ jumps across the $SU(2)$ WZNW lines from $-1$ to $+1$, together with a jump in the Berry phase $\gamma_{b}=(\gamma/2)(1+\text{sgn}(r))$ from $0\rightarrow \pi$ (eq.\eqref{berry_phase_dirac}). These changes are associated with a modification of the c.o.m Hamiltonian from $H^{0}_{cm}=P_{cm}^{2}/2M$ to
\begin{equation}
H_{cm}=\frac{P_{cm}^{2}}{2M} +\Delta E_{s/c}\cos\left(\frac{2\pi}{\gamma} X_{c/s}\right)~,\label{comHam}
\end{equation}
where $M$ is the total mass of the system and the cosine potential arises from the pseudospin-flip operation (eq.\eqref{spin-flip operation}). This changes the gauge symmetry of the state manifold from $Z_{N}\simeq S_{1}$~(in the thermodynamic limit $N\rightarrow \infty$) to $Z_{2}$ (where $Z_{2}$ is the $X_{c/s}:0\rightarrow \pi$ symmetry of the cosine potential). The translation- and gauge-invariant wavefunction basis $\mathcal{B}_{\gamma_{b}=0}$ for the metal (described by $H^{0}_{cm}$) and basis $\mathcal{B}_{\gamma_{b}=\pi}$ for the (spin/charge) gapped insulator (described by $H_{cm}$) are given by
\begin{eqnarray}
\mathcal{B}_{0}&=&\lbrace |k\rangle =\frac{1}{\sqrt{N}}\sum_{j}e^{ikj}|X_{c/s}=j\rangle , k=\frac{2\pi j}{N}\rbrace~,~
\mathcal{B}_{\pi} = \lbrace |P_{cm}=0,\pi\rangle\rbrace\nonumber\\
 |P_{cm}=0,\pi\rangle &=&\sqrt{\gamma/ 2\pi}\left[|X_{c/s}=0\rangle \pm |X_{c/s}=\gamma\rangle\right]\rbrace~,\label{c.o.m basis states}
\end{eqnarray}
where $|P_{cm}=0,\pi\rangle$ represents the two degenerate states in c.o.m spectrum. That this degeneracy is not lifted due to tunneling (via the kinematic term $P_{cm}^{2}/2M$) arises from the destructive interference in the manifold of quantum states within $\mathcal{B}_{\pi}$ (eq.\eqref{destructive-interference}). The Hamiltonian $H_{cm}$ projected onto the basis $\mathcal{B}_{\pi}$ vanishes, where the projection operator $P_{\pi}=|P_{cm}=0\rangle\langle P_{cm}=0|+|P_{cm}=\pi\rangle\langle P_{cm}=\pi|$. Thus, $P_{\pi}H_{cm}P_{\pi}=0$ leads to purely topological dynamics governed by a $0+1D$ WZNW term (of the form shown in eq.\eqref{chern}) with Chern coefficient $C_{cm}=1$ for $\gamma=\pi$. The two-fold ground state degeneracy ($d=2$) is directly related to the Chern coefficient as $d=C+1$. We will shortly see how this establishes topological order. 
\\
\par\noindent{\bf \textit{Two spectral gaps and topological order}}
\pin
We have seen already that the $\Theta$-term, $\gamma_{b}=\gamma$, leads to an increasing $J_{\perp}$ coupling under RG, eventually ending at an asymptotically safe stable fixed point $J_{||}(\Lambda_{n*})=4$ in the IR. The fixed point theory resides in the low-energy window $L_{n^{*}}:[-\Lambda_{n*},\Lambda_{n*}]$ with a width given by the many-body gap ($\Delta E_{MB}$)
\begin{eqnarray}
\Delta E_{MB}=2\Lambda_{n*}= 2\Lambda_{0}\left(\frac{J^{0}_{||}}{4}\right)^{1/4}\exp\left[\frac{1}{4}- \frac{1}{J^{0}_{||}}\right]~,\label{non-perturbative_gap}
\end{eqnarray}
where $J^{0}_{||}$ is the bare coupling. This gap is the analogue of that obtained from the one-loop BKT RG for the 1D superconducting Luther-Emery liquid and 1D Mott insulator~\cite{giamarchi2004quantum,gogolin2004bosonization}. The asymptotically free nature of the one-loop RG leads, in contrast, to an exponentially small gap $\Delta E^{\text{1-loop}}_{MB}\sim\exp[-1/J^{0}_{||}]$~. 
\par\noindent
Within the window $L_{n^{*}}$, the FS pseudo-spin degrees of freedom ($F$) undergo backscattering and lead to an \textit{additional (spin/charge) gap} between the two states $|\mathbf{A}_{a}+\mathbf{A}_{b}=(1,0)\rangle =1/\sqrt{2}\left[|\uparrow_{a}\downarrow_{b}\rangle \pm |\downarrow_{a}\uparrow_{b}\rangle\right]$ around the Fermi energy $E_{F}=0$. In the previous section, we have already obtained the effective Hamiltonians for the LE and MI phases. From there, we determine the spin/charge gap in ($\Delta E_{s/c}=J^{*}_{\perp}/N_{\Lambda^{*}}$) between the pseudospin singlet and triplet states ($|\mathbf{A}_{\Lambda_{n*}a}+\mathbf{A}_{\Lambda_{n*}b}|^{2}=0,2$) as 
\begin{equation}
\Delta E_{s/c}= \left(\frac{4}{J^{0}_{||}}\right)^{1/4}\frac{(\sqrt{J^{*}_{||})^{2}+r^{2}}}{2\Lambda_{0}}\exp\left[\frac{1}{J^{0}_{||}}-\frac{1}{4}\right],
\end{equation}
where $J^{*}_{||}$ is determined from the Table.\ref{table-coup} for the MI and LE phases. The spin/charge gap at the intermediate-coupling fixed point ($J_{n^{*}||}=1/S^{2}$) is equivalent to the gap between the states $|\mathbf{A}_{a}+\mathbf{A}_{b}=1\rangle$ and $|\mathbf{A}_{a}+\mathbf{A}_{b}=0\rangle$ at the FS, and can be attributed to the fact that the Hamiltonians eq.\eqref{FP_anisotropic} and eq.\eqref{anisotropic_form} possess the same couplings. Following Ambegaokar et al.~\cite{ambegaokar1980dynamics}, we can now define the coherence length $\xi_{coh}$ for the BKT transition in terms of the RG invariant 
\begin{eqnarray}
\xi_{coh}(J_{\perp}^{*}) =l_{c}e^{1/\sqrt{r}}=l_{c}e^{(16-J_{\perp}^{*2})^{-1/4}}~,
\end{eqnarray} 
where $l_{c}$ is the vortex core size and we have used the relation between the RG invariant $r$ and the final fixed point coupling $J^{*}_{\perp}$.
\begin{figure}[h!]
\includegraphics[width=\textwidth]{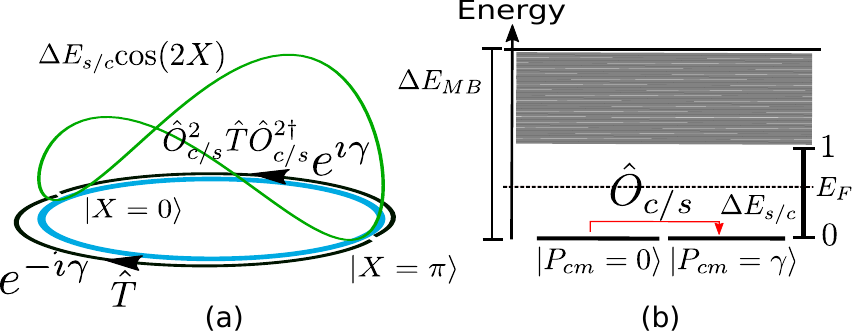}
\caption{Constructive interference phenomenon for scattering processes on the center of mass Hilbert space described by basis $\mathcal{B}_{\pi}$ in eq.\eqref{c.o.m basis states} (left). The interference shows up between two opposing trajectories involving the twist ($\hat{O}^{2}$) and translation ($\hat{T}$) operations of the c.o.m. position $X$, with phases $\pm\gamma$ respectively. This leads to a \textit{spin/charge gap} ($\Delta E_{s/c}$) around the Fermi energy $E_{F}$, along with the standard many-body gap $\Delta E_{MB}$ demarcating the low energy window $L_{n^{*}}$ (right). The green curve within the left figure corresponds to the interaction induced potential ($\Delta E_{s/c}\cos(2X)$) within the c.o.m. Hamiltonian. See main text for discussions.}\label{topological_order}
\end{figure}
\par\noindent  
The modified c.o.m PHS has a emergent $Z_{2}$ symmetry and associated doubly-degenerate ground states 
\begin{eqnarray}
|P_{cm}=0,A_{a+b}=0\rangle && ~,~
\hat{O}_{c/s}|P_{cm}=0,A_{a+b}=0\rangle = |P_{cm}=\gamma,A_{a+b}=0\rangle~,\nonumber\\
\hat{O}_{c/s}^{2}|P_{cm}=0,A_{a+b}=0\rangle &=& |P_{cm}=2\gamma,A_{a+b}=0\rangle~,~~~~~
\end{eqnarray}
separated from the lowest-lying excited states with $|A_{a+b}=1\rangle$ by the (spin/charge) gap $\Delta E_{s/c}$ given above. We recall that the commutation relation between the twist ($\hat{O}_{c/s}$) and translation ($\hat{T}$) operators was shown in Sec.(\ref{sec2}) to be
\begin{eqnarray}
\hat{O}_{c/s}\hat{T}=\hat{T}\hat{O}_{c/s}e^{\gamma[X_{c/s},P_{cm}]}~,
\end{eqnarray}
where $[X_{c/s},P_{cm}]=i~,~\gamma =\pi$. This leads to destructive interference in a multiply-connected geometry ($S^{1}\equiv [0,2\pi)$) involving path-1 ($X_{c/s}=[0,\pi]$ traversed in clockwise sense) and path-2 ($X_{c/s}=[0,\pi]$ traversed in anti-clockwise sense ) between the two degenerate states $|\Psi_{1}\rangle=|X_{c/s}=0,A_{a+b}=0\rangle$  and $|\Psi_{2}\rangle=|X_{c/s}=\pi,A_{a+b}=0\rangle$,
\begin{eqnarray}
|\Psi_{1}|^{2}+|\Psi_{2}|^{2}+2|\Psi_{1}||\Psi_{2}|\cos\gamma =0~,\label{destructive-interference}
\end{eqnarray}
where $\Psi_{1}=e^{i\gamma}\Psi_{2}$. The Aharonov-Bohm phase $e^{i\gamma}$ arises from charge $e/2$ objects encircling the Dirac string in the FS PHS.  Indeed, the finding of topological degeneracy with a gap $\Delta E_{s/c}$ for $\gamma=\pi$ corresponds to one of the possible conclusions of the LSM-type criterion~\cite{lieb1961two} discussed in Sec.\ref{sec4}. The zero mode of the square-root of the vertex operator $\hat{O}_{c/s}^{2}$~\cite{nakamura2002lattice} is a nonlocal gauge transformation with an associated $Z_{2}$ Wilson loop given by eq.\eqref{wilson_loop_operator}. This Wilson loop operator reveals the presence of a vortex condensed topological state of matter at the intermediate coupling fixed point $J_{\perp}^{*}$~\cite{fradkin1979phase,hansson2004superconductors}. The modified Hilbert space $\mathcal{B}_{\pi}$, topological degeneracy, topological excitations and \textit{spin/charge} gap are depicted pictorially in Fig.(\ref{topological_order}).
On the other hand, for $\gamma_{b}=0$, the gapless TLL metal corresponds to the second conclusion of the LSM-type criterion (see Sec.\ref{sec4}). 
\\
\par\noindent{\bf \textit{Observables for changes in Hilbert space geometry and topology}}
\pin
We now identify observables associated with the c.o.m twist operator $\hat{O}_{c/s}$ (eq.\eqref{twist_op_s_c}) that probe changes in Hilbert space geometry and topology across the RG phase diagram. The modification in c.o.m Hilbert space as the sign of the RG invariant $\text{sgn}(r)$ changes from $-1$ to $+1$ can be tracked by the cumulants $\mathcal{C}_{n}$ of the c.o.m. position operator $X_{c/s}$. The cumulants $\mathcal{C}_{n}$ are defined as the $n$th derivative of the gauge-invariant cumulant generating function $G(\lambda)$
\begin{eqnarray}
G(\lambda)=\ln\langle\hat{O}^{\lambda}_{s/c}\rangle ~,~ \mathcal{C}_{n}=\frac{(-i)^{n}}{(2\pi)^{n}}\lim_{L\rightarrow\infty}\frac{d^{n}}{d\lambda^{n}}G(\lambda)|_{\lambda =0}~.\label{cumulants}
\end{eqnarray} 
The first cumulant, 
\begin{equation}
\mathcal{C}_{1}=\frac{\gamma_{b}}{2\pi}=\frac{\gamma}{4\pi}(1+sgn(r))sgn(J_{||})
\end{equation}
is related to the $\Theta$-term associated with the Dirac string in the PHS~\cite{souza2000polarization,nakamura2002lattice}. Now, for $r>0$, as $J_{||}$ changes sign, $\gamma_{b}:\gamma \rightarrow -\gamma$~. Nelson and Kosterlitz~\cite{nelson1977} have shown that for crossover RG flows $r>0$, the superfluid stiffness $\chi$ for the classical 2D XY model is related to the coupling $J_{||}$ in the Hamiltonian eq.\eqref{FP_anisotropic}
\begin{equation}
\chi=\frac{m^{2}k_{B}T}{\hbar^{2}\rho_{s}}=\frac{\pi}{\pi J_{||}-2}~.
\end{equation} 
They then employ the asymptotically free nature of the weak coupling perturbative RG for the BKT transition to show that for $r>0$, $J_{||}: 0\to \infty$, there exists a universal jump in the superfluid stiffness across the transition given by $\Delta \chi =\chi(J_{||}=0)-\chi(J_{||}\to\infty)=\pi/2$~\cite{nelson1977}. Our asymptotically safe RG equations eq.\eqref{anisotropic_RG_eqns} show that for $r>0$, $J_{||}: 0\to 4\pi^{2}/\gamma^{2}=4=1/S^{2}$ and $\Delta \chi =\chi(J_{||}=0)-\chi(J_{||}=1/S^{2})=\pi/(4\pi -2)$~. In this light, the asymptotic safety of our RG results arise from the intermediate coupling fixed point at $S=1/2$ discussed earlier and are indicative of fermionic criticality, while the asymptotically free BKT RG equations are obtained for the limit of $S\to 0$. Importantly, we see that the crossover RG flow is also associated with a jump in the Berry phase $\Delta\gamma_{b}=2\pi$,  as well as a jump in the first cumulant $\mathcal{C}_{1}$
  \begin{eqnarray}
  \Delta \mathcal{C}_{1}=\frac{\Delta\gamma_{b}}{2\pi}=1~.
  \end{eqnarray}
 \pin
 The second cumulant $\mathcal{C}_{2}$ is related to the localization length ($\xi$)~\cite{souza2000polarization} 
\begin{equation}
\mathcal{C}_{2}=\xi^{2}=\langle\hat{X}^{2}_{c/s}\rangle -\langle\hat{X}_{c/s}\rangle^{2}~.
\end{equation} 
For $J_{||}<0$, the change in the localization length is $\xi: \infty \to (\gamma/2\pi)$, as the change in $sgn(r)$ depicts the transition from a metal to a (spin/charge) gapped state. The localization length $\xi$ has an interpretation as a geometric distance on the c.o.m. Hilbert space. To see this, we define the quantum metric of the c.o.m. Hilbert space as
\begin{equation}
g_{\lambda\lambda}=\langle \partial_{\lambda}\Psi(\lambda)|\partial_{\lambda}\Psi(\lambda)\rangle -|\langle\Psi(\lambda)|\partial_{\lambda}\Psi(\lambda)\rangle |^{2}~,
\end{equation} 
where $|\Psi(\lambda)\rangle \in \hat{O}\mathcal{B}_{\pi}$ or $|\Psi(\lambda)\rangle \in \hat{O}\mathcal{B}_{0}$. By performing an average over the gauge $\lambda$, we obtain   
\begin{eqnarray}
d\xi^{2}= \frac{1}{4\pi^{2}}d\lambda^{2}g_{\lambda\lambda}  \Rightarrow \xi =\frac{1}{2\pi}\int_{0}^{2\pi}d\lambda \sqrt{g_{\lambda\lambda}}~.
\end{eqnarray}
We can now relate the imaginary part of the (spin/charge) conductivity at zero frequency $\lim_{\omega \to 0}\text{Im}~\sigma_{s/c}(\omega)$  to the fluctuations encoded in the quantum metric~\cite{provost1980riemannian} $g_{\lambda\lambda}$ via the fluctuation-dissipation and Kramers-Kronig relations 
\begin{eqnarray}
g_{\lambda\lambda}=\lim_{\omega \rightarrow 0}\text{Im}~\sigma_{s/c}(\omega)=\int_{0}^{\infty} d\omega' \frac{\text{Re}~\sigma_{s/c}(\omega')}{\omega'}~.
\end{eqnarray}
Thus, the drastic jumps in both $\xi$ and Im~$\sigma_{s/c}(\omega =0)$ are both topological in origin.
The cumulant generating function $G(\lambda)$ in the (spin/charge) gapped phase $J_{||}<0,r<0$ is given by $G(\lambda)=\ln(\gamma /2\pi)+\ln(1+e^{i2\gamma\lambda})$. In the light of the arguments presented above, we can conclude that all cumulants ($M_{n}$) are functions of the quantum metric $g_{\lambda\lambda}$ and Berry phase $\gamma$ of the c.o.m. Hilbert space. Further, they will all show universal jumps across the transition.
\\
\par\noindent{\bf \textit{Entanglement entropy and noise in the center of mass Hilbert space}}
\pin
The center of mass Hilbert space $\mathcal{H}_{cm}$ for $\gamma_{b}=\gamma$ is spanned by a $SU(2), S=1/2$ representation basis $\mathcal{B}_{\pi}$ (eq.\eqref{c.o.m basis states}) embedded on a circle ($S_{1}$) $S_{cm}:X_{c/s}=[0,2\pi)$. In order to compute the entanglement entropy, the circle $S_{cm}$ is cut at diametrically opposite points
\begin{eqnarray}
S_{cm}&=&S_{1cm}\oplus S_{2cm}~,~
S_{1cm}=\lbrace \pi/2, \ldots, 0, \ldots, 3\pi/2]~,~
S_{2cm}=[\pi/2, \ldots, \pi, \ldots, 3\pi/2]~.~~~~~
\end{eqnarray}
This results in a surgery of the c.o.m. Hilbert space $\mathcal{H}_{cm}\equiv SU(2)$ into $\mathcal{H}_{1}\otimes\mathcal{H}_{2}$, where
\begin{eqnarray}
&&\mathcal{H}_{1}:\mathcal{B}_{1}=\lbrace |0_{X_{1}=0}\rangle , |1_{X_{1}=0}\rangle , X_{1}\in S_{1cm}\rbrace ~,~
\mathcal{H}_{2}:\mathcal{B}_{2}=\lbrace |0_{X_{2}=\pi}\rangle , |1_{X_{2}=\pi}\rangle , X_{2}\in S_{2cm}\rbrace~,~~~~~ 
\end{eqnarray} 
where $|0_{0,\pi}\rangle, |1_{0,\pi}\rangle$ are labelled by eigenvalues of the number operators $\hat{n}_{0}$and $\hat{n}_{\pi}$. As both $\mathcal{H}_{1}$ and $\mathcal{H}_{2}$ represent two-level systems ($SU(2),S=1/2$), the combined Hilbert space $\mathcal{H}_{1}\otimes \mathcal{H}_{2}$ is isomorphic to $SU(2)\otimes SU(2)$. The c.o.m. coordinate in $\mathcal{H}_{1}\otimes\mathcal{H}_{2}$ is represented as $\hat{X}_{c/s}=\hat{n}_{0}X_{1}+\hat{n}_{\pi}X_{2}=\gamma\hat{n}_{\pi}$. Within this formalism, the basis $\mathcal{B}_{\pi}\subset\mathcal{B}_{1}\times\mathcal{B}_{2}$ is subject to the constraint $\hat{n}_{0}+\hat{n}_{\pi}=1$~. The basis $\mathcal{B}_{\pi}$ can then be represented in terms of entangled states
\begin{equation}
\hspace*{-0.2cm}|P_{cm}=0,\pi\rangle =\sqrt{\frac{\gamma}{2\pi}}|1_{X_{1}}0_{X_{2}}\rangle \pm \sqrt{\frac{2\pi-\gamma}{2\pi}}|0_{X_{1}}1_{X_{2}}\rangle~.~\label{basis}
\end{equation} 
Recall that $\gamma =\pi$ is the first Chern class on the c.o.m. torus $\mathcal{T}^{2}$ (eq.\eqref{1stchernclass}). 
Now, by partial tracing over $\mathcal{H}_{2}$, the reduced density matrix for $\mathcal{H}_{1}$ is given by 
\begin{eqnarray}
\rho_{\mathcal{H}_{1}}=\frac{\gamma}{2\pi}|0_{X_{1}}\rangle\langle 0_{X_{1}}|+\left(1-\frac{\gamma}{2\pi}\right)|1_{X_{1}}\rangle\langle 1_{X_{1}}|~.
\end{eqnarray}
The entanglement entropy, $S=-\text{Tr}(\rho_{\mathcal{H}_{1}}\ln\rho_{\mathcal{H}_{1}})$, is thus dependent on the c.o.m. Hilbert space topology and is given by
\begin{eqnarray}
S=-\frac{\gamma}{2\pi}\log\frac{\gamma}{2\pi}-\left(1-\frac{\gamma}{2\pi}\right)\log\left(1-\frac{\gamma}{2\pi}\right)~.~
\end{eqnarray}
Clearly, the entanglement entropy is $S_{top}=\ln 2$~\cite{ding2009entanglement} for $\gamma=\pi$, where the $2$ within the log refers to the two-fold topological degeneracy ($d=2$) on the torus $\mathcal{T}_{2}$ observed earlier. We recall that the same topological entanglement entropy was derived in an earlier section by the scattering dynamics at the Fermi surface. As seen earlier, there exists a relation~\cite{klich2009quantum} between the even cumulants $\mathcal{C}_{2m}$ in eq.\eqref{cumulants} (i.e. fluctuation associated with the center of mass position vector) and the entanglement entropy 
\begin{equation}
S_{top}=\sum_{m>0}\frac{\alpha_{2m}}{(2m)!}\mathcal{C}_{2m}~,
\end{equation}
where $\alpha_{2m}=(2\pi)^{2m}|B_{2m}|$, and $B_{2m}$ are the Bernoulli numbers. The cumulant generating function (eq.\eqref{cumulants}) can therefore be written in terms of basis states of eq.\eqref{basis} 
\begin{eqnarray}
G(\lambda)=\ln\langle\exp\left[i\lambda\gamma\hat{n}_{\pi}\right]\rangle~.
\end{eqnarray}
This shows that changing boundary conditions via operator $\hat{O}^{\lambda}$ tracks the number fluctuations associated with the state of the c.o.m. position $X_{c/s}=\pi$.
\section{Discussions and Outlook}\label{sec9}
\noindent
In summary, we have applied the unitary renormalization group (URG) technique to the problem of interacting electrons in 1D, unveiling thereby the role played by emergent topological features in guiding the RG flow towards critical (TLL) and stable (LE and MI) fixed points. While the nature of critical RG flows comprising the BKT RG phase diagram is preserved, the nonperturbative nature of the URG formalism shows the emergence of non-Abelian constraints leading to a family of gapped intermediate coupling fixed points. We also perform a quantum mechanical analysis of scattering processes at the Fermi surface, and compute topological terms induced by interactions. In doing so, we obtain a skeletal phase diagram that shares the essential features characteristic of the BKT phase diagram. This skeletal phase diagram highlights regions possessing different symmetries, and which are separated by transitions involving changes in topological quantities related to the projected Hilbert space at the Fermi surface. In this way, we observe an interaction driven Fermi surface topology-changing (Lifshitz) transition of the 1D metal across the critical point, similar to the transitions observed in the 2D Hubbard~\cite{anirbanmotti,anirbanmott2} and other models of correlated lattice electrons~\cite{anirbanurg1,anirbanurg2}. Given the simplicity of the two-point Fermi surface at hand, we expect that these insights into the nature of fermionic criticality are universal, i.e., they shed light also on the instabilities of regular connected~\cite{anirbanurg2}, as well as Dirac point-like~\cite{pal2019}, Fermi surfaces in higher dimensions. In this sense, fermionic criticality appears to be shaped by the global (topological) features of the Fermi surface.  
\par\noindent
The many-particle entanglement renormalization comprising the EHM tensor network show distinct features for the critical and gapped phases. Thus, the topological phase transition across the critical point of the BKT phase diagram also corresponds to a entanglement phase transition. By verifying the Ryu-Takayanagi holographic entropy bound for the EHM network, we demonstrate the emergence of two distinct spacetimes across the critical point in the dual gravity theory (i.e., corresponding to the gapless TLL and gapped LE/MI phases respectively), such that the critical point appears to act as a horizon in the holographic dual spacetime generated from entanglement RG flow~\cite{lee2016horizon}. We also demonstrate the important role played by the degrees of freedom at the Fermi surface in guiding the entanglement RG flows from UV to IR for both the gapped and gapless phases. It is tempting to speculate that these are again universal findings, and will be explored in future works.
\par\noindent
The MI and LE phases emerge respectively from the charge gapping and spin gapping of the Fermi surface and its neighbourhood. We find that these phases possess the essential features of topological order: ground state degeneracy (for the system placed on a periodic manifold) protected by a many body gap, charge fractionalization etc. An important feature of emergent phases is the phenomena of dynamical spectral weight transfer between the emergent pairs of electronic states and the fundamental electrons. Using the URG analysis of the scattering phase shift, we compute nonperturbatively the net Friedel phase shift and quantify the dynamical spectral weight transferred from UV to IR. Finally, by treating the degrees of freedom at the two-point Fermi surface as a ``quantum impurity" coupled to other (``bath") degrees of freedom at higher energies, we present a scattering matrix thought-experiment for the topological and entanglement features of the Fermi surface. In this way, we track the entanglement entropy generated by isolating the Fermi surface from the rest of the system. We also suggest ways of measuring this entanglement entropy by studying the quantum noise and higher order cumulants generated by two-electron scattering between the Fermi surface and the excitations. An experimental verification of the results obtained from our gedanken would pave the way towards systematic studies of many-particle entanglement and the Fermi surface.
\acknowledgments
The authors thank R. K. Singh, A. Dasgupta, A. Bhattacharya, A. Taraphder, N. S. Vidhyadhiraja, P. Majumdar, S. Jalal and M. Patra for several discussions and feedback. A. M. and S. P. thank the CSIR, Govt. of India for funding through research fellowships. S. L. thanks the DST, Govt. of India for funding through a Ramanujan Fellowship during which a part of this work was carried out.
\appendix
\section{RG equation for the BCS instability}\label{BCS_RG}
Starting from the Hamiltonian RG flow eq.\eqref{Hflow}, and accounting only for two-particle vertices in the Hamiltonian given in eq.\eqref{H-BCS}, we obtain the renormalisation of the Hamiltonian as
\begin{eqnarray}
\Delta H_{(j)} &=& \sum_{\kappa,\kappa_{1},p,\eta}\bigg[c^{\dagger}_{\kappa(p,\eta)}c^{\dagger}_{\kappa'(p,\eta)}c_{\kappa'_{1}(p,\eta)}c_{\kappa_{1}(p,\eta)}\frac{(K^{(j)}(p))^{2}\tau_{\kappa_{j}
(p,-\eta)}\tau_{\kappa'_{j}
(p,-\eta)}}{[G_{j, p}]^{-1}-V^{(j)}(p)\tau_{\kappa_{j}
(p,-\eta)}\tau_{\kappa'_{j}
(p,-\eta)}}\nonumber\\
&+&c^{\dagger}_{\kappa(p,\eta)}c^{\dagger}_{\kappa'(p,\eta)}c_{\kappa'_{1}(p,-\eta)}c_{\kappa_{1}(p,-\eta)}\frac{K^{(j)}(p)V^{(j)}(p)\tau_{\kappa_{j}
(p,-\eta)}\tau_{\kappa'_{j}
(p,-\eta)}}{[G_{j, p}]^{-1}-V^{(j)}(p)\tau_{\kappa_{j}
(p,-\eta)}\tau_{\kappa'_{j}
(p,-\eta)}}\bigg]~.\label{vertex_H_Flow}
\end{eqnarray}
Here, 
\begin{eqnarray}
G_{j, p}^{-1}=\omega - \epsilon_{\kappa_{j}}\tau_{\kappa_{j}
(p,-\eta)}-\epsilon_{\kappa'_{j}}\tau_{\kappa'_{j}
(p,-\eta)}~.
\end{eqnarray}
Importantly, we note that in obtaining the denominator of the above RG equation, we have limited the study of the quantum fluctuation scale $\hat{\omega}_{(j)}$ to only the one-particle dispersion
\begin{eqnarray}
\hat{\omega}_{(j)}&=&\sum_{i}(\epsilon_{i}+\Sigma^{(j-1)}_{i})\left(\hat{n}_{i}-\frac{1}{2}\right)+\sum_{a,b}\Gamma^{4,(j-1)}_{a,b}\left(\hat{n}_{a}-\frac{1}{2}\right)\left(\hat{n}_{b}-\frac{1}{2}\right)+\ldots+\Delta H^{X}_{(j)}\nonumber\\
&\approx &\sum_{i}(\epsilon_{i}+\Sigma^{(j-1)}_{i})\left(\hat{n}_{i}-\frac{1}{2}\right)~.
\end{eqnarray}
We interpret the various quantum fluctuation eigenvalues $\omega$ of the operator $\hat{\omega}_{(j)}$ as an inherent quantum parameter arising out of the renormalization of the interaction vertices, and study thereby the RG equations as a function of $\omega$.
\par\noindent
The vertex RG flow equations for the forward ($K$) and backscattering ($V$) vertices are then obtained as
\begin{eqnarray}
\Delta K^{(j)}(p)&=&\frac{4K^{(j)}(p)V^{(j)}(p)\tau_{\kappa_{j}
(p,-\eta)}\tau_{\kappa'_{j}
(p,-\eta)}}{[G_{j, p}]^{-1}-V^{(j)}(p)\tau_{\kappa_{j}
(p,-\eta)}\tau_{\kappa'_{j}
(p,-\eta)}}~,\nonumber\\
\Delta V^{(j)}(p)&=&\frac{4(K^{(j)}(p))^{2}\tau_{\kappa_{j}
(p,-\eta)}\tau_{\kappa'_{j}
(p,-\eta)}}{[G_{j, p}]^{-1}-V^{(j)}(p)\tau_{\kappa_{j}
(p,-\eta)}\tau_{\kappa'_{j}
(p,-\eta)}}~.\label{vertexRGeqnBCS}
\end{eqnarray}

\bibliographystyle{JHEP}
\bibliography{netbib}
\end{document}